\documentclass[11pt]{emulateapj}
\usepackage{natbib}
\usepackage{graphicx}
\usepackage{amsmath}
\usepackage{amssymb}

\def\jgr{J. Geophys. Res. }

\def\ao{Appl. Opt. }
\def\aj{Astron. J.}

\def\chisq{$\chi^2$ }
\def\chisqx{$\chi^2$}

\def\icm{cm$^{-1}$ }
\def\icmx{cm$^{-1}$}

\def\deg{$^\circ$ }
\def\degx{$^\circ$}

\def\mum{$\mu$m }
\def\mumx{$\mu$m}
\def\nht{NH$_3$ }
\def\nhtx{NH$_3$}
\def\nhfsh{NH$_4$SH }
\def\nhfshx{NH$_4$SH}

\begin{document}
\title{The source of widespread 3-\mum absorption
in Jupiter's clouds: Constraints from 2000 Cassini VIMS observations\footnotemark[\dag]} 
\author{L.A. Sromovsky\altaffilmark{1} and P.M. Fry\altaffilmark{1}}
\altaffiltext{1}{University of Wisconsin - Madison, Madison WI 53706}
\altaffiltext{\dag}{Partly based on observations obtained from the
 data archive at the Space Telescope Science Institute. 
STScI is operated by the Association of Universities for Research in
Astronomy, Inc. under NASA contract NAS 5-26555.}
\slugcomment{Journal reference: Icarus 210 (2010) 230-257.}

\begin{abstract}

%\hyphenation{small sig-ni-fi-cant} 
The Cassini flyby of Jupiter in
2000 provided spatially resolved spectra of Jupiter's atmosphere using
the Visual and Infrared Mapping Spectrometer (VIMS).  A prominent
characteristic of these spectra is the presence of a strong absorption
at wavelengths from about 2.9 $\mu$m to 3.1 $\mu$m, previously noticed
in a 3-\mum spectrum obtained by the Infrared Space Observatory (ISO)
in 1996.  While Brooke et al. (1998, Icarus 136, 1-13) were able to
fit the ISO spectrum very well using ammonia ice as the sole source of
particulate absorption, Sromovsky and Fry (2010, Icarus 210,
211-229), using significantly revised \nht gas absorption models,
showed that ammonium hydrosulfide (\nhfshx) provided a better fit to
the ISO spectrum than \nhtx, but that the best fit was obtained when
both \nht and \nhfsh were present in the clouds. Although the large
FOV of the ISO instrument precluded identification of the spatial
distribution of these two components, the VIMS spectra at low and
intermediate phase angles show that 3-\mum absorption is present in
zones and belts, in every region investigated, and both low- and
high-opacity samples are best fit with a combination of \nhfsh and
\nht particles at all locations.  The best fits are obtained with a
layer of small ammonia-coated particles ($r\sim0.3$ $\mu$m) overlying
but often close to an optically thicker but still modest layer of much
larger NH$_4$SH particles ($r\sim 10$ $\mu$m), with a deeper optically
thicker layer, which might also be composed of NH$_4$SH.  Although
these fits put NH$_3$ ice at pressures less than 500 mb, this is not
inconsistent with the lack of prominent \nht features in Jupiter's
longwave spectrum because the reflectivity of the core particles
strongly suppresses the \nht absorption features, at both near-IR and
thermal wavelengths. Unlike Jupiter, Saturn lacks the broad 3-\mum
absorption feature, but does exhibit a small absorption near 2.965
\mumx, which resembles a similar Jovian feature and suggests that both
planets contain upper tropospheric clouds of sub-micron particles
containing ammonia as a minor fraction.

\end{abstract}


\maketitle
\shortauthors{Sromovsky and Fry}
\shorttitle{Jupiter's 3-\mum absorber}

\section{Introduction}

Analysis of Pioneer and groundbased observations of Jupiter,
summarized by \cite{West1986}, led to an expected Jovian cloud
structure that included an upper ammonia cloud layer starting near
700-mb and a putative NH$_4$SH cloud top near 2 bars, which was
thought to be optically thick outside the hot spot regions.
The putative ammonia cloud was thought to have two particle
populations: a vertically compact layer of large particles (of 3 to
100 $\mu$m in radius) and a vertically diffuse component of small
particles ($r\sim 1 \mu$m) extending up to 200-300 mb in low latitude
regions.  However, the more diffuse component should have produced
prominent spectral signatures at 9.4 $\mu$m and 26 $\mu$m, which were
not observed \citep{Orton1982}.  After considering possible masking of
these features by the likely tetrahedral shapes of these particles,
\cite{West1989} concluded that these particles could not be primarily
composed of ammonia ice.  It was also the case that neither Voyager
Infrared Interferometer Spectrometer (IRIS) observations
\citep{Carlson1993neb}, nor microwave observations
\citep{DePater1986Icar}, ever found an ammonia vapor profile that
would support a 700-mb condensation level. Carlson's derived
condensation pressure was closer to 500 mb, while a later analysis of
microwave observations \citep{DePater2001Icar} suggested NH$_3$
condensation near 600 mb.

Only in the last decade or so has there been even a hint of the
spectral signatures expected of NH$_3$ ice clouds. From an analysis of
a 3-$\mu$m absorption anomaly in a central-disk spectrum of Jupiter,
\cite{Brooke1998} inferred the existence of a layer of ammonia ice
particles of 10 $\mu$m in radius, beginning at 550 mb with a scale
height of 30\% of the gas scale height.  The wide field of view
covered by the observation (roughly a quarter of the Jovian disk)
suggested that the ammonia ice was widely distributed.
\cite{Irwin2001BZ} found a similar absorption anomaly in analysis of
observations by the Galileo Near Infrared Mapping Spectrometer (NIMS),
but concluded that it was not due to NH$_3$ ice because a key spectral
signature at 2.0 $\mu$m was missing.  The subsequent
\cite{Baines2002Icar} detection of spectrally identifiable ammonia
clouds (SIACs) in other NIMS observations was based on depressed
reflectivity at 2.7 $\mu$m as well as at 2.0 $\mu$m, but these
detections covered a very tiny fraction ($<1$\%) of Jupiter's cloud
features.  On the other hand, \cite{Wong2004} inferred more widely
distributed ammonia ice, at least in some latitude bands, 
from a detection of a 9.4-\mum spectral feature in Jovian spectra
obtained from the Cassini Composite Infrared Spectrometer (CIRS).  The
model calculations of \cite{Wong2004} implied that the 9.4-$\mu$m ice
feature could only be detected when the aerosols were at pressures
$\leq$500 mb and the particle effective radius was within a factor of
two of 1 $\mu$m.

New band models for \nht absorption developed by \cite{Bowles2008} and
recent additions to the HITRAN line data base motivated
\cite{Sro2010iso} to update exponential sum models for \nht
absorption.  They found substantial changes relative to the absorption
models used in the analysis by \cite{Brooke1998} and undertook a
reanalysis of the ISO spectrum.  They found that the best fit to the
3-\mum absorption feature was obtained not with \nht as the sole
absorber, but instead when particles of \nhfsh and \nht were both
present, with the predominant opacity provided by \nhfshx. Two types
of solutions were found to provide good fits to the ISO spectrum: one
had relatively large \nht and \nhfsh particles at nearly the same
pressure, near 500 mb, and the other had small \nhtx-coated particles
closer to 350 mb, with large \nhfsh particles again near 500 mb. The
large FOV of the ISO measurement did not permit any conclusions
concerning the spatial distribution of these two components, i.e.
whether some clouds had only \nhfsh particles, and others had only
\nht particles, or whether the two types existed in the same physical
location.

In the following analysis we present new results based on near-IR
spatially resolved spectral observations of the Cassini Visual and
Infrared Mapping Spectrometer (VIMS).  Our analysis takes advantage of
recently improved models of methane and ammonia gas absorption, and uses
refractive index properties of putative cloud materials in
constraining model structures and composition.
We first describe the VIMS observations, then the radiation transfer
methods, the modeling results, how they compare with other models
based on other observations at visible, near-IR, and thermal
wavelengths, and finally summarize their implications for Jovian cloud
structure.  We show that VIMS observations display a widely
distributed strong absorption in the 3.1-$\mu$m region of the
spectrum, and evidence for a much weaker absorption near 2 $\mu$m, both of
which are qualitative features expected from a cloud consisting of
ammonia ice particles.  However, it is only the darker cloud features,
with smaller optical depths, that can be well fit with ammonia ice as the
sole absorber.  We find that brighter clouds require more absorption
at 3 \mum without increased absorption at 2 \mum and that this can
be provided by NH$_4$SH.

\section{Observations.}

\subsection{VIMS Overview}

\begin{figure*}[!htb]\centering
\includegraphics[width=6.5in]{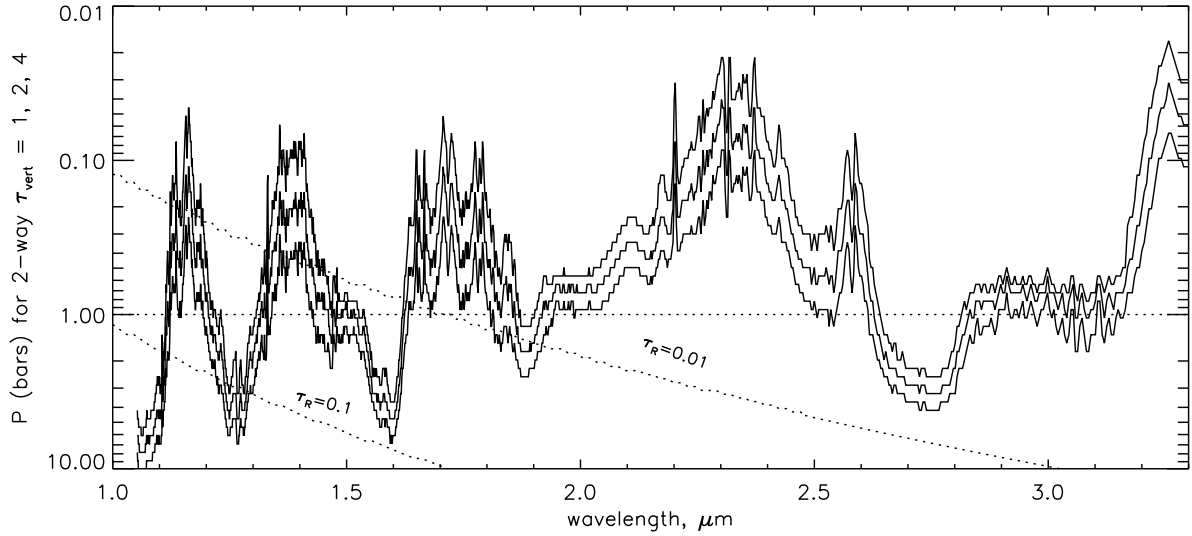}
\caption{Penetration depth of photons in a clear Jovian atmosphere,
indicated by pressures at which a unit change in albedo of a
reflecting layer results in an external I/F change of 1/e, 1/e$^2$,
and 1/e$^4$ (three solid curves), for NH$_3$ mixing ratios typical of
the NEB. The sloping dotted lines define the location at which 2-way
optical depths for Raleigh scattering reach 0.1 and 0.01.}
\label{Fig:pendepth}
\end{figure*}

\begin{figure*}[!htb]\centering
\includegraphics[width=6.5in]{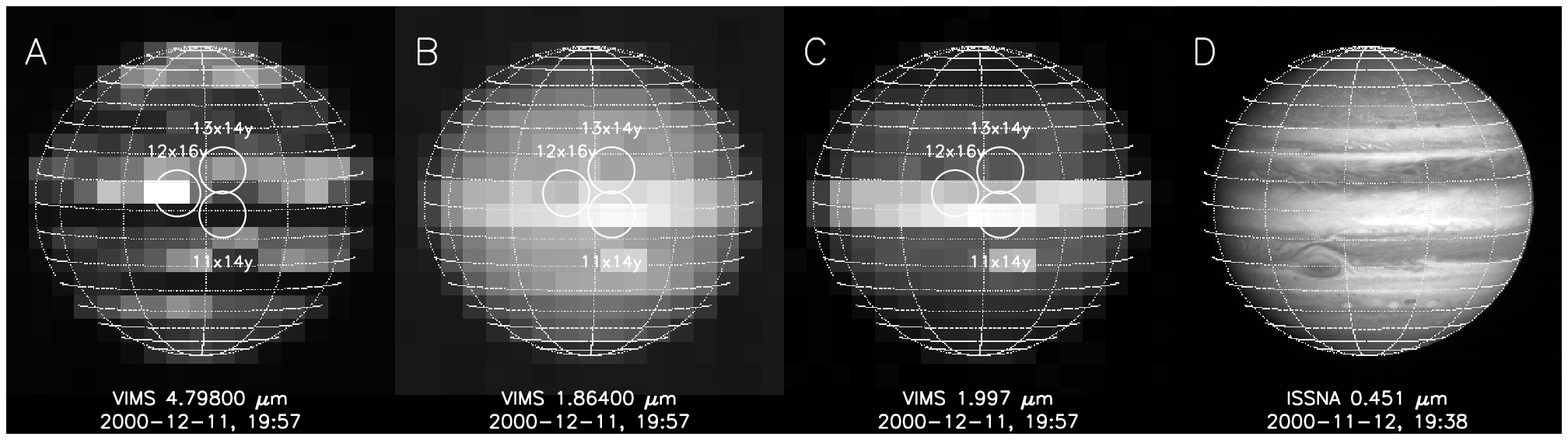}
\caption{Cassini VIMS images from 11 December 2000 at 4.798 $\mu$m
(A), 1.864 $\mu$m (B), and 1.997 $\mu$m (C) compared to an ISS image
acquired roughly one month earlier (D).  Circles in A-C, locate pixels
from which we extracted spectra for analysis. These are located at
6\degx N, 290\degx E (12x16y), 1.2\degx S, 306\degx E (11x14y), and
14\degx N, 306\deg E (13x14y). An additional spectrum (not shown) was
extracted from 1.5\degx S, 298\degx E (11x15y). Here y increases to the left
and x upward.}
\label{Fig:imvims1}
\end{figure*}

\begin{figure*}[!htb]\centering
\includegraphics[width=6.5in]{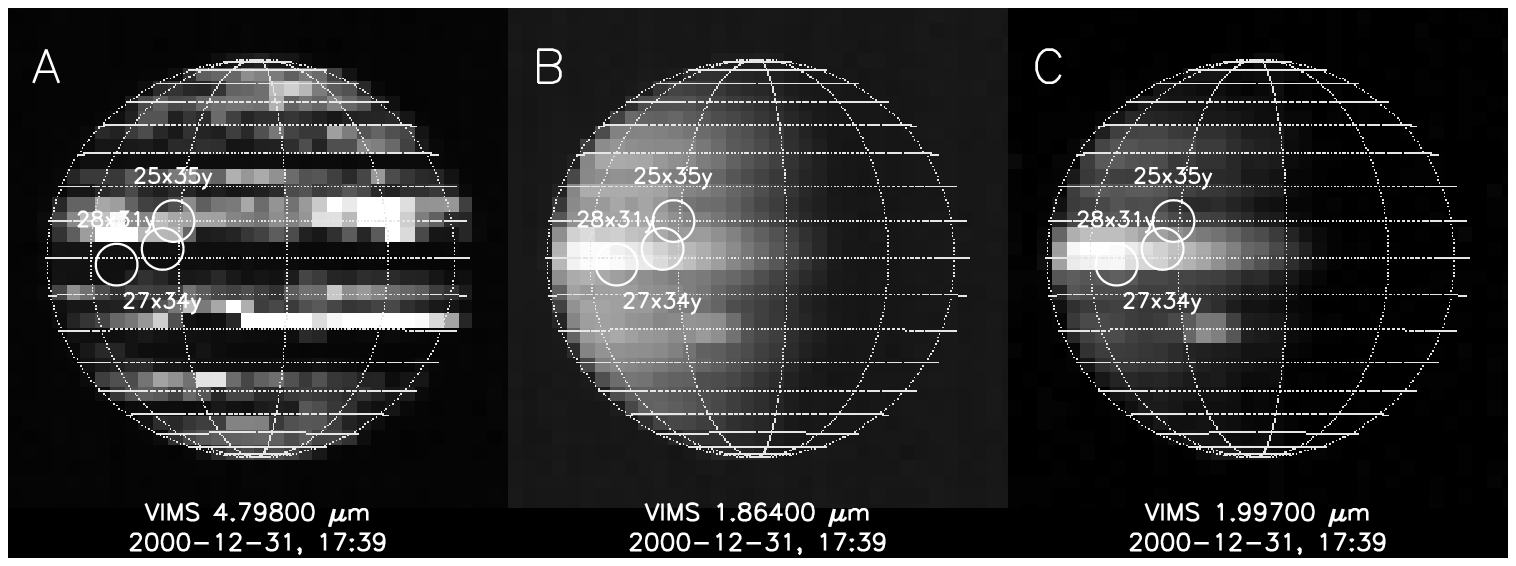}
\caption{Cassini VIMS images from 31 December 2000 at 4.798 $\mu$m
(A), 1.864 $\mu$m (B), and 1.997 $\mu$m (C).  Circles in A-C, locate
pixels from which we extracted spectra for analysis. These are located
at 9.9\deg N (centric), 298\deg E (25x,35y), 2.4\degx N, 295\degx E (27x34y), and
1.8\degx S, 280\degx E (28x31y). An additional spectrum was extracted at
6.3\degx N, 280\degx E (26x31y). Here y increases to the right
and x downward. }
\label{Fig:imvims2}
\end{figure*}

For constraining models of low-latitude Jovian cloud structure we used
spectra acquired by VIMS during Cassini's flyby of Jupiter in December
2000.  The VIMS instrument and investigation are described by
\cite{Miller1996SPIE} and \cite{Brown2004SSR}. The instrument's
spectral range is 0.35-5.1 $\mu$m, with an effective pixel size of 0.5
milliradians on a side and a near-IR spectral resolution of
approximately 15 nm (sampled at intervals of approximately 16 nm). Our
analysis focuses on the 1-3.2 \mum interval where scattering by
aerosols predominates over Rayleigh scattering and where a wide range
of atmospheric pressures can be sampled and discriminated by means of
variations in gas absorption with wavelength (Fig.\
\ref{Fig:pendepth}). In this wavelength range the contribution of
Jupiter's thermal emission is everywhere less than 0.1\% of the observed I/F.
To provide an approximate comparison with prior low phase angle
observations
we chose VIMS data cube CM\_1355256529, obtained on 11 December 2000
at a phase angle of 2.5\deg.  This provides a pixel size of 9925
km at the sub-spacecraft point.
In Fig.\ \ref{Fig:imvims1} VIMS images extracted from this cube at
4.798 $\mu$m, 1.864 $\mu$m and 1.997 $\mu$m are compared to a Cassini
ISS (Imaging Science Subsystem) image obtained at 0.451 \mumx.  The
encircled pixels in the figure are locations from which we extracted
VIMS spectra, taking samples from the North Equatorial Belt (NEB), and
from the Equatorial Zone (EZ).  The pixel labeled 12x16y, which is
brightest at 4.798 $\mu$m, is located in a region where deep cloud
opacity is at a relative minimum, corresponding to a relatively low
reflectivity at 1.864 $\mu$m.  To obtain samples at higher spatial
resolution and less sensitive to details of backscatter phase
functions, we also selected data cube CM\_1356976257, obtained on 31
December 2000 at a phase angle of 67.8\degx,
with a sub-spacecraft pixel size of 4887 km, which is about twice
the resolution obtained from the low phase angle cube. Sample images
and target selections from this cube are shown in Fig.\
\ref{Fig:imvims2}.

\subsection{Image Processing and Navigation}

VIMS image cubes were processed using the pipeline processing code
downloaded from the Planetary Data System, described by
\cite{McCord2004}.  The cubes were navigated by manual adjustments of
planet center coordinates using visual comparisons of model and image
limbs at window wavelengths to find an approximate best fit, which is
estimated to be within a fraction of a pixel.  This provided
sufficient definition of view angles and latitudes near the disk
center to allow a meaningful comparison of spectral characteristics in
broad latitude regions, as appropriate for the relatively low spatial
resolution of these observations.

\subsection{Photometry}

\begin{figure*}[!htb]\centering
\includegraphics[width=5.5in]{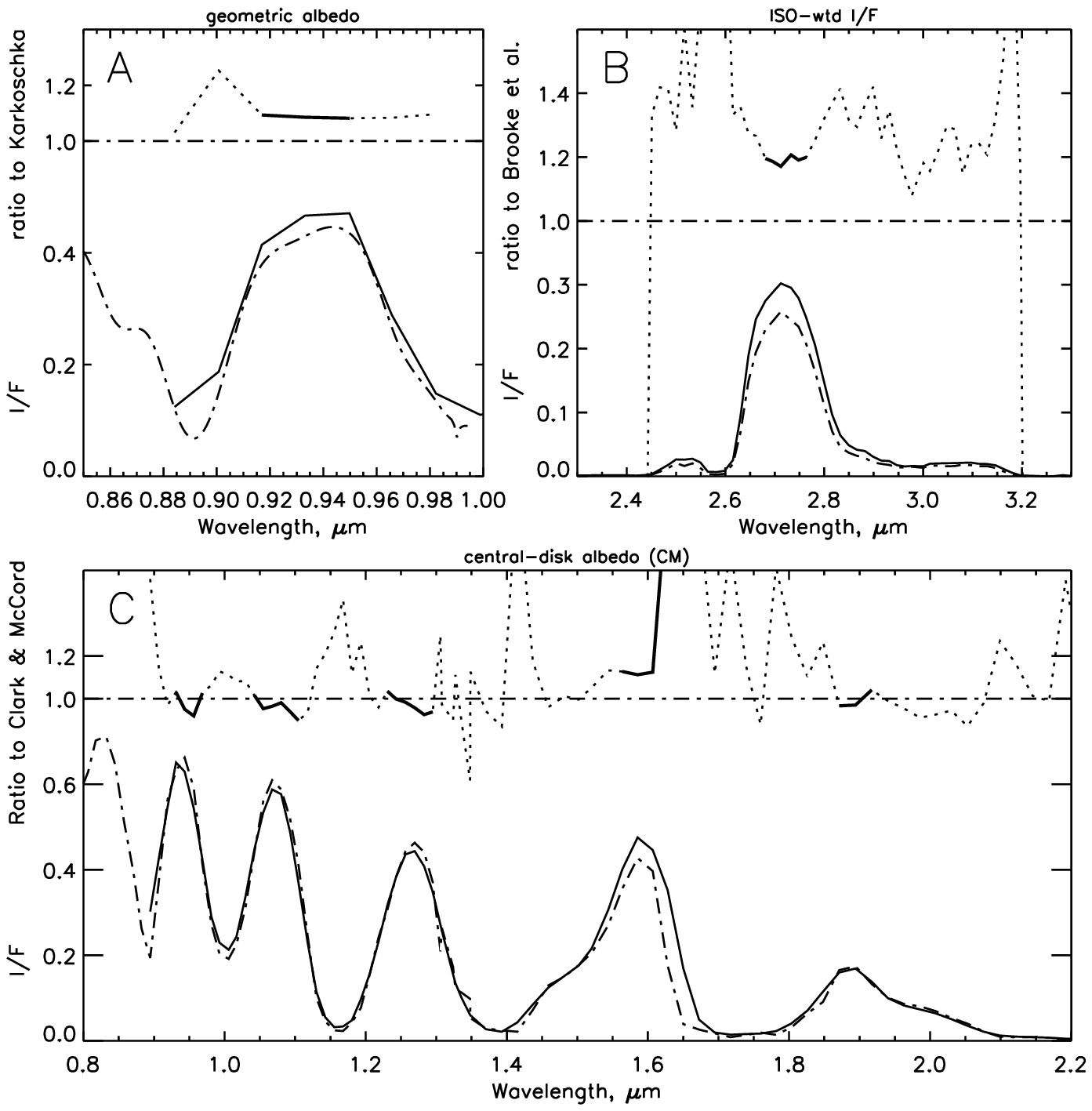}
\caption{VIMS I/F spectra (solid lines) compared to the
 \cite{Kark1998Icar} 1995 geometric albedo spectrum of Jupiter (A),
 the \cite{Brooke1998} ISO 1996 spectrum (B), and the 1976
 \cite{Clark1979jup} central-disk spectrum (C). In (A) and (B) the
 reference spectra (dot-dash) are averaged to VIMS resolution.  In (C)
 the VIMS spectrum (solid) is averaged to the resolution of the
 reference spectrum (dot-dash).  Ratio plots are dotted, except where
 heavy black lines are used in low-absorption regions that are not
 much affected by upper level aerosols.}
\label{Fig:vimscalcomp}
\end{figure*}

The VIMS calibration obtained by the pipeline processing has an
uncertainty that has not yet been documented, but is at least as large
as the near-IR solar spectral irradiance uncertainty of $\sim$3\%-5\%
\citep{Colina1996}.  It appears to be less than 10\% uncertain, based
on window-region comparisons with independent spectra of Jupiter,
shown in Fig.\ \ref{Fig:vimscalcomp}.  In the 0.9-0.98 $\mu$m region
the disk-integrated VIMS spectrum is $\sim$8\% brighter than that of
\cite{Kark1998Icar}.  In the 0.9-2.2 $\mu$m region we computed a
central-disk VIMS spectrum to compare with the groundbased
central-disk spectrum of \cite{Clark1979jup}.  In the brighter window
regions, where thin upper-level hazes and low-level artifacts have
much reduced impact, the VIMS observations are within $\sim$5\% of the
\cite{Clark1979jup} spectrum, except near 1.6 $\mu$m, where an
order-sorting filter joint in the VIMS detector array apparently
results in poor response corrections.
The VIMS spectrum averaged over the same angular FOV as a 1996 ISO
spectrum \citep{Brooke1998} is found to be about 20\% brighter, which
might be entirely due to temporal variations or even phase angle variations (the VIMS
and ISO phase angles in this comparison  are 2.5\deg and 11\deg respectively). Another
possible contribution to the VIMS - ISO difference is the ISO calibration uncertainty,
which could be as large as 12\% for wavelengths less than 3.8 \mum
\citep{Schaeidt1996}.
Overall, there seems to be a tendency for the VIMS spectra to be on
the high side, but how much of the differences are due to calibration
errors as opposed to temporal variation or phase-angle differences is
uncertain.

\subsection{VIMS artifacts.}\label{Sec:artifact}

\begin{figure}[!htb]\centering
\includegraphics[width=3.5in]{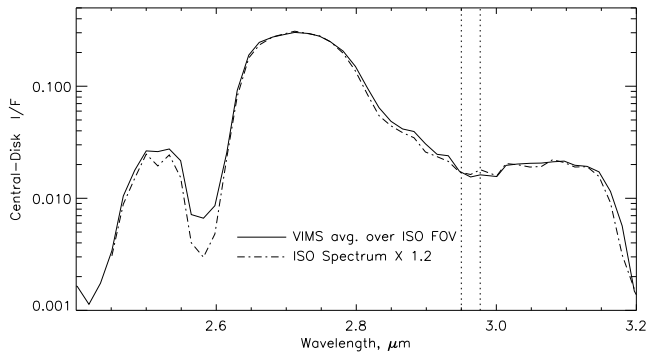}
\caption{Comparison of central-disk I/F spectra from VIMS (solid) and
ISO$\times$1.2 (dot-dash) with comparable fields
of view covering about 1/4 of the disk. The ISO spectrum was converted
from the flux spectrum of \cite{Brooke1998}. Vertical dotted lines
mark the locations of an absorption feature of \nht ice at 2.95 \mum and
the VIMS filter joint at 2.98 \mumx.}
\label{Fig:isocomp}
\end{figure}

\begin{figure}[!htb]\centering
\includegraphics[width=3.5in]{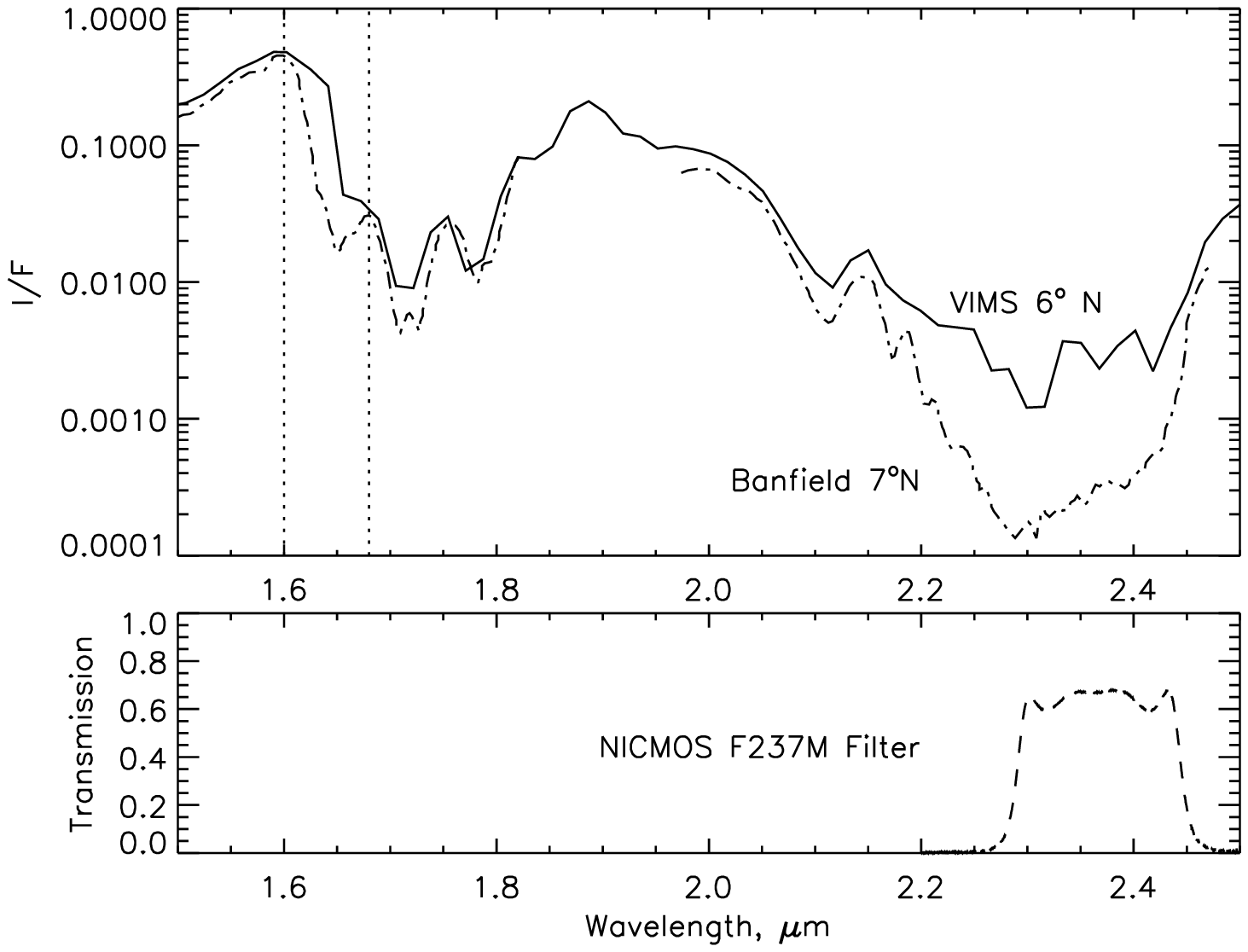}
\caption{Top: Comparison of VIMS 6\deg N spectrum (solid) with a 7\deg
N groundbased spectrum (dot-dash) from
\cite{Banfield1998NIR}. Vertical dotted lines at 1.6 and 1.68 \mum
bound the region in which VIMS spectra are in significant
error. Lower: transmission spectrum of the NICMOS F237M filter.}
\label{Fig:banfieldcomp}
\end{figure}

There are two important artifacts that need to be considered in the
interpretation of VIMS spectra.  The first has to do with responsivity
corrections at wavelengths in close proximity to joints between order
sorting filters overlaid on the VIMS linear detector arrays. The
joints of concern here appear at wavelengths of 1.64 \mum and 2.98
\mumx.  The required responsivity correction factors to compensate for
shading effects
are very large for the former ($\sim$300) and relatively small for the
latter ($\sim$2).  To test the correction near 2.98 \mum we used a
comparison between the ISO 3-\mum spectrum of \cite{Brooke1998} and a
central disk VIMS spectrum covering approximately the same field of
view.  When the ISO spectrum is empirically scaled by a factor of 1.2, the two
spectra are in generally excellent agreement (Fig.\ \ref{Fig:isocomp}), including the
region of the filter joint near 2.98 \mumx.  However, we do see a
significant disagreement near 2.58 \mumx, where the ISO spectrum
reaches a deeper minimum. This may be related to the second kind of
artifact, which is discussed later. There is also a small discrepancy
near 2.52 \mumx, where the ISO spectrum has a local minimum that is
not seen in the VIMS spectrum. A similar discrepancy is also seen
between model spectra and the VIMS observations in this region.
The rather large responsivity correction near 1.64 \mum does have
significant errors, which can be seen from a comparison between a
groundbased spectrum of \cite{Banfield1998NIR} and a VIMS spectrum
from a nearby latitude region (Fig.\ \ref{Fig:banfieldcomp}). The
discrepant region is from about 1.60 to 1.68 \mumx.  This can also
be seen from a comparison to the \cite{Clark1979jup} spectrum shown in
Fig.\ \ref{Fig:vimscalcomp}C.  It is also apparent in the striking
disagreement between model and observations in this region, as
discussed in later sections.

\begin{figure}[!htb]\centering
\includegraphics[width=3.2in]{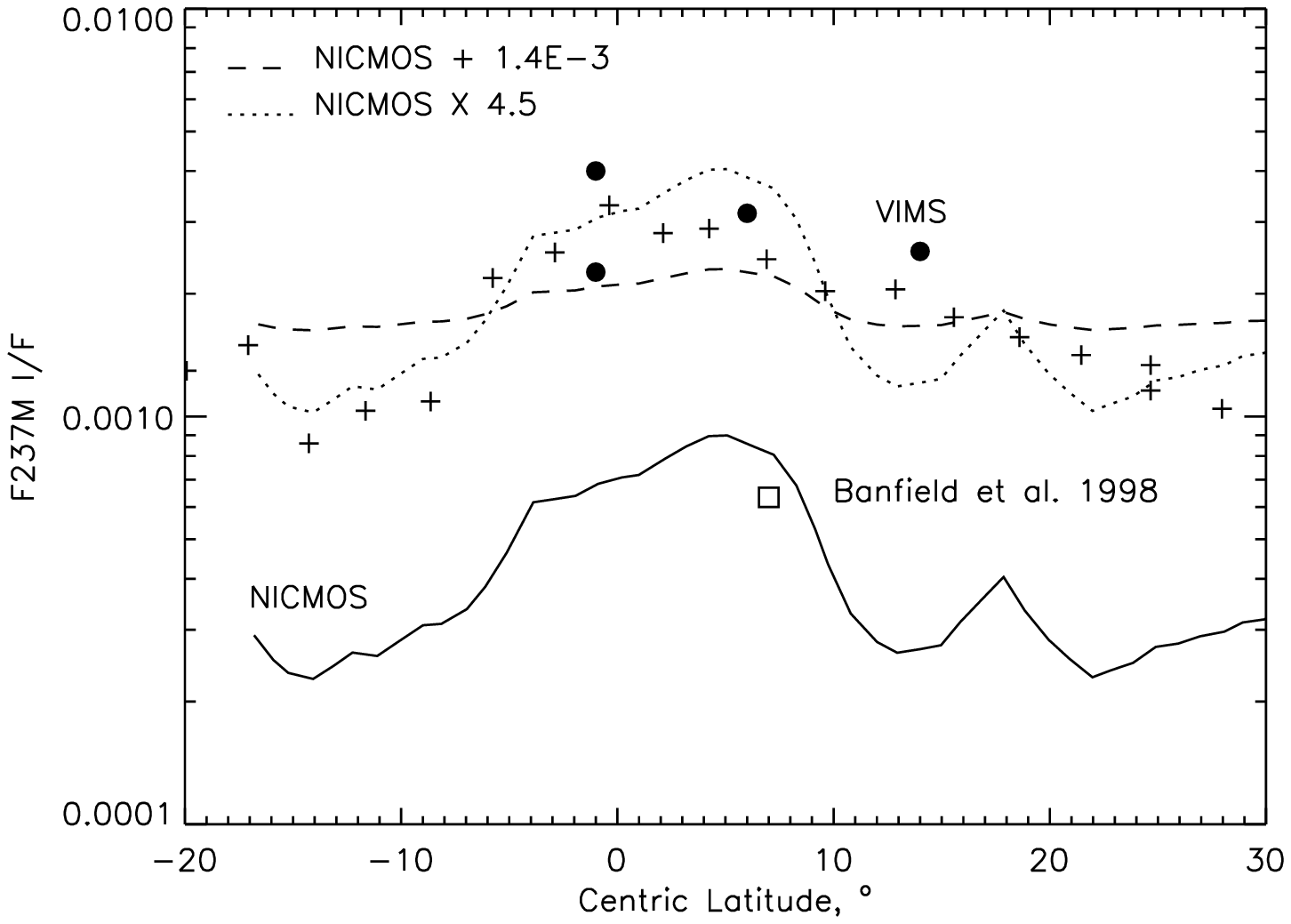}
\caption{Comparison of NICMOS F237M central meridian I/F vs latitude
(solid) with synthetic photometry band-pass results computed from VIMS
spectra and a 7\deg N spectrum from \cite{Banfield1998NIR}. The filled
circles are from VIMS locations identified in Fig.\
\ref{Fig:imvims1}. The plus symbols are for a central meridian
latitude scan with a 3-pixel longitudinal average.  Offset (dashed)
and scaled (dotted) versions of the NICMOS scan are also shown.}
\label{Fig:f237mcomp}
\end{figure}

The second type of artifact is evident in the 2.3-\mum spectral region
where the groundbased spectrum in Fig.\ \ref{Fig:banfieldcomp} has a
minimum that is a factor of 10 lower than that of the VIMS spectrum.
This cannot be attributed to local variations because the latitude
difference is small and the I/F disagreement is typical of several
comparisons.  To resolve which of these two measurements is more
believable we used 1997 NICMOS observations (HST Program 7445, Reta
Beebe PI), selecting an image (n49o01mvs) made with the F237M filter,
which heavily weights the region of disagreement, as shown by the
filter transmission plot in Fig.\ \ref{Fig:banfieldcomp}. We computed
synthetic F237M I/F values from the groundbased I/F spectrum and from
VIMS I/F spectra along a central meridian scan using weighted
integrals over the spectra. The weighting function was the product of
filter throughput and solar spectral irradiance, normalized to yield a
unit integral over the same spectral range.  As shown in Fig.\
\ref{Fig:f237mcomp}, the synthetic photometry result for the
\cite{Banfield1998NIR} spectrum is in good agreement with the NICMOS
scan.  But synthetic F237M values obtained from VIMS spectra are
$\sim$4.5 times as large as the NICMOS values.  The artifact is not a
simple offset error, because that does not provide the correct
spectral shape.  It seems more likely that the artifact is due to
light scattered by the grating, which allows light in bright regions of
the spectrum to contaminate measurements of dark regions. This may
also be responsible for the discrepancy with the ISO spectrum near
2.58 \mum evident in Fig.\ \ref{Fig:isocomp}.  In any case, the defect
in the VIMS measurements near 2.3 \mum has the effect of distorting
our model fits in the stratosphere, suggesting far more particulate
reflectivity than is actually present.

\subsection{VIMS noise characteristics.}\label{Sec:vimsnoise}

For most applications at moderate signal levels, the random noise
level of VIMS is very small, nominally less than one digital
quantization number (DN) (Kevin Baines, personal communication,
January 2010). As there is no published equation defining how VIMS
measurement noise depends on signal level, exposure time, and
wavelength, we tried to better characterize the noise by comparing
spectra in cubes that imaged the same region of Jupiter with as small
a time difference as possible.  We chose CM\_1355309574 and
CM\_1355309697 cubes, which were exposed only two minutes apart.
These results were consistent within about 1.2 DN RMS for a single
measurement, even for signal levels up to 2000 DN or more.
For the cubes we used in our detailed spectral analysis, which had
exposures much shorter than these comparison cubes, 
the noise level would still be only 0.14\% at 2.7 \mum and thus not a
significant error source. However, at signals 100 times smaller, as
found near 2.3 \mumx, the noise becomes a much larger fraction of the
total signal. This is
evident from the rapid variations with wavelength observed in that
region of the spectrum, which are up to 50\% or more of the signal
level.
This can be roughly characterized as an offset noise in I/F, which we
crudely estimated as $\sim$5$\times 10^{-4}$ from the on-disk spectra.
Other possible sources of uncertainty, though not strictly random
noise, are scattered light inside the spectrometer, wavelength errors,
line-spread uncertainties, and variable absolute calibration and
wavelength calibration errors. We lumped all these potential
uncertainties into a somewhat arbitrary 2.5\% error proportional to
signal, and an I/F offset error of $\sim$5$\times 10^{-4}$. We did
look into the possibility of wavelength errors by making fits with
shifted wavelength scales, and determining $\chi^2$ as a function of
shift. We found that the minimum $\chi^2$ was within 1 nm of the nominal
wavelength scale and that no revision was needed.

\section{Radiation transfer calculations}

\subsection{Atmospheric structure and composition}

\begin{figure}[!htb]\centering
\includegraphics[width=3.5in]{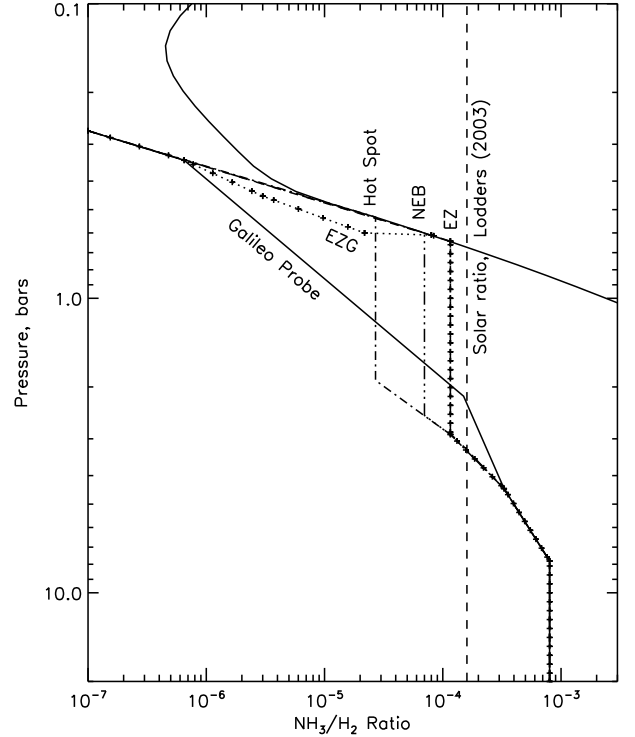}\\
\caption{Model NH$_3$ profiles, adapted from \cite{Sro2010iso}. All profiles
overlap above the point near 330 mb where they intersect the Galileo Probe profile (lower
solid curve) and below the intersection near 4.2 bars. At intermediate
pressures the Hot Spot (dot-dash), NEB (triple-dot dash), EZ (dash+),
and EZG (dotted+) follow separate paths until intersecting the saturation
mixing ratio (upper solid curve), at which point all but the EZG profile
follow the same line to the upper intersection with the Probe profile.}
\label{Fig:nh3mix}
\end{figure}

We followed the same approach described by \cite{Sro2010iso}.  In summary,
we used the tabulated results of \cite{Seiff1998} for Jupiter's temperature
structure down to the 22-bar level, and  assumed an atmospheric
composition of He/H$_2$=0.157$\pm$0.003 \citep{VonZahn1998}
and CH$_4$/H$_2$= 2.1E-3$\pm$0.4E-3 \citep{Niemann1998}, which are
expressed as number density ratios.  
We used model mixing ratio profiles (Fig.\ \ref{Fig:nh3mix}) ranging
from what should be typical of the equatorial zone to hot-spot values.
The basis for these models is discussed by \cite{Sro2010iso}. The deep
mixing ratio is from \citep{Folkner1998}, the upper level Probe result
is from \citep{Sro1998}.  The EZ (Equatorial Zone, $\pm$7\deg
latitude), NEB (North Equatorial Belt, $\approx$8\degx N-18\degx N),
Hot Spot ($\approx$7\degx N), and EZG ( $\pm$7\deg latitude) profiles
were based on combinations of NH$_3$ mixing ratios inferred from ISO
observations by \cite{Fouchet2000} and on microwave results of
\cite{Sault2004} and \cite{Gibson2005Icar}. Upper level mixing ratios were also guided by
results of \cite{Edgington1999}, \cite{Lara1998}, and
\cite{Achterberg2006}.

\subsection{Gas absorption models}

We used the k-distribution models of hydrogen-broadened methane
absorption described by \cite{Irwin2006ch42e}, which employ new
two-term models of temperature dependence \citep{Sro2006ch4}. 
For ammonia we used the combined correlated-k absorption model described by
\cite{Sro2010iso}, which is based on the Goody-Lorentz band model of
\cite{Bowles2008}, and also makes use of 1996 and 2004 (or equivalently, 2008) HITRAN
spectral line compilations.   
he basis for and details of how these results were combined are
described by \cite{Sro2010iso}.
Collision-induced absorption (CIA) for H$_2$ and H$_2$-He was
calculated using programs downloaded from the Atmospheres Node of the
Planetary Data System, which are documented by \cite{Borysow1991h2h2f,
Borysow1993errat} for the H$_2$-H$_2$ fundamental band,
\cite{Zheng1995h2h2o1} for the first H$_2$-H$_2$ overtone band, and by
\cite{Borysow1992h2he} for H$_2$-He bands. We followed the
\cite{Birnbaum1996} parameter recommendations 
to avoid sudden drops in wing absorption in the 1.58-$\mu$m
window.
We assumed equilibrium mixing ratios
of ortho and para hydrogen in all our model fitting calculations and
later show that the resulting errors are not significant to our main
conclusions.

\subsection{Reflecting layer methods}

\subsubsection{Linear combination equations}

 If all the cloud layers were composed of broken cloud fields of
opaque elements, a linear combination of reflecting layer
contributions would be an ideal model for representing the combined
I/F of the vertical cloud structure.  By fitting parameters of such a
model to obtain agreement with the observed I/F spectra or variation
of I/F with angle, it can also be used to obtain a rough idea of the
distribution of scatterers when the cloud layers are optically thin.
Its main virtue is the great speed of the inversion, because it makes
use of precomputed arrays.  Our model equations for a multilayer
atmosphere are based on Eq. 4 of \cite{Sro2007struc}, which can be
written in the following form in a three-layer model:
 \begin{eqnarray} I(\vec{\theta})&=&
I_0(P_1,\vec{\theta}) + f_1\times[I_1(P_1,\vec{\theta})-I_0(P_1,\vec{\theta})]
\nonumber\\ 
& & + \Big[I_0(P_2,\vec{\theta})-I_0(P_{1},\vec{\theta}) + f_2\times \nonumber\\
& & [I_1(P_2,\vec{\theta})- I_0(P_2,\vec{\theta})]\Big]\times T_1(\vec{\theta}) \nonumber\\
 & & + \Big[I_0(P_3,\vec{\theta})-I_0(P_{2},\vec{\theta}) + f_3\times \nonumber\\
& & [I_1(P_3,\vec{\theta})- I_0(P_3,\vec{\theta})]\Big]\times T_1(\vec{\theta}) T_2(\vec{\theta}) 
\nonumber\\ && +
\Big[ I_b(\vec{\theta}) -I_0(P_3)\Big] \times T_1(\vec{\theta})  T_2(\vec{\theta})  T_3(\vec{\theta}) 
\label{Eq:lincomb2}\end{eqnarray}
where $I_0(P,\vec{\theta})$ is the I/F observed at the top of the
atmosphere for viewing geometry $\vec{\theta}$ when a zero-albedo surface is
placed at pressure $P$ and $I_1(P,\vec{\theta})$ is the
top-of-atmosphere I/F for a unit-albedo surface placed at the same
level. Here $\vec{\theta}$ denotes the viewing geometry vector
[$\theta$, $\theta_0$, $\phi$], which refer to viewer and solar zenith
angles and azimuth angle, respectively.  

The first line in the above equation contains one contribution from
the atmosphere above the first reflecting layer ($I_0$) and a second
differential contribution from the reflecting layer itself
($f\times[I_1-I_0]$), where $f$ can be interpreted as the fraction of area
covered by the reflecting layer, for a pure broken cloud model, or as
the reflectivity of the cloud, for models approximating translucent
clouds. Successive lines have a similar structure, except that the
atmospheric contribution is only from scattering between a layer and
the next highest layer, and the contributions from both cloud and
atmosphere are attenuated by a transmission factor $T_1T_2..T_i$. In
the pure broken cloud model this is just the fraction that is not
blocked by the opaque elements in the previous layers,
i.e. $(1-f_1)(1-f_2)..(1-f_i)$.  In the translucent approximation we
use $T_i\approx \exp[ - \tau'_i(1/\mu+1/\mu_0)]$, where $\mu$ and
$\mu_0$ are cosines of observer and solar zenith angles, $\tau'_i =
(1-\frac{1}{2}\varpi_i (1+g_i))\tau_i$, $g_i$ is the scattering
asymmetry parameter for the $i$th layer, and $\varpi_i$ is the
corresponding single-scattering albedo.  

The formula for the effective attenuation optical depth for each layer
($\tau'_i$) is an empirical result optimized for low opacity layers
considered by \cite{Sro2007struc}.  Here, we make the simplifying
assumption that the overlying layers are conservative ($\varpi =1$)
and symmetric scatterers ($g=0$), so that $\tau'_i =\tau_i/2$.  Under
the same approximation we have $\tau_i \approx 4\times f_i$
\citep{Sro2007struc}, so that $\tau'_i \approx 2 \times f_i$.
Because our analysis of each spectrum is at a fixed viewing geometry,
we ignore possible angular variation in effective cloud
fractions. However, we do need to parameterize variations with
wavelength, which is accomplished by setting
$f_i=f_0\times(\lambda/\lambda_0)^{n_i}$, where large particle clouds
would likely have exponents near zero, and very small particles would
likely approach the Rayleigh limit of $n_i=-4$.

\subsubsection{Reflecting layer I/F computations}

A comparison of single-scattering and multiple-scattering reflecting
layer calculations shows that atmospheric multiple scattering is
insignificant over the range 1.1 $< \lambda < 4$ $\mu$m, and even the
1.05 $\mu$m window adds no more than a few percent to the I/F obtained
from single-scattering.
 The computed
clear-atmosphere I/F values throughout this range are well below the
observed values by 1-3 orders of magnitude, and thus particulate
scattering is the dominant influence at all wavelengths considered in
this analysis. For our reflecting layer calculations, atmospheric
scattering plays a minor role that is adequately modeled with single
scattering. 

We computed reflecting layer I/F values for unit-albedo and
zero-albedo reflecting layers placed at 91 pressures logarithmically
distributed between 0.5 mb and 40 bars, using the single-scattering
equations \begin{eqnarray}
I_j^a&=&\frac{1}{4} \sum_{i=1}^{j} \varpi_i P_R(\theta_S) \exp(-m\tau_{i-1})
  \nonumber \\
& & \times (1-\exp[-m \Delta\tau_i])/(m\mu) \nonumber\\
& & + a \exp(-m\tau_i)\mu^{K-1}\mu_0^{K}\label{Eq:ss}\\
\Delta\tau_i&=&\tau_i-\tau_{i-1}\\
\varpi_i&=&(\tau_{i,scat}-\tau_{i-1,scat})/(\tau_{i}-\tau_{i-1})\\
m&=&(1/\mu+1/\mu_0)
\end{eqnarray}
\begin{eqnarray}
P_R(\theta_S)&=&\frac{3}{2}\frac{[1+\delta +(1-\delta)\cos^2(\theta_S)]}{2+\delta}
\end{eqnarray}
\begin{eqnarray}
\cos(\theta_S)&=&\cos(\phi)\sqrt{1-\mu^2}\sqrt{1-\mu_0^2}-\mu\mu_0
\end{eqnarray}
where $\mu$ and $\mu_0$ are observer and solar zenith angle cosines,
$\phi$ is the azimuth angle between incident and scattered directions,
$\varpi_i$ is the effective single-scattering albedo of the gas
between pressures $p_i$ and $p_{i-1}$, $a$ denotes normal albedo of
the reflecting layer (both 0 and 1 are used), $K$ defines the limb
darkening character of the layer (assumed to have the
Minnaert form in which reflected radiance at zenith cosine $\mu$ is
proportional to incident irradiance times $\mu^{K-1}\mu_0^K$),
$P_R(\theta_S)$ is the scalar Rayleigh phase function for anisotropic
randomly oriented molecules, evaluated at depolarization factor
$\delta=0.020$ \citep{Sro2005pol} and scattering angle $\theta_S$,
$\tau_i$ is the vertical extinction optical depth from the top of the
atmosphere to pressure $p_i$, and $\tau_{i,scatt}$ is the
corresponding scattering optical depth. Our calculations cover the
spectral range from 1.05 \mum to 3.3 \mum with a uniform
step size of 5 \icmx, the sampling interval of the CH$_4$
absorption model. For wavelengths where both methane and ammonia
absorptions are important, opacities are computed 100 times for each
wavelength (for each of the 10 terms of the correlated-k model of
CH$_4$, the calculation is done for each of the 10 terms of the
correlated-k model of NH$_3$). The most accurate I/F is obtained by
solving the radiation transfer problem for each opacity combination
and then computing the weighted sum of the 10 or 100 calculations per
wavelength.

\subsection{Multiple scattering methods}

For multiple scattering calculations, which are needed to accurately
model translucent clouds, we followed \cite{Sro2010iso}. In brief,
we used the doubling and adding code described by \cite{ Sro2005raman,Sro2005pol},
We used a grid of 44 pressure levels from 0.5 mb to 10 bars, distributed
roughly in equal log increments, except with finer spacing in the
ammonia condensation region. 
We generally used 10 zenith angle quadrature points per hemisphere
(NQUAD) and handled the sharp forward scattering peak of larger
particles using the $\delta$-Fit procedure of \cite{Hu2000}.  The
backscatter phase function of larger particles is not well
characterized with NQUAD of 10, so that calculations for low phase
angles are subject to errors, as shown in Section \ref{Sec:candidate}.
However, it is unlikely that these larger particles are spherical in
any case, so that a truncation of the spherical backscatter peak may
actually be more realistic than keeping it. Reduced backscatter
(relative to spheres) is characteristic of aggregate particles
\cite{West1991}, and many randomly ordered crystals and spheroids
\citep{Yang2000}, including the prolate spheroidal shapes favored by
\cite{Wong2004}.

The most significant difference between the approach we used and that
of \cite{Sro2010iso} is that in modeling VIMS spectra we made an
additional approximation to account for the VIMS spectral resolution
being less than our model resolution.  We approximate the line-spread
function of the VIMS instrument as a Gaussian of FWHM that is
wavelength dependent (Kevin Baines, private communication 2010), and
is typically about 0.015 $\mu$m. We then collect all the opacity
values within $\pm$FWHM of a VIMS sample wavelength, weight those
according to the relative amplitude of the line-spread function, then
sort and refit to ten terms in the exponential sums representing
transmission.

\subsection{Fitting cloud models to observations}

Our cloud models were constructed as an assemblage of discrete compact
layers.  Model parameters were adjusted to minimize $\chi^2$ using a
form of the Levenberg-Marquardt algorithm, as described by
\cite{Press1992}. The parameter uncertainties obtained from this
method are based on the assumption of normally distributed errors, and
thus cannot be absolutely relied on.  
For the reflecting layer models, we have
supplemented these estimates by varying each fitted parameter about
its best fit value, refitting all the other parameters to minimize
$\chi^2$, then setting the parameter uncertainty equal to the
deviation needed to increase $\chi^2$ by one unit.  This alternate
method generally agreed with that obtained from the
Levenberg-Marquardt algorithm to within a factor of two.  It was
not practical to use this alternate method for multiple scattering
fits because of the enormously increased computation time that would
be required.

In computing $\chi^2$ we tried to account for measurement errors as
well as errors in the modeling of atmospheric opacity.
For the measurements we assigned a random error for all sources a
value of 2.5\% of the VIMS I/F value, with an additional I/F offset
error of $5\times 10^{-4}$, as discussed in Sec.\ \ref{Sec:vimsnoise}.
A more important error source, which is not strictly random, but can
vary from wavelength to wavelength, is the uncertainty in radiation
transfer modeling due to the uncertainty in opacity calculations. The
transmission curves on which the methane opacity models are based have
an assumed fractional error of 0.05 \citep{Irwin2006ch42e}.  To
convert this uncertainty to an I/F uncertainty, we assume that the
derived optical depths themselves have a fractional error of similar
size.  We assigned a value of 6\% to the error in computed optical
depth, but converted this to I/F under the crude assumption that I/F
would be proportional to the negative exponential of optical
depth. This leads to a fractional I/F error estimate of
1-(I/F)$^\alpha$, where $\alpha$ is the fractional error in optical
depth.  Thus a 6\% error in optical depth would lead to errors that range
from nearly zero at low absorption optical depths to 6\% near $\tau=1$
and to 100\% for $\tau\gg 1$.

To judge whether one fit is significantly better than another on the
basis of a difference in \chisqx, we need to know how that difference
compares to the uncertainty in \chisq itself.  The latter uncertainty
arises from measurement and gas opacity modeling errors, and is easily
determined from a simple Monte-Carlo calculation.  For our reflecting
layer models, we fit 94 measurements using 10 fitted parameters
(typical), which implies an expected \chisq of 84 and an expected
standard deviation of $\sigma_{\chi^2}$=13.  The expected values for
the multiple scattering fits are increased to \chisq = 110 and
$\sigma_{\chi^2}$=15 (more spectral samples are fit). Deviations from
the expected value of \chisq are a result of a combination of factors:
statistical variability in \chisq itself, errors in estimating
uncertainties, and errors in the physical features of the model.

\section{Evidence for Jupiter's 3-$\mu$m absorption}

\begin{figure*}[!htb]\centering
\includegraphics[width=5.5in]{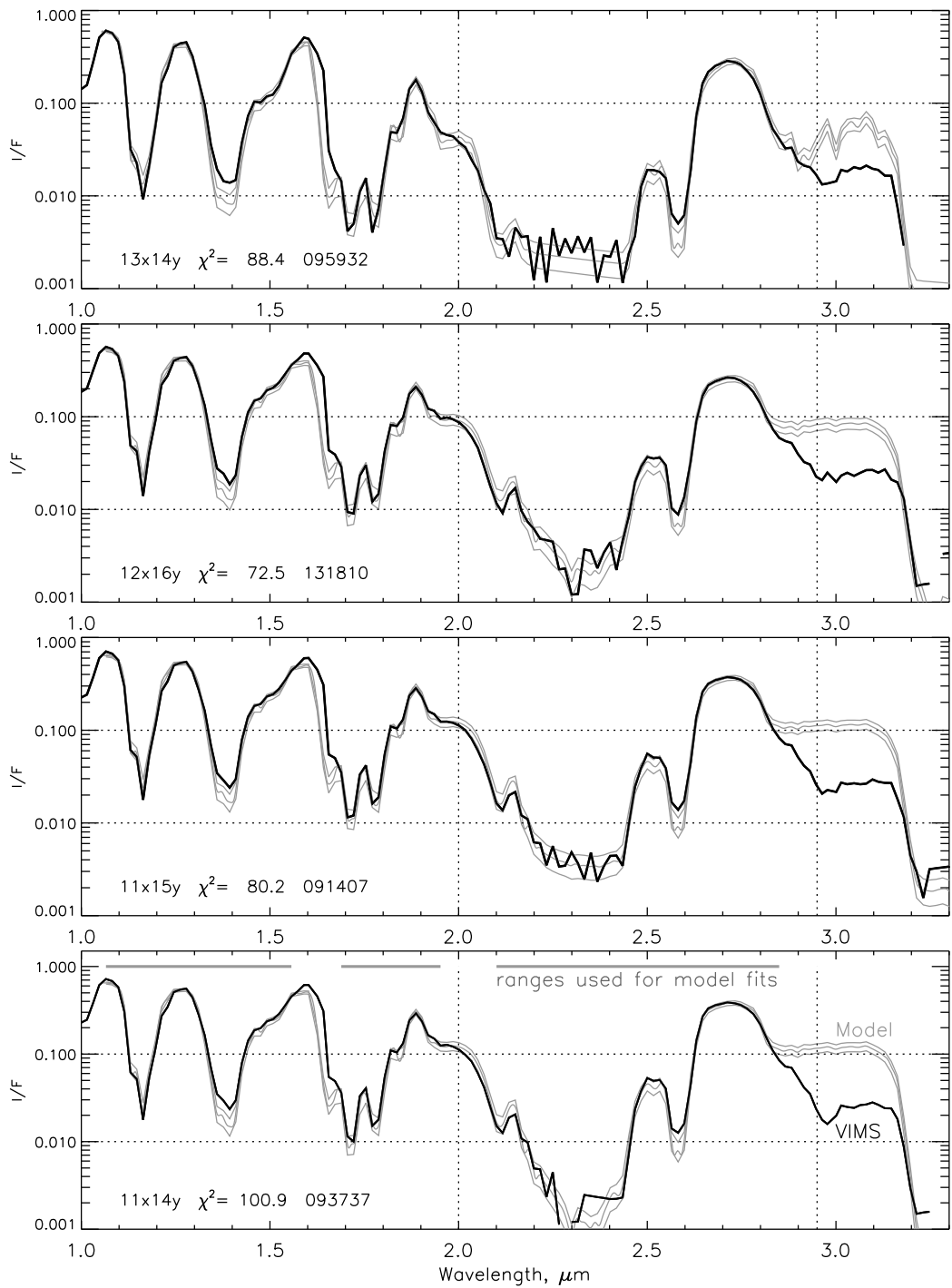}
\caption{Selected low phase angle VIMS spectra (black) compared to
reflecting layer model spectra (gray). Pairs of lines surrounding the
model spectra indicate uncertainties in opacity calculations;
measurement uncertainties (not shown for clarity) are similar but
slightly smaller. The gray horizontal bars in the bottom panel
indicate the ranges over which observations were used to constrain the
models.  Each panel legend provides pixel coordinates in the VIMS
cubes (as in Fig.\ \ref{Fig:imvims1}) and $\chi^2$ values (80$\pm$13 is
expected). Vertical dotted lines indicate pure \nht ice absorption
features at 2.0 and 2.95 \mumx.}
\label{Fig:rlfitspec1}
\end{figure*}

\begin{figure*}[!htb]\centering
\includegraphics[width=6in]{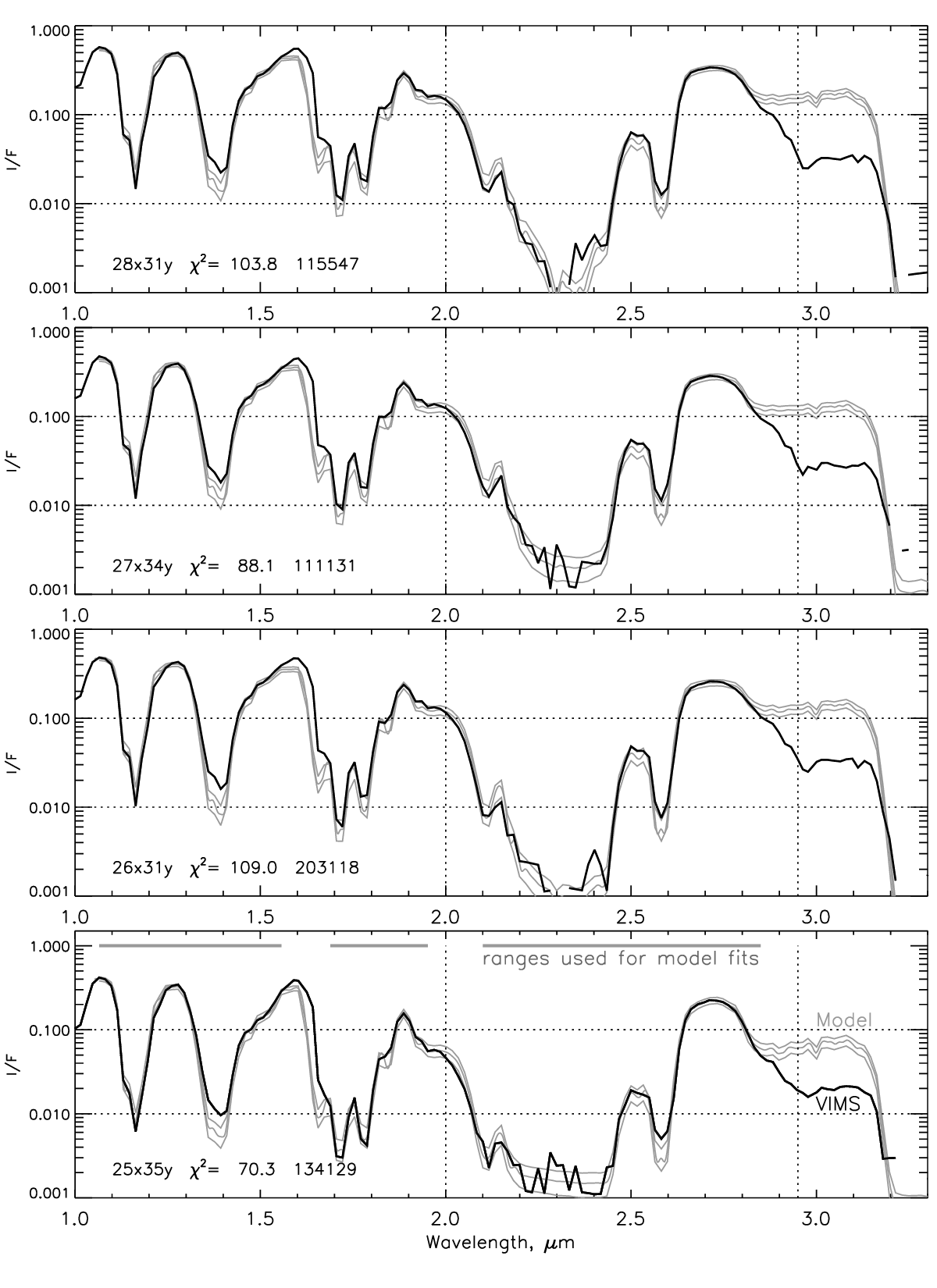}
\caption{As in Fig.\ \ref{Fig:rlfitspec1}, except that the spectra are for medium
phase angles (as in Fig.\ \ref{Fig:imvims2}).}
\label{Fig:rlfitspec2}
\end{figure*}

To clearly display the need for a 3-\mum absorber, we first fit VIMS
spectra with aerosol models that did not contain such an absorber. We
chose the low-latitude VIMS spectra from pixel locations given in
Fig.\ \ref{Fig:imvims1} (low phase angles) and in Fig.\
\ref{Fig:imvims2} (intermediate phase angles).  The model spectra are
compared with observation in Figs. \ref{Fig:rlfitspec1} and
\ref{Fig:rlfitspec2}, respectively.  In these figures the observed
spectra are shown in black, with models shown in gray. The three
spectral regions used to constrain the fits are indicated by gray bars
plotted in the bottom panels.  The region from 2.85 to 3.3 $\mu$m is
omitted from the fits because it contains local absorption and
including it greatly worsens the fit in other regions.  The region
near 2 $\mu$m is avoided because of a potential connection between
absorption at 2 $\mu$m and absorption near 3 $\mu$m if the cloud
absorber is ammonia ice.  The 1.60-1.68 $\mu$m region is also excluded
because of the previously noted VIMS responsivity correction error in
this region.  The fits outside these excluded regions are generally
excellent, with $\chi^2$ values close to expected values.

Each layer in our 3-layer model has three parameters: a pressure $p$,
a fractional reflectivity $f$ at 2 $\mu$m and a wavelength dependence
exponent, $n$.  The fit parameters are listed in
Table\ \ref{Tbl:rlfits} for each of the spectra, and the listed
uncertainties show that the parameters are well constrained by the
observations. The need for significant absorption in the 2.85-3.2
$\mu$m region is apparent from the large ratios between model and
observed spectra in this region (the ratios between the gray and black
curves in Figs.\ \ref{Fig:rlfitspec1} and \ref{Fig:rlfitspec2}).  The
quantitative ratios at 2.04 $\mu$m, 2.96 $\mu$m, and 3.0-3.1 $\mu$m
are listed in Table\ \ref{Tbl:rlratios}.
  We see that as clouds brighten (I/F at 1.96 $\mu$m increases), the
absorption needs to increase in the 3-$\mu$m region, while the
fractional absorption near 2 $\mu$m decreases slightly, suggesting
that the two absorptions are not both due to the same cloud component
(or same material). What appears to be happening here is that as the
absorbing cloud layer thickens, it gets brighter in spectral regions
where it doesn't absorb much, but it can't brighten in regions where
it absorbs strongly (there its reflectivity saturates at a very low optical
depth).  Also, we find that the peak in the 2-$\mu$m absorption is
nearer to 2.04 $\mu$m than it is to the NH$_3$ peak at 2.0 $\mu$m, and
the local dip in reflectivity near 2.97 \mum is shifted from the minimum
single-scattering albedo of pure \nhtx, which is located at 2.95 \mumx (the
peak in the imaginary index is actually located at 2.965 \mumx).

\begin{table}\centering
\caption{Best fit parameter values for 3-layer reflecting-layer fits to selected VIMS spectra.}
%\vspace{0.2in}
\begin{tabular}{|c c c c c| }
\hline
Spectrum & Parameter & Layer 1 & Layer 2 & Layer 3 \\
\hline %\\[-0.05in]
11x14y& {\it p} (bars) & 0.0450$^{+0.014}_{-0.015}$& 0.392$^{+0.02}_{-0.01}$ & 0.919$^{+0.05}_{-0.04}$\\[0.1in]
11x15y& {\it p} (bars) & 0.0020$^{+0.018}_{-0.001}$& 0.380$^{+0.02}_{-0.02}$ & 0.935$^{+0.06}_{-0.05}$\\[0.1in]
12x16y& {\it p} (bars) & 0.0335$^{+0.013}_{-0.015}$& 0.411$^{+0.03}_{-0.03}$ & 0.976$^{+0.13}_{-0.11}$\\[0.1in]
13x14y& {\it p} (bars) & 0.0005& 0.602$^{+0.01}_{-0.01}$ & 1.624$^{+0.16}_{-0.14}$\\[0.1in]
25x35y& {\it p} (bars) & 0.0005& 0.462$^{+0.02}_{-0.02}$ & 1.023$^{+0.16}_{-0.14}$\\[0.1in]
26x31y& {\it p} (bars) & 0.0179$^{+0.019}_{-0.017}$& 0.397$^{+0.02}_{-0.02}$ & 0.754$^{+0.03}_{-0.03}$\\[0.1in]
27x34y& {\it p} (bars) & 0.0019$^{+0.026}_{-0.001}$& 0.343$^{+0.01}_{-0.01}$ & 0.812$^{+0.10}_{-0.09}$\\[0.1in]
28x31y& {\it p} (bars) & 0.0558$^{+0.016}_{-0.014}$& 0.377$^{+0.03}_{-0.03}$ & 0.772$^{+0.09}_{-0.08}$\\[0.1in]
11x14y& {\it f} & 0.0070$^{+0.0022}_{-0.0022}$& 0.183$^{+0.01}_{-0.01}$ & 0.624$^{+0.06}_{-0.05}$\\[0.1in]
11x15y& {\it f} & 0.0058$^{+0.0006}_{-0.0005}$& 0.177$^{+0.02}_{-0.02}$ & 0.591$^{+0.02}_{-0.02}$\\[0.1in]
12x16y& {\it f} & 0.0082$^{+0.0003}_{-0.0003}$& 0.146$^{+0.01}_{-0.01}$ & 0.368$^{+0.02}_{-0.02}$\\[0.1in]
13x14y& {\it f} & 0.0030$^{+0.0003}_{-0.0002}$& 0.209$^{+0.02}_{-0.02}$ & 0.601$^{+0.23}_{-0.17}$\\[0.1in]
25x35y& {\it f} & 0.0024$^{+0.0002}_{-0.0002}$& 0.138$^{+0.01}_{-0.01}$ & 0.373$^{+0.04}_{-0.04}$\\[0.1in]
26x31y& {\it f} & 0.0026$^{+0.0000}_{-0.0000}$& 0.184$^{+0.02}_{-0.02}$ & 0.248$^{+0.04}_{-0.04}$\\[0.1in]
27x34y& {\it f} & 0.0037$^{+0.0019}_{-0.0011}$& 0.192$^{+0.00}_{-0.00}$ & 0.422$^{+0.04}_{-0.04}$\\[0.1in]
28x31y& {\it f} & 0.0090$^{+0.0010}_{-0.0010}$& 0.214$^{+0.01}_{-0.01}$ & 0.465$^{+0.04}_{-0.04}$\\[0.1in]
11x14y& $n$ (in $\lambda^n$) &-2.86$^{+0.24}_{-0.25}$&-0.92$^{+0.16}_{-0.19}$ &-0.15$^{+0.23}_{-0.18}$\\[0.1in]
11x15y& $n$ (in $\lambda^n$) &-2.06$^{+0.35}_{-0.34}$&-1.00$^{+0.19}_{-0.21}$ &-0.19$^{+0.11}_{-0.10}$\\[0.1in]
12x16y& $n$ (in $\lambda^n$) &-1.94$^{+0.46}_{-0.44}$&-1.19$^{+0.10}_{-0.10}$ &-0.24$^{+0.33}_{-0.39}$\\[0.1in]
13x14y& $n$ (in $\lambda^n$) &-2.54$^{+0.12}_{-0.12}$&-0.65$^{+0.11}_{-0.13}$ &-0.51$^{+0.57}_{-0.55}$\\[0.1in]
25x35y& $n$ (in $\lambda^n$) &-1.99$^{+0.26}_{-0.26}$&-0.91$^{+0.16}_{-0.15}$ &-0.26$^{+0.38}_{-0.38}$\\[0.1in]
26x31y& $n$ (in $\lambda^n$) &-2.86$^{+0.49}_{-0.41}$&-0.85$^{+0.16}_{-0.17}$ &-0.11$^{+0.08}_{-0.08}$\\[0.1in]
27x34y& $n$ (in $\lambda^n$) &-2.17$^{+0.29}_{-0.30}$&-0.85$^{+0.10}_{-0.10}$ & 0.15$^{+0.39}_{-0.47}$\\[0.1in]
28x31y& $n$ (in $\lambda^n$) &-2.34$^{+0.25}_{-0.24}$&-0.75$^{+0.07}_{-0.07}$ & 0.05$^{+0.59}_{-0.55}$\\[0.1in]
\hline%\\[-0.06in]
\end{tabular}\label{Tbl:rlfits}%\par
\parbox{3.3 in}{NOTE: Layer-1 pressures without uncertainties had best fit values at the lower limit of the fit range.}
\end{table}

\begin{table*}\centering
\caption{Fitted spectra compared to VIMS spectra in regions of enhanced absorption.}
%\vspace{0.1in}
\begin{tabular}{|c | c  | c c c| }
\hline 
 &  Measured & \multicolumn{3}{c}{Ratio of model I/F to measured I/F}\\
 spectrum  &  I/F at 1.95 $\mu$m & at 2.04 $\mu$m & at 2.96 $\mu$m & at 3.0-3.1 $\mu$m\\[0.05in]
\hline &&&& \\[-0.06in]
  13x14y &    0.048 &    1.300 &    3.44 &    2.76\\[0.1in]
  25x35y &    0.058 &    1.261 &    4.32 &    3.35\\[0.1in]
  12x16y &    0.095 &    1.191 &    4.03 &    3.40\\[0.1in]
  11x15y &    0.124 &    1.233 &    5.45 &    4.24\\[0.1in]
  11x14y &    0.125 &    1.211 &    6.65 &    4.68\\[0.1in]
  26x31y &    0.133 &    1.200 &    5.43 &    4.06\\[0.1in]
  27x34y &    0.137 &    1.180 &    4.55 &    4.36\\[0.1in]
  28x31y &    0.163 &    1.143 &    6.54 &    5.02\\[0.1in]
\hline
\end{tabular}\label{Tbl:rlratios}\par
\par
%\vspace{0.1in}
\parbox[l]{3.9 in}{Note: above rows are in order of increasing I/F at 1.95 \mumx.}\par
\end{table*}

\begin{figure}[!htb]\centering
\includegraphics[width=3.5in]{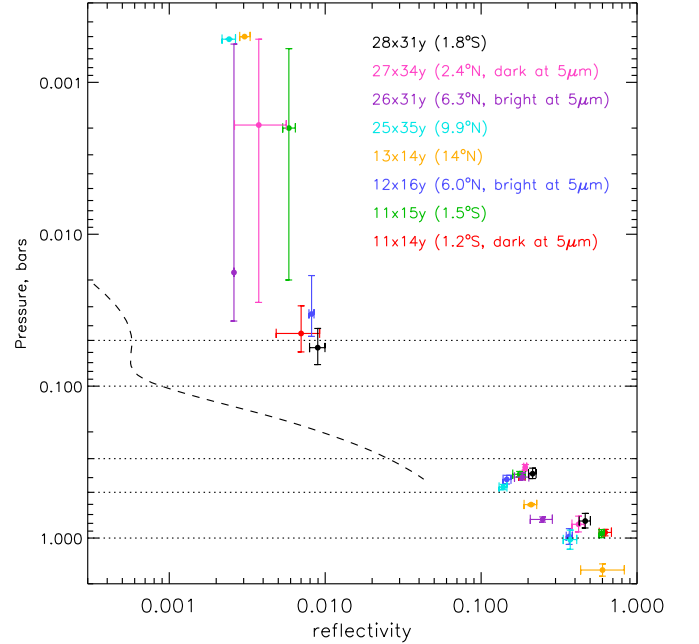}
\caption{Pressure and reflectivity parameters for 3-layer reflecting
layer models that provide best fits to VIMS spectra, as given in
Table\ \ref{Tbl:rlfits}. The dashed curve displays the vertically
integrated reflectivity derived by \cite{Banfield1998NIR} for the GRS
but is broadly representative of other regions as well.}
\label{Fig:rlfitpars}
\end{figure}

The overall character of the 3-layer model fit parameters is
illustrated in Fig.\ \ref{Fig:rlfitpars}.  A high altitude aerosol is
needed to account for the I/F in the 2.3-$\mu$m region, but there
are two problems with the result.  The effective pressure of this layer is
only well constrained for the 11x14y, 12x16y, and 28x31y spectra, for
which P $\sim$50 mb.  At that level the reflectivity is $\sim$0.008
($\tau\sim$0.03 for conservative symmetric scatterers), which is
nearly 20 times the integrated value obtained by
\cite{Banfield1998NIR}.  Banfield et al. measured low-latitude I/F values of
2-3$\times 10^{-4}$ at 2.3 $\mu$m, while the VIMS values in Fig.\
\ref{Fig:rlfitspec1} are $\sim$10 times larger and are likely due in
part to a VIMS measurement artifact, possibly due to scattering of
light from the VIMS grating, as described in Sec.\ \ref{Sec:artifact}. 
Our previous comparison with NICMOS observations (Fig.\
\ref{Fig:f237mcomp}) also indicates that a simple offset error is not
consistent with the observations.  The substantial difference between
the \cite{Banfield1998NIR} reflectivity and our value in the 400-mb
layer might also be a result of excessive VIMS I/F values at low
levels of I/F.

The reflecting layer fits put the second layer in the 340-600 mb
 region, with a reflectivity of 15-21\%.  The third layer, located in
 the 770-1600 mb region has a relatively high reflectivity of 40-60\%.
 The best-fit wavelength dependence exponents are roughly in accord
 with expectations.  The high altitude contributions have an exponent
 of -2 to -2.9, indicating relatively small particles.  This might
 have been closer to -4 if the excess I/F near 2.3 \mum could have
 been removed.  The next layer has best-fit exponents in the -0.75 to
 -1 range, which indicates larger particles, and the deepest layer has
 even smaller exponents (-0.5 to +0.15), indicating even larger
 particles.  We also tried fitting the observations with gas opacity
 profiles other than the nominal EZ profile, but even for the least
 opaque region sampled in the NEB, the best fit was obtained with the
 EZ profile.

\begin{figure}[!htb]\centering
\includegraphics[width=3.5in]{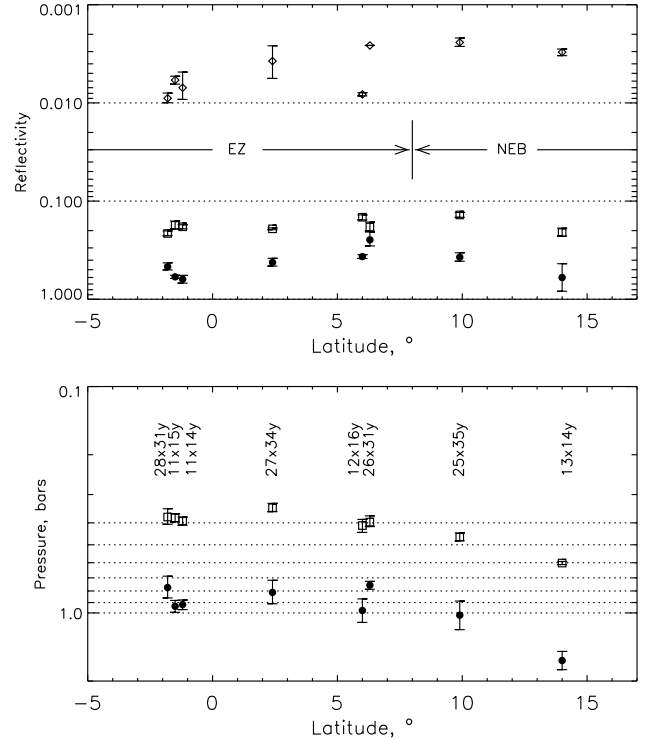}
\caption{Pressure and reflectivity parameters versus latitude for
best-fit 3-layer reflecting layer models given in Table\
\ref{Tbl:rlfits}. Parameters are plotted for layer 1 (diamonds), layer
2 (open squares), and layer 3 (filled circles).}
\label{Fig:rlfitvlat}
\end{figure}

The latitude dependence of the fit parameters (Fig.\
\ref{Fig:rlfitvlat}) displays an increase in pressure towards the
north equatorial belt for both layers 2 and 3. (The Cassini VIMS
0.45-\mum image in Fig.\ \ref{Fig:imvims1}D indicates that the NEB
extended from about 8\degx N to 18\degx N in 2000.) The trends in
Layer 1 pressures is not shown because it is likely not meaningful (several
fits yielded best-fit pressures at the lower limit of the allowed range).
The pressure changes seem to have more impact on the observed I/F than
the changes in optical depths.

The pressure level at which the 3-\mum absorber resides is roughly
that of layer 2 in the 3-layer model, which is in the 300-600 mb range
where \nht ice is a plausible condensate. The location is inferred
from the cloud reflectivity and the penetration depth profile given in
Fig.\ \ref{Fig:pendepth}. Vertical penetration depths at 2 \mum and 3
\mum are comparable and limited to pressures less than 1 bar or so.
The relatively high reflectivity seen at 2 \mum (compared to that at 3
\mumx) is produced by contributions mainly from layer 2, which would
be seen even better at 3 \mumx. It thus follows that layer 2 must be
strongly absorbing at 3 \mum to account for its much lower I/F at that
wavelength.  However, the layer-2 and layer-3 fractions in this model
are too large for the reflecting layer equations to provide physically
meaningful parameter values if the clouds are actually uniform and
translucent rather than broken and opaque.  For the case of uniform
translucent clouds, we need multiple scattering fits, which are
presented after first identifying a plausible composition for the
particles in such clouds.

\section{Modeling the 3-\mum absorption.}

\subsection{Candidate 3-$\mu$m absorbers.}\label{Sec:candidate}

\begin{figure*}[!htb]\centering
\includegraphics[width=6in]{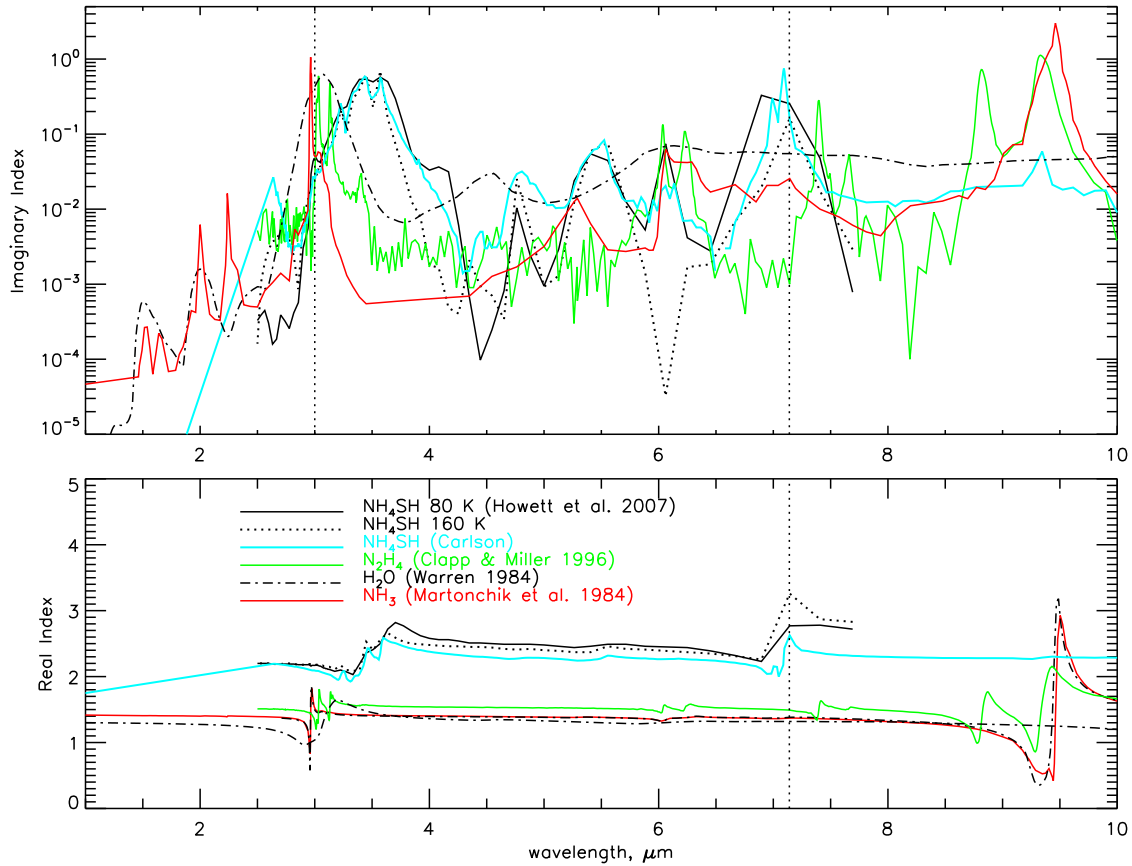}
\caption{Real (bottom) and imaginary (top) components of the
refractive index vs. wavelength for candidate 3-$\mu$ absorbers,
including H$_2$O results of \cite{Warren1984}, \nht results of \cite{Martonchik1984},
NH$_4$SH results of \cite{Howett2007} and Barbara Carlson (personal
communication), and N$_2$H$_4$ results of \cite{Clapp1996}.}
\label{Fig:indexplot}
\end{figure*}

Candidate cloud materials expected in Jupiter's atmosphere that also
absorb light in the 3-$\mu$m region include NH$_3$, NH$_4$SH (ammonium
hydrosulfide), N$_2$H$_4$ (hydrazine), and water ice.  Their real and
imaginary refractive indexes are shown vs. wavelength in Fig.\
\ref{Fig:indexplot}.  Because of its very high imaginary index, water
ice cannot produce the absolute reflectivity level needed at 3 \mum
and the transparency needed at 2.7 \mumx. Trial fits with water ice
resulted in very large $\chi^2$ values.
%In fact, when we tried to use water ice in our
%models, our fitting algorithm reduced the opacity of the water ice
%layer to nearly zero, and the resulting $\chi^2$ was triple the value
%obtained with our best candidate absorber. 
Water is also a poor candidate because of its extremely low mixing
ratio in the pressure range where the 3-\mum absorber seems to be
located.  Hydrazine has a pair of strong absorption peaks between 3
and 3.2 $\mu$m, which were very apparent in in preliminary model
spectra, and thus seemed inconsistent with observations.  Among the
more plausible compounds, only NH$_3$ has a known absorption at 2
$\mu$m (the 2.25 $\mu$m absorption feature of \nht would not be
visible due to overlying gas absorption).
% The strong ammonia absorption near 2.95 $\mu$m is also muted in
%larger particles.  
\nhfsh has an absorption that is roughly comparable to that of NH$_3$
at 3 $\mu$m (except for the sharp \nht feature at 2.95 \mumx), but
continues to increase beyond 3.1 $\mu$m where NH$_3$ absorption drops
significantly.  In the following detailed discussion we only include
ammonia and ammonium hydrosulfide as viable candidates (see
\cite{Sro2010iso} for an evaluation of a broader range of materials).

\begin{figure*}[!htb]\centering
\hspace{-0.3in}
\includegraphics[width=3.5in]{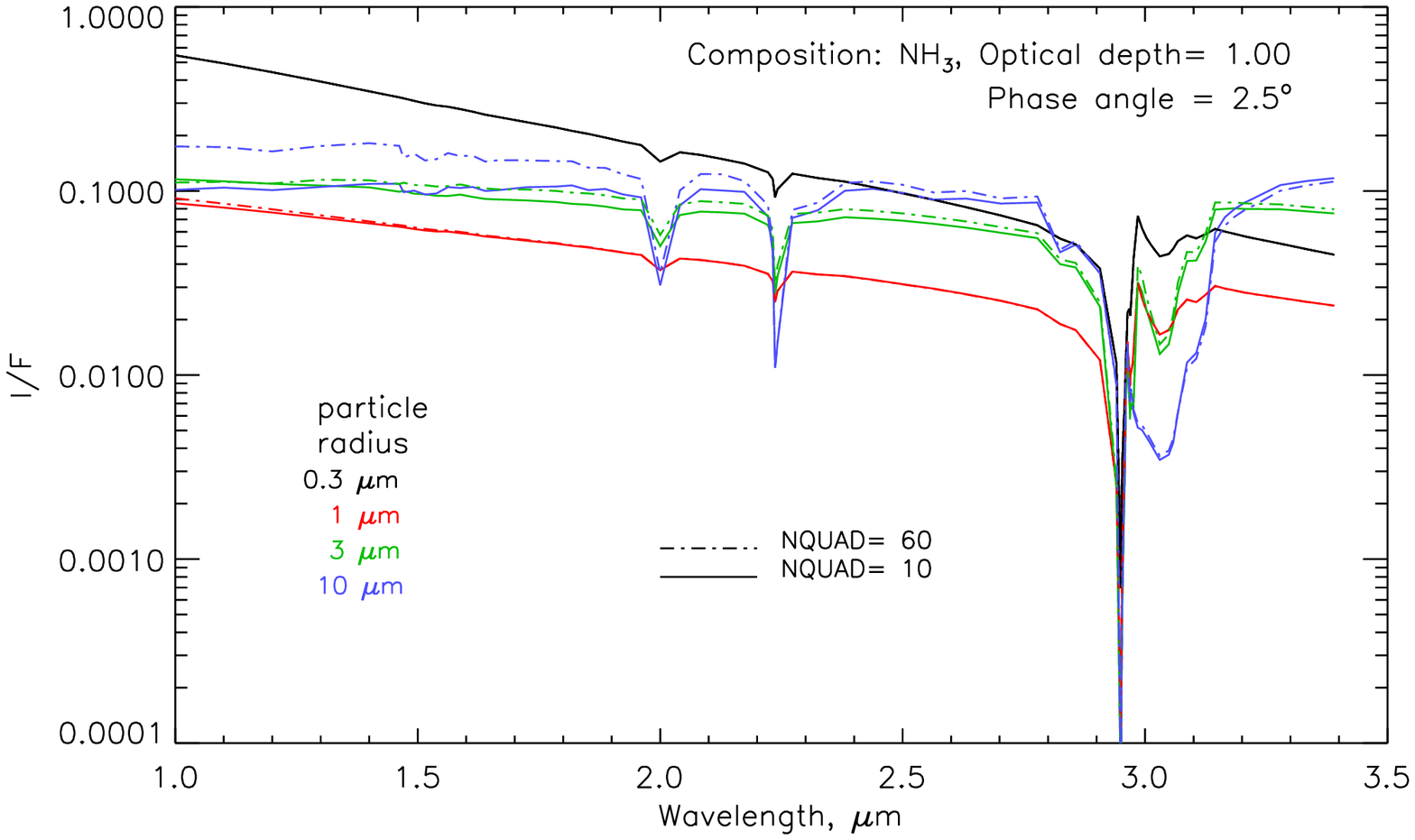}
\includegraphics[width=3.5in]{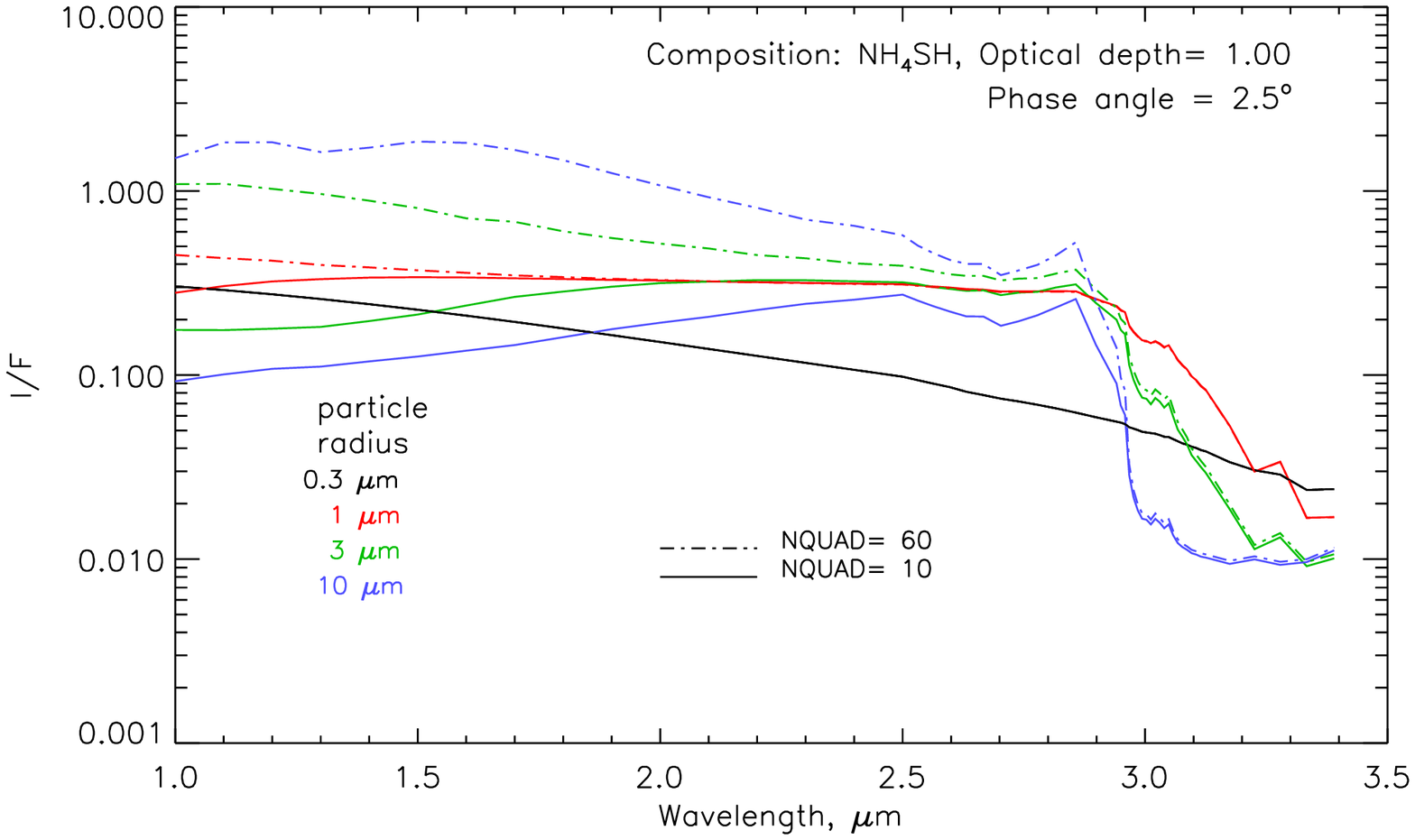}\par
\hspace{-0.3in}
\includegraphics[width=3.5in]{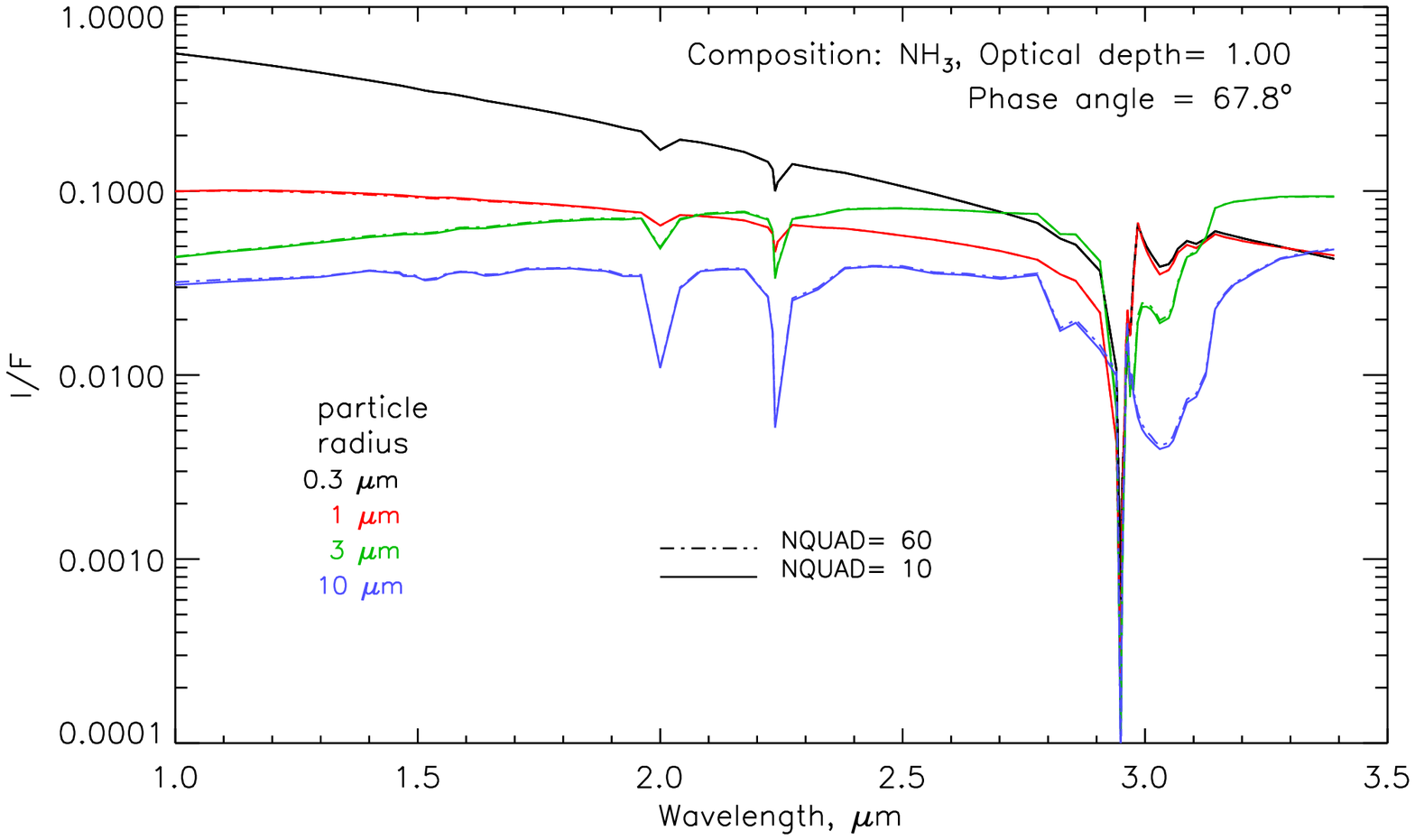}
\includegraphics[width=3.5in]{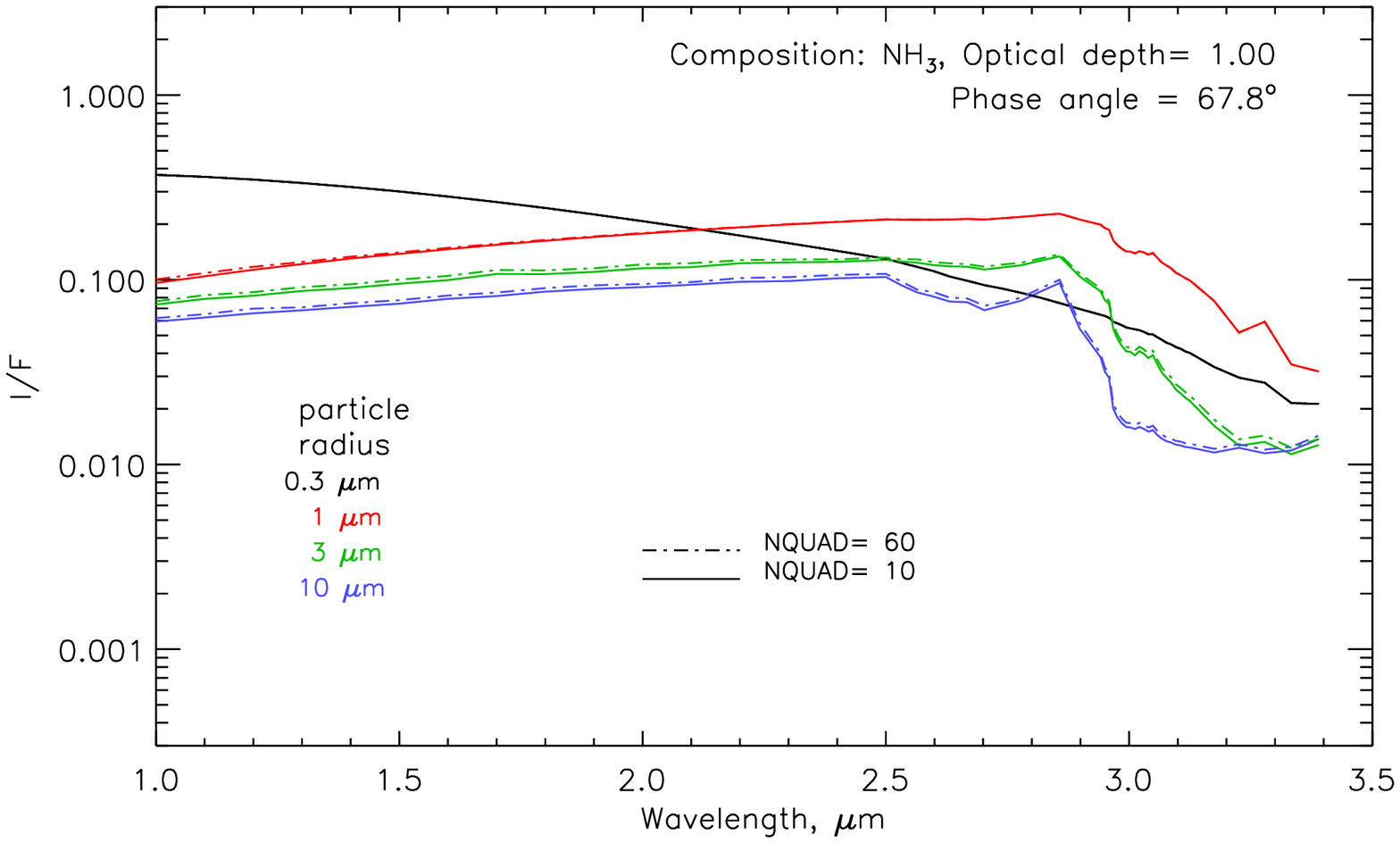}\par
\caption{I/F vs. wavelength for unit optical depth layers of spherical
NH$_3$ (left) and NH$_4$SH (right) particles of radii from 0.3 $\mu$m
to 10 $\mu$m. Results for two phase angles are shown: 2.5\deg (top
panels) and 67.8\deg (bottom panels) and for NQUAD values of 10 (solid)
and 60 (dot-dash). Note that the \nhfsh calculations
between 1 and 2.5 \mum are based on interpolation of refractive index
measurements at 1 and 2.5 \mumx.  Frost reflection spectra
\citep{Fanale1977} show broad absorption features that may be
significant between 1 and 2 \mumx, but no refractive index values are
available.}
\label{Fig:reflections}
\end{figure*}

The cloud layers at which the 3-$\mu$m and 2 -$\mu$m absorptions
contribute do so primarily through their reflected intensities.  Thus,
it is useful to consider the spectral reflectivity of unit optical
depth layers of NH$_3$ and NH$_4$SH for a variety of particle sizes
and for low and medium phase angle observations. As shown in Fig.\
\ref{Fig:reflections}, the reflectivity plots provide clues as to what
sorts of layers can provide the appropriate shortwave reflectivity and
the needed 3-$\mu$m absorption.  Fig.\ \ref{Fig:reflections} also
displays the effect of NQUAD (quadrature angles per hemisphere) on the
accuracy the computed reflectivity.  The difference between
calculations for NQUAD values of 10 and 60 show that as particle size
increases much beyond a few microns model computations for low phase
angles are much more demanding than for medium phase angles. This
occurs because the scattering phase function is very smooth at middle
phase angles (and scattering angles) but becomes quite sharp and more
strongly wavelength dependent for large size parameters in near
backscatter configurations (low phase angles).

A prominent feature in the reflectivity of ammonia layers is the deep
and sharp minimum at 2.95 $\mu$m. The lack of a such a feature in the
VIMS spectra, or in the higher resolution ISO spectrum
\citep{Sro2010iso} implies that the top cloud layer on Jupiter cannot
be made of pure \nht ice.  If only ammonia is considered for the
middle layer (500-700 mb), then only relatively large particles
provide roughly the correct spectral variation in reflectivity. But
even then, to fill in the deep minimum at 2.95 \mum some additional
reflectivity must be provided by a gray particle layer extending above
the ammonia layer.  Small particle layers of ammonia (0.25-1 $\mu$m in
radius) are relatively too bright just beyond 3 \mumx, and they also
contribute too much reflection at short wavelengths. Large ammonia
particles produce a relatively strong 2-$\mu$m feature that is too
strong to be compatible with observations, as pointed out by
\cite{Irwin2001BZ}. The \nhfsh layers provide a flatter absorption
between 3 and 3.4 $\mu$m, with a slope between 3 and 3.1 $\mu$m that
depends on particle size.  Note that \nhfsh particle layers are
brighter than \nht layers of the same optical depth and particle size,
a result of the higher real index of \nhfshx.

The VIMS spectra display a significant variation in brightness at 1.95
$\mu$m, without a significant variation in brightness at 3 $\mu$m, and
the fractional absorption at 2 $\mu$m remains relatively constant,
while the fractional absorption at 3 $\mu$m changes by a factor of
two.  Another feature of the spectra is the small dip at 2.97 $\mu$m,
which is close to the 2.95-\mum dip that is characteristic of an
ammonia ice layer.  These facts suggest that both absorbers are
present at pressure levels that are visible at 2 $\mu$m and 2.9-3.3
$\mu$m, with NH$_4$SH providing most of the cloud mass and most of the
3-$\mu$m absorption, and ammonia ice contributing a small amount of 2
$\mu$m absorption and helping to produce the small dip at 2.97 $\mu$m.
These suggestions are quantitatively evaluated in the following
section of multiple scattering model fits.

\subsection{Models with an ammonia cloud layer.}

\begin{figure*}[!htb]\centering
\includegraphics[width=6in]{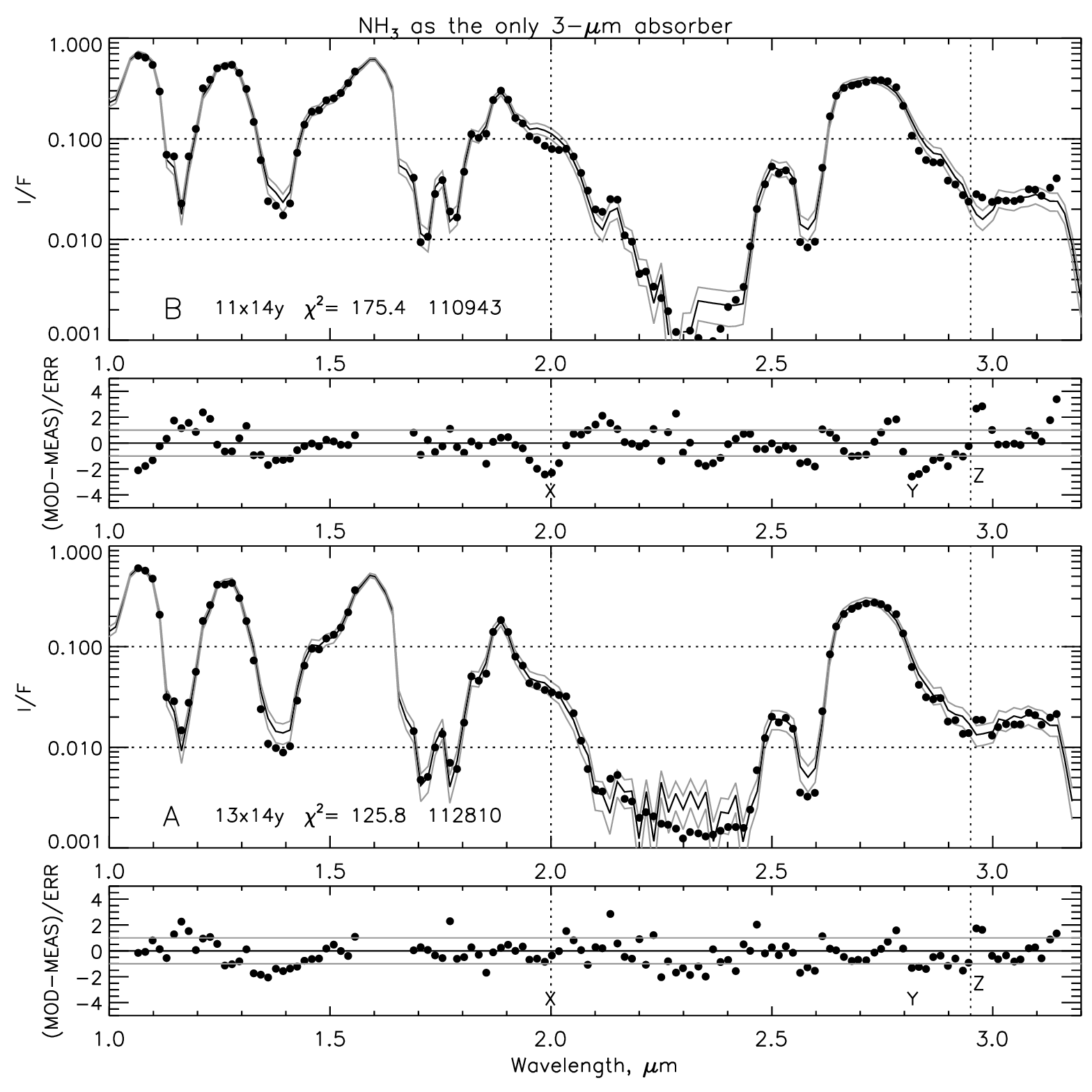}
\caption{Multiple-scattering 4-layer fits to lowest (A) and
highest opacity (B) cloud structures at low phase angles.  These fits use
NH$_3$ Mie particles for the third layer, which fits well only for low
opacity clouds. Model spectral points are displayed as filled circles,
measurements as solid lines with gray lines indicating
uncertainties. Vertical dotted lines at 2.0 and 2.95 \mum mark
locations of minima in the single scattering albedo \nht ice for
particle radii from 0.2 to 20 \mumx. Companion plots show the difference
between model and measurement divided by the expected combined error.
Letters X, Y, and Z mark locations of noteworthy differences.}
\label{Fig:msnh3fits}
\end{figure*}

We first considered the possibility that \nht could provide all the
needed 3-\mum absorption. Our initial multiple scattering models
consisted of four scattering layers. The top layer was a small
particle haze characterized by symmetric Henyey-Greenstein (H-G) phase
function, a $\lambda^{-3}$ wavelength dependence, an adjustable
optical depth, and a fixed pressure of either 15 mb or 50 mb, with the
choice guided by the reflecting layer fits.  This top layer is needed
to provide reflectivity in the 2.3-$\mu$m region.  The second layer is
a Mie-scattering layer of conservative particles with adjustable
particle size, pressure and optical depth. The third layer consists of
spherical NH$_3$ ice particles, with adjustable particle size,
pressure and optical depth.  The fourth layer was modeled as a nearly
conservative layer ($\varpi$ = 0.997) using the H-G phase function of
\cite{Sro2002jup} and a $\lambda^{n}$ wavelength dependence, with
fit-adjustable pressure, optical depth, and $\lambda$ exponent $n$.
We also found comparable fits treating this as a Mie-scattering layer
of NH$_4$SH particles, although in that case a stronger backscatter
peak is needed to provide enough reflectivity. Without this deeper
layer it appears impossible to match the spectral shape of the
reflectivity peak near 1.23 \mumx.  This structure, with \nht as the
only aerosol absorber, provides an excellent fit to the lower opacity
regions, such as the 13x14y spectrum, as shown in Fig.\
\ref{Fig:msnh3fits}A,B.  In this example the ammonia layer pressure
(673 mb) is high enough that its spectral features at 9.6 $\mu$m and
26 $\mu$m would likely be hidden from view in observed thermal emission
spectra, where such features are not easily observed.

In regions of higher opacity, such as location 11x14y in the low phase
angle cube, the spectra cannot be as accurately fit with this
structure.  As shown in Fig.\ \ref{Fig:msnh3fits}C and D, there is an
absorption at 2 $\mu$m in the model spectra that does not appear in
the observed spectra.  The best-fit parameter values for both
fits are listed in Table \ref{Tbl:3compfits}.

\begin{table*}\centering
\caption{Best-fit cloud structures with single and multiple 3-$\mu$m absorbers.}
%\vspace{0.1in}
\begin{tabular}{|c | c c c | c c c|}
\hline
Parameter & \multicolumn{3}{c}{Low Opacity Spectrum (13x14y)} & \multicolumn{3}{c}{High Opacity Spectrum (11x14y)} \\[0.06in]
\hline %& & & & %\\[-0.1in]
Layer-2 Composition &  1.40 + 0i    &    1.40 + 0i & NH$_3$ shell*  & 1.40 + 0i  & 1.40 + 0i   &  NH$_3$ shell*\\[0.06in]
Layer-3 Composition &  NH$_3$    &  NH$_4$SH & NH$_4$SH&  NH$_3$  &  NH$_4$SH  &  NH$_4$SH \\[0.06in]
\hline %& & & & %\\[-0.1in]
$p_1$ (bars)         &   0.015      &   0.015  &0.015  &   0.050 & 0.050 & 0.050\\[0.06in]
$p_2$ (bars)         &   0.517      &   0.553  &0.546  &   0.303 & 0.307 & 0.336\\[0.06in]
$p_3$ (bars)         &  0.628       &   0.608  &0.566  &   0.625 & 0.646 & 0.635\\[0.06in]
$p_4$ (bars)         &  1.261       &   1.158  &1.128  &   1.616 & 1.285 & 1.252\\[0.06in] 
\hline %& & & & \\[-0.1in]
\large$\tau_1$\normalsize & 0.014 & 0.014 &0.014   &   0.023 &  0.023 & 0.025\\[0.06in]
\large$\tau_2$\normalsize & 0.270 & 0.369 &0.414   &   0.306 &  0.291 & 0.420\\[0.06in]
\large$\tau_3$\normalsize & 1.569 & 0.386 &0.231   &   6.63  &  2.83  & 2.10\\[0.06in]
\large$\tau_4$\normalsize & 4.66  & 5.51  & 5.19   &  14.50  &  20.0  & 10.64\\[0.06in] 
\hline %& & & & %\\[-0.1in]
Layer-1 $\lambda$ exponent & -3   &-3     & -3     & -3      & -3     & -3 \\[0.06in]
$r_2$ ($\mu$m)      &  0.289       & 0.293   & 0.303  &   0.265 &  0.237 & 0.259\\[0.06in]
$r_3$ ($\mu$m)      &  6.98        & 13.35   & 6.86   &   6.51  &  12.54 & 10.80 \\[0.06in]           
Layer-4 $\lambda$ exponent & -1.75&-1.21  & -1.25  &   -1.28 & -0.625   & -1.12 \\[0.06in]
\hline
% & & & & \\[-0.1in]
$\chi^2$        &  125.79       &129.70   &118.06  &  175.41 & 160.47 & 124.44 \\[0.06in]
%REFERENCE(P,LOG)  129,I        128,I     8,III        126,I     127,I   3,II
\hline
\end{tabular}\label{Tbl:3compfits}\par
\par
%\vspace{0.1in}
\parbox[l]{5.1 in}{Notes: For layer 1 we used a symmetric phase
function and fixed pressures and $\lambda$ exponents. For layer 4 we
used the double Henyey-Greenstein phase function of
\cite{Sro2002jup}. Parameter uncertainties are given for the composite
fit in Table\ \ref{Tbl:msfits}. Optical depths are evaluated at 2
$\mu$m. The \nht shell* is a simulation of an \nhtx-coated particle as
described in the text. \chisq differences less than 15 are of marginal
significance.}\par
\end{table*}

\subsection{Models with NH$_4$SH as the only 3-$\mu$m absorber.}

\begin{figure*}[!htb]\centering
\includegraphics[width=6in]{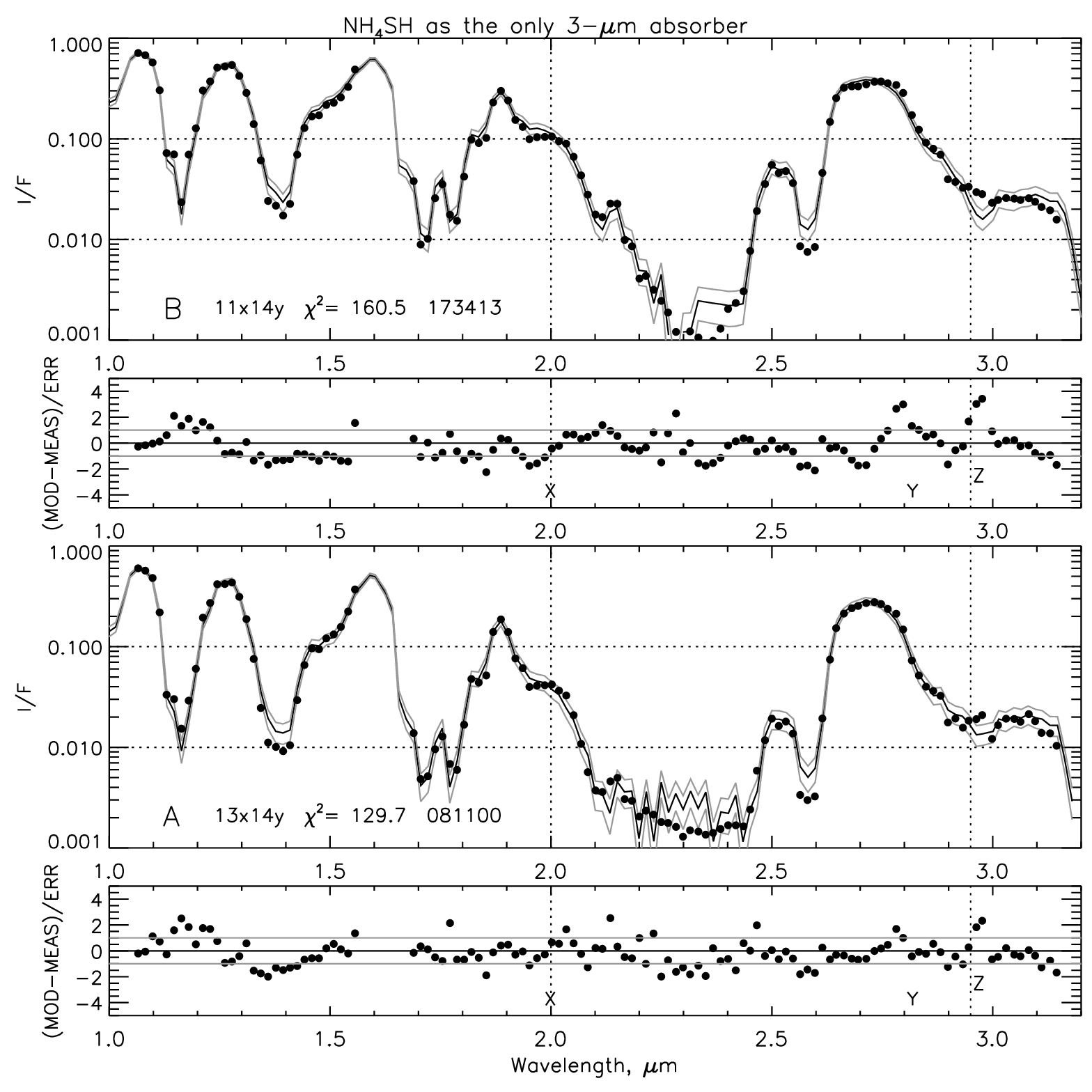}
\caption{
As in Fig.\ \ref{Fig:msnh3fits}, except these 4-layer multiple
scattering fits use NH$_4$SH Mie particles for the third
layer. }
\label{Fig:msnh4shfits}
\end{figure*}

We next considered the possibility that NH$_4$SH could provide all the
absorption at 3 $\mu$m.  This would certainly eliminate the problem of
excessive absorption at 2 $\mu$m, although it would not be able to fit
the feature at 2.97 $\mu$m.  For this study we used a model similar to
that used for the NH$_3$ analysis, with the third layer consisting of
NH$_4$SH instead of NH$_3$. We used a composite refractive index
function for \nhfshx: at wavelengths where \cite{Howett2007}
observations exist we used the average of their 80 K and 160 K
measurements; elsewhere we used Barbara Carlson's compilation
(personal communication 1994).  Again we tried the two extreme cases
(13x14y for low cloud opacity, and 11x14y for high cloud opacity).
The best-fit spectra are displayed in Fig.\ \ref{Fig:msnh4shfits} and
the corresponding parameters given in Table\ \ref{Tbl:3compfits}. The
low-opacity fit is slightly worse than the \nhtx-only fit (by only
0.26$\times \sigma_{\chi^2}$), but the high opacity fit is better (by
1$\times \sigma_{\chi^2}$) than we obtained using NH$_3$ as the sole
3-$\mu$m absorber.  
%But these fits also have deficiencies: as
%expected, they don't fit the feature at 2.97 \mumx, and still don't
%provide close fits in the 1.9-2.0 $\mu$m region.  
However, the NH$_4$SH fits have specific local deficiencies: as expected,
they don't fit the feature at 2.97 \mumx, they still don't provide
close fits in the 1.9-2.0 $\mu$m region, and they have problems fitting the
2.7-$\mu$m peak and its long-wavelength shoulder.

There is also some question concerning the appropriate \nhfsh
absorption model to use.  The standard equilibrium chemistry model
places \nhfsh condensation near 160 K, and particles formed at that
level and lofted to higher altitudes will retain the same crystal
structure, as evident from laboratory measurements of
\cite{Ferraro1980}.  On the other hand if particles formed at somewhat
lower temperatures, a different crystal structure would apply and the
absorption would be close to that measured at 80 K.  By picking an
average of measurements at 80 K and 160 K, we are allowing for a 50:50
mix of formation temperatures above and below 160K.  A trial spectral
fit to the 11x14y spectrum, assuming that all the \nhfsh was formed at
high temperatures and that \nhfsh was the sole absorber, was
somewhat degraded: \chisq increased by
1.5$\times \sigma_{\chi^2}$. A trial spectral fit using the 80 K
absorption model for \nhfshx, was also worse, but less significantly so
(\chisq increased by 0.8$\times \sigma_{\chi^2}$).  Thus the
observations slightly favor a mix of formation temperatures.

We see that neither NH$_3$ alone nor NH$_4$SH alone provides accurate
fits to spectra over the full range of cloud structures.  \nhfsh seems
useful for fitting the high opacity cases and either \nht or \nhfsh
are equally useful for fitting the low-opacity cases (the improvement
of \nht over \nhfsh in those cases is really insignificant).  It is
worth noting however, that \nhfsh and \nht have somewhat complementary
fitting defects: (1) \nht tends to be low near 2.85 \mum where \nhfsh
tends to be slightly higher (see location Y in Figs.\ \ref{Fig:msnh3fits} and
\ref{Fig:msnh4shfits}); (2) \nht tends to be increasingly high beyond
3.05 \mum where \nhfsh tends to be increasingly low.  A combination of
the two absorbers thus offers the possibility of reducing the net
fitting defects in these areas.  But the most compelling reason to
keep both \nhfsh and \nht particles in the model is that when \nht is
combined with a non-absorbing material it becomes possible to fit the
2.97-\mum feature (location Z in Figs.\ \ref{Fig:msnh3fits} and
\ref{Fig:msnh4shfits}). While the \nht reflectivity minimum for pure
\nht particles is displaced from this feature, a shift towards the
peak in the \nht imaginary index (at 2.965 \mumx) can be accomplished if
\nht is diluted by 
coating or mixing it with a conservative material, or by condensing it
as a coating on a conservative core, as will be shown in the following
section.

\subsection{Models with both NH$_3$ and NH$_4$SH layers.}

We first considered a structure similar to the previous models except
that we inserted a layer of \nht particles between a conservative Mie
layer and a Mie layer of \nhfsh particles, which is essentially the same structure used
by \cite{Sro2010iso} to successfully fit the ISO spectrum.  This
approach produced relatively good fits for the low-opacity spectra,
such as 13x14y and 25x35y, with best fits obtained with all three
particle layers at nearly the same pressure.   However, this
structure did poorly at fitting the high-opacity spectra, such as 11x14y,
for which the result was a poor match to the 2.97-$\mu$m
feature and too low an I/F in the 1.95-2.1 \mum region.  This suggests that the
structure inferred from the ISO spectrum does indeed not apply at all
locations, which is not surprising, given the large FOV of the ISO
spectrum and the wide variety of clouds averaged together.

\begin{figure*}[!htb]\centering
\includegraphics[width=6in]{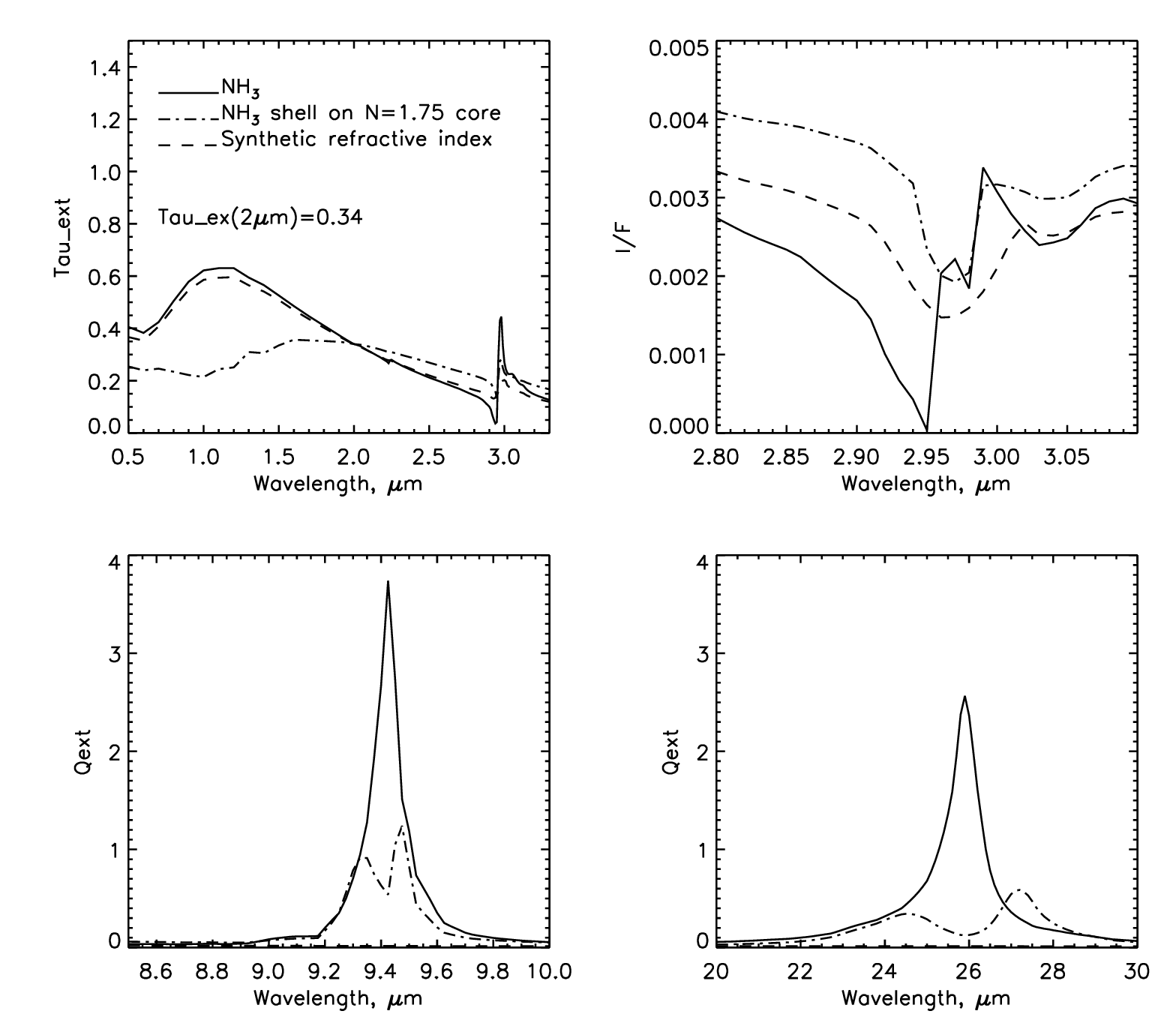}
\caption{Comparison of scattering properties of 0.8-\mum radius spherical
particles of pure \nht (solid curves), N=1.75 cores coated with \nht shells of 20\% of
total radius (dot-dash), and particles with a synthetic refractive index described
in the text (dashed). Shown are extinction optical depth normalized to
0.34 at 2 \mum (upper left), extinction efficiency near 9.4 \mum (lower left)
and 26 \mum (lower right), and reflectivity of a thin layer in the backscatter
direction near 2.95 \mum (upper right). }
\label{Fig:coatex}
\end{figure*}

We then considered the possibility raised by \cite{Sro2010iso} that an
\nhtx-coated particle might provide a better fit, either as a coating
on the larger \nhfsh particles, or as a coating on the smaller
conservative particles usually found near 350 mb.  We computed
scattering properties of coated spheres using the Fortran subroutine
DMiLay, based on original work by \cite{Toon1981}, and further refined
by W. Wiscombe (program code and documentation available at
ftp://climate1.gsfc.nasa.gov/wiscombe/). Although the 2.95-\mum
absorption feature of pure \nht is not seen in the VIMS spectra, the
feature near 2.97 \mum could still be due to \nht when it is applied
as a coating.  An example is provided in Fig.\ \ref{Fig:coatex}. The
upper right panel of that figure displays the reflectivity of a thin
layer of spherical particles of 0.8 \mum in radius, viewed in a
backscatter configuration.  When the particles are composed of pure
\nhtx, a deep minimum is seen in the layer I/F, exactly at 2.95 \mum
(shown as the solid curve).  But when the particles are composed of
N=1.75 material out to 80\% of the radius, and the remainder consists
of \nhtx, the I/F minimum is shallow and shifted close to the
2.97-\mum peak in the imaginary index of \nhtx. A similar effect is
seen when an ammonia core is covered by a heavy coat of benzene
(thickness equal to core radius), as shown in Fig. 5 of
\cite{Kalogerakis2008}.  It is also noteworthy that the coated
particle has a greatly reduced extinction efficiency relative to pure
\nht particles at thermal wavelengths, making them less likely to be
detected in thermal spectra.

Also shown in Fig.\ \ref{Fig:coatex} is the I/F for uniform particles
with a synthetic refractive index that roughly simulates the
reflectivity of the composite particle. This allows us to incorporate
the essence of a coated particle into the Mie calculations contained
in our multiple scattering model without the time penalty of carrying
out the more complex and time-consuming coated sphere calculations.  We created the
synthetic imaginary index by inserting a Gaussian absorption feature
of amplitude 0.08 and FWHM of 0.06 \mum at 2.98 \mumx, and using 1/2
of the \nht imaginary index, wherever that exceeded the wings of the
absorption feature.  We then using the Kramers-Kronig relation,
following \cite{Howett2007}, to compute a spectrally varying real index
that was consistent with the imaginary index.
%, starting with a real index of 1.4.  
The synthetic index provides an absorption feature that crudely
matches the coated-sphere feature. However, the detailed character of
this absorption depends on particle size, core material, shell
thickness, and probably particle shape.  Thus, the example shown in
Fig.\ \ref{Fig:coatex} should be considered suggestive, rather than
typical, of the absorption feature that can be created by coatings of
\nht on other materials.  We tried to create absorption features of
this type on larger (10-15 \mumx) particles with \nhfsh as the core
material, but were unsuccessful, largely because of the rapid increase
in absorption by \nhfsh on the long-wavelength side of the feature.
We thus find that the most plausible means of creating an absorption
feature at 2.97 \mum is by coating small conservative particles with
\nht (or by coating \nht cores with a conservative shell).  We thus
proceeded to carry out fits using small particles made of the
synthetic index material that roughly simulated these composite
particles.

\begin{figure*}[!htb]\centering
\includegraphics[width=6in]{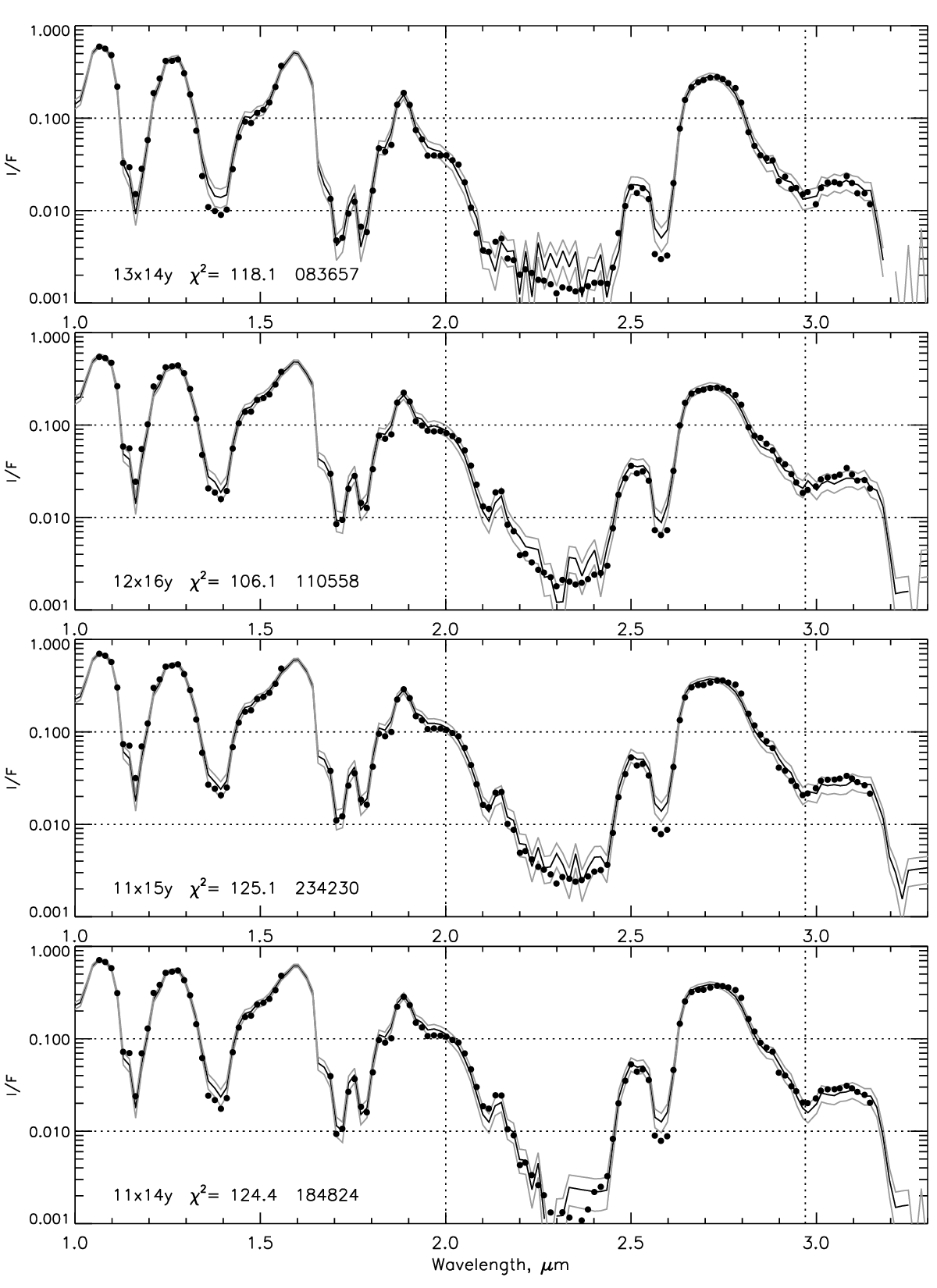}
\caption{Selected low phase angle VIMS spectra (lines) compared to
translucent layer multiple scattering model spectra (filled circles)
using both simulated \nhtx-coated and \nhfsh cloud layers. Light gray lines
indicate combined uncertainty bands.  The VIMS spectra from 1.58 to
1.68 $\mu$m are not used for constraining fits because of a
responsivity error. Each panel legend provides pixel coordinates in
the VIMS cubes (as in Fig.\ \ref{Fig:imvims1}) and $\chi^2$ values (a
value of 110$\pm$15 is expected). Here a vertical dotted line marks
a wavelength of 2.97 $\mu$m.}
\label{Fig:msfitspec1}
\end{figure*}

\begin{figure*}[!htb]\centering 
\includegraphics[width=6in]{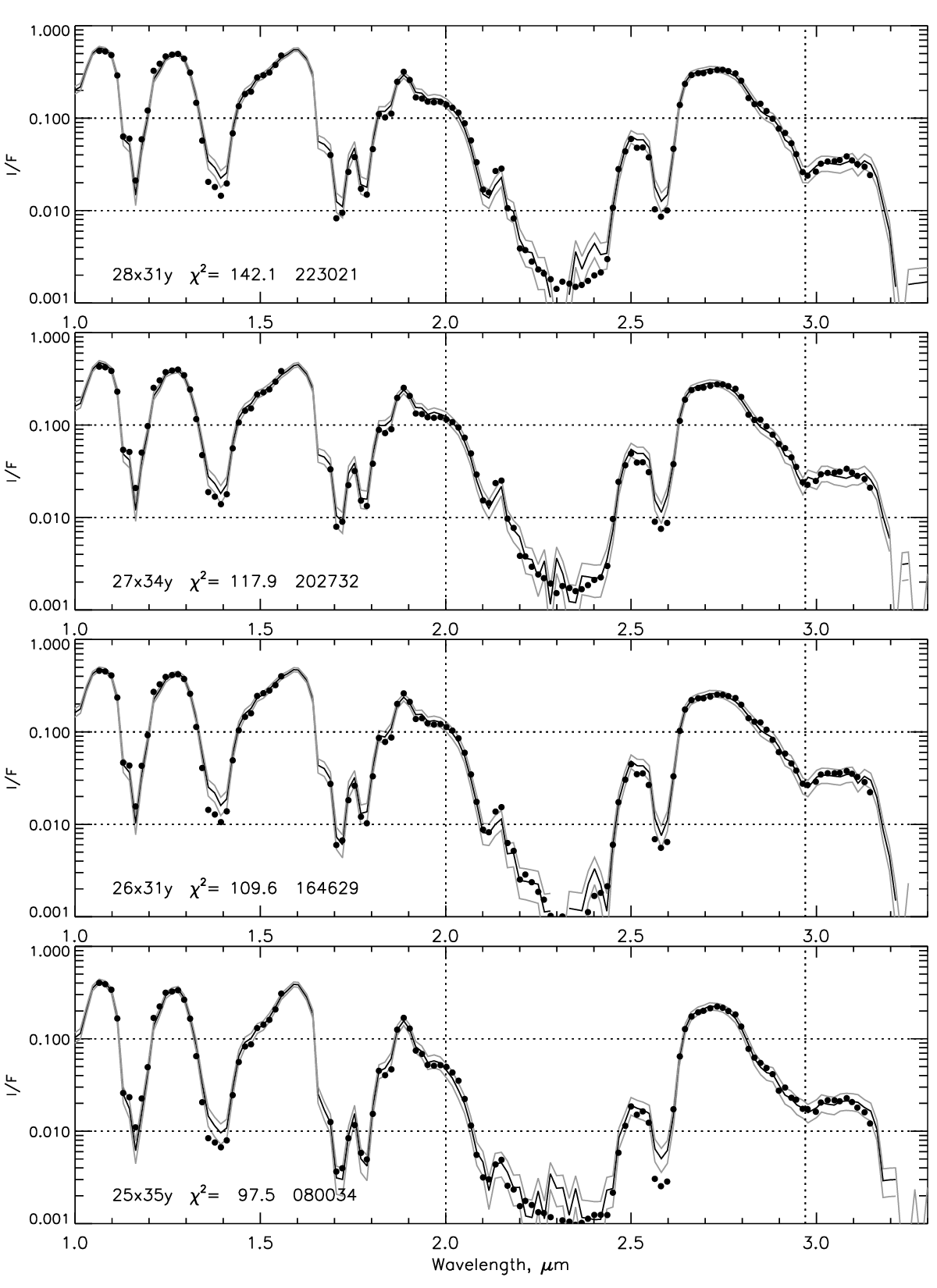}
\caption{As in Fig.\ \ref{Fig:msfitspec1}, except that the spectra are for medium
phase angles (as in Fig.\ \ref{Fig:imvims2}).}
\label{Fig:msfitspec2}
\end{figure*}

Our revised structure placed the synthetic index particles in layer 2,
to simulate the absorption by an ammonia coating on a conservative
core.  Fit-adjusted parameters included the pressures of the bottom
three layers, the optical depths of all four layers, and the particle
sizes of layers 2 and 3.  From the comparisons of model and measured
spectra, shown in Figs.\ \ref{Fig:msfitspec1} and
\ref{Fig:msfitspec2}, we see that these model fits are much better
than were obtained when only a single 3-$\mu$m absorber was used . The
shape of the 2.7-$\mu$m peak and its shoulder down to the 2.97-$\mu$m
NH$_3$ feature is especially well matched, and overall $\chi^2$ values
are much improved as well, as can be seen in Table
\ref{Tbl:3compfits}, which compares parameter values and fit quality
for all three absorber models (the improvement in \chisq is 2.4-3.4
$\times \sigma_{\chi^2}$ for the high opacity case and 0.5-0.8 $\times
\sigma_{\chi^2}$ for the low opacity case) . For the 13x14y spectrum,
we see that layer 2 and layer 3 are closer together than for the
single absorber models, the layer 3 optical depth is less, and the
bottom cloud pressures are very similar.  For the 11x14y spectrum,
which is from a much cloudier region, the pressures of layers 2 and 3
are similar to those for the \nht case, but the optical depth for
layer 2 is about 40\% more than for either of the single-absorber
cases, and the pressure of the bottom layer has been reduced from
$\sim$1.6 bars to 1.25 bars.

The best-fit parameter values for the two-absorber model are listed
for all eight fitted spectra in Table\ \ref{Tbl:msfits}. We find that
the preferred sizes of the NH$_4$SH particles are relatively large
($r\sim$ 7-16 $\mu$m), which is comparable to the sizes we inferred
for models in which only one absorber is present. On the other hand,
the preferred size of the NH$_3$-coated particles is quite small, $\sim$0.3
$\mu$m. In these models the NH$_4$SH layer provides most of the
3-$\mu$m absorption, but NH$_3$ provides additional spectral shaping
that helps to achieve a better match to the observed spectra.  
%Both
%absorbers work together to create the 2.97-\mum feature in the I/F
%spectrum (see later discussion).  
Because we did not include sufficient quadrature points to
characterize the complex backward scattering phase functions of the
larger particles (if treated as spheres), the larger particle sizes
are mainly constrained by the spectral shape of the 3-\mum absorption
feature, rather than the large scale wavelength dependence. For models
in which there was a close proximity of layers containing \nht and
\nhfsh particles, cross-correlations resulted in very poorly
constrained uncertainty estimates. This was handled by forcing the
layer 2 pressure to follow the layer 3 pressure by a fixed offset
(usually 20-30 mb).

\begin{table*}\centering
\caption{Best fit parameter values for 4-layer multiple-scattering 2-absorber fits to VIMS spectra.}
%\vspace{0.2in}
\begin{tabular}{|c c c c c c|}
\hline
         &           & Layer 1       & Layer 2        & Layer 3   & Layer 4  \\ [0.01in]
Spectrum & Parameter & ($\lambda^n$) & (NH$_3$ shell) &(NH$_4$SH) & ($\lambda^n$) \\ [0.05in]
\hline 
& & & & &\\[-0.05in]
  11x14y& $p$ (bars) & 0.0500  & 0.336$^{+0.014}_{-0.016}$ & 0.635$^{+0.042}_{-0.071}$& 1.252$^{+0.04}_{-0.04}$\\[0.1in]
  11x15y& $p$ (bars) & 0.0150  & 0.350$^{+0.003}_{-0.003}$ & 0.619$^{+0.013}_{-0.013}$& 1.268$^{+0.19}_{-0.17}$\\[0.1in]
  12x16y& $p$ (bars) & 0.0150  & 0.381 ($p_3$-0.030)   & 0.411$^{+0.006}_{-0.006}$& 0.885$^{+0.04}_{-0.04}$\\[0.1in]
  13x14y& $p$ (bars) & 0.0150  & 0.546 ($p_3$-0.020)   & 0.566$^{+0.001}_{-0.001}$& 1.128$^{+0.01}_{-0.01}$\\[0.1in]
  25x35y& $p$ (bars) & 0.0150  & 0.460 ($p_3$-0.060)   & 0.520$^{+0.011}_{-0.011}$& 1.007$^{+0.01}_{-0.01}$\\[0.1in]
  26x31y& $p$ (bars) & 0.0500  & 0.409 ($p_3$-0.030)   & 0.439$^{+0.010}_{-0.010}$& 0.954$^{+0.04}_{-0.04}$\\[0.1in]
  27x34y& $p$ (bars) & 0.0150  & 0.345 ($p_3$-0.013)   & 0.358$^{+0.028}_{-0.023}$& 0.885$^{+0.04}_{-0.03}$\\[0.1in]
  28x31y& $p$ (bars) & 0.0150  & 0.341 ($p_3$-0.030)   & 0.371$^{+0.001}_{-0.001}$& 0.789$^{+0.02}_{-0.01}$\\[0.1in]
  11x14y& \Large $\tau$ \normalsize & 0.0250$^{+0.005}_{-0.004}$& 0.420$^{+0.04}_{-0.03}$ &  2.10$^{+0.54}_{-0.49}$& 10.64$^{+ 1.37}_{- 1.26}$\\[0.1in]
  11x15y& \Large $\tau$ \normalsize & 0.0260$^{+0.001}_{-0.001}$& 0.450$^{+0.01}_{-0.01}$ &  2.10$^{+0.25}_{-0.24}$& 19.99$^{+ 1.33}_{- 1.33}$\\[0.1in]
  12x16y& \Large $\tau$ \normalsize & 0.0204$^{+0.002}_{-0.002}$& 0.370$^{+0.04}_{-0.04}$ &  0.23$^{+0.02}_{-0.02}$&  3.87$^{+ 0.12}_{- 0.12}$\\[0.1in]
  13x14y& \Large $\tau$ \normalsize & 0.0143$^{+0.003}_{-0.002}$& 0.414$^{+0.05}_{-0.05}$ &  0.23$^{+0.01}_{-0.01}$&  5.19$^{+ 7.23}_{- 3.74}$\\[0.1in]
  25x35y& \Large $\tau$ \normalsize & 0.0101$^{+0.002}_{-0.001}$& 0.284$^{+0.01}_{-0.01}$ &  1.20$^{+0.06}_{-0.06}$&  8.03$^{+ 0.70}_{- 0.68}$\\[0.1in]
  26x31y& \Large $\tau$ \normalsize & 0.0168$^{+0.003}_{-0.002}$& 0.374$^{+0.04}_{-0.04}$ &  1.75$^{+0.24}_{-0.24}$&  6.80$^{+ 0.57}_{- 0.63}$\\[0.1in]
  27x34y& \Large $\tau$ \normalsize & 0.0160$^{+0.019}_{-0.009}$& 0.387$^{+0.17}_{-0.14}$ &  1.39$^{+0.35}_{-0.31}$& 10.90$^{+ 4.46}_{- 3.66}$\\[0.1in]
  28x31y& \Large $\tau$ \normalsize & 0.0129$^{+0.001}_{-0.001}$& 0.378$^{+0.06}_{-0.05}$ &  1.48$^{+0.22}_{-0.20}$& 12.10$^{+ 3.38}_{- 3.03}$\\[0.1in]
  11x14y& $r$ ($\mu$m) or $n$ & $n$= -3 &$r$= 0.259$^{+0.11}_{-0.07}$ &$r$= 10.80$^{+0.29}_{-0.29}$&$n$= -1.12$^{+0.50}_{-0.45}$\\[0.1in]
  11x15y& $r$ ($\mu$m) or $n$ & $n$= -3 &$r$= 0.280$^{+0.05}_{-0.04}$ &$r$= 16.39$^{+1.18}_{-1.20}$&$n$=  0.00  \\[0.1in]
  12x16y& $r$ ($\mu$m) or $n$ & $n$= -3 &$r$= 0.270$^{+0.01}_{-0.01}$ &$r$= 16.48$^{+0.13}_{-0.13}$&$n$= -0.82$^{+0.18}_{-0.17}$\\[0.1in]
  13x14y& $r$ ($\mu$m) or $n$ & $n$= -3 &$r$= 0.303$^{+0.42}_{-0.15}$ &$r$=  6.86$^{+0.33}_{-0.32}$&$n$= -1.25$^{+0.03}_{-0.03}$\\[0.1in]
  25x35y& $r$ ($\mu$m) or $n$ & $n$= -3 &$r$= 0.364$^{+0.01}_{-0.01}$ &$r$= 11.83$^{+0.39}_{-0.40}$&$n$=  0.43$^{+0.11}_{-0.12}$\\[0.1in]
  26x31y& $r$ ($\mu$m) or $n$ & $n$= -3 &$r$= 0.355$^{+0.06}_{-0.05}$ &$r$= 13.84$^{+0.23}_{-0.23}$&$n$=  1.73  \\[0.1in]
  27x34y& $r$ ($\mu$m) or $n$ & $n$= -3 &$r$= 0.333$^{+0.03}_{-0.03}$ &$r$= 11.30$^{+0.38}_{-0.37}$&$n$=  1.56  \\[0.1in]
  28x31y& $r$ ($\mu$m) or $n$ & $n$= -3 &$r$= 0.304$^{+0.14}_{-0.11}$ &$r$= 10.93$^{+5.02}_{-4.07}$&$n$=  1.64$^{+0.33}_{-2.25}$\\[0.1in]
\hline
\end{tabular}\label{Tbl:msfits}\par
\vspace{0.15in}
\parbox{4.75 in}{NOTE: Only formal uncertainties from the fits are
given; see text for other sources of uncertainty.  Layer 2 pressures
without uncertainties follow layer 3 pressures with fixed
offsets. Some Layer 4 wavelength exponents were also fixed to
stabilize uncertainty estimates.}
\end{table*}

\begin{figure}[!htb]\centering
\includegraphics[width=3.5in]{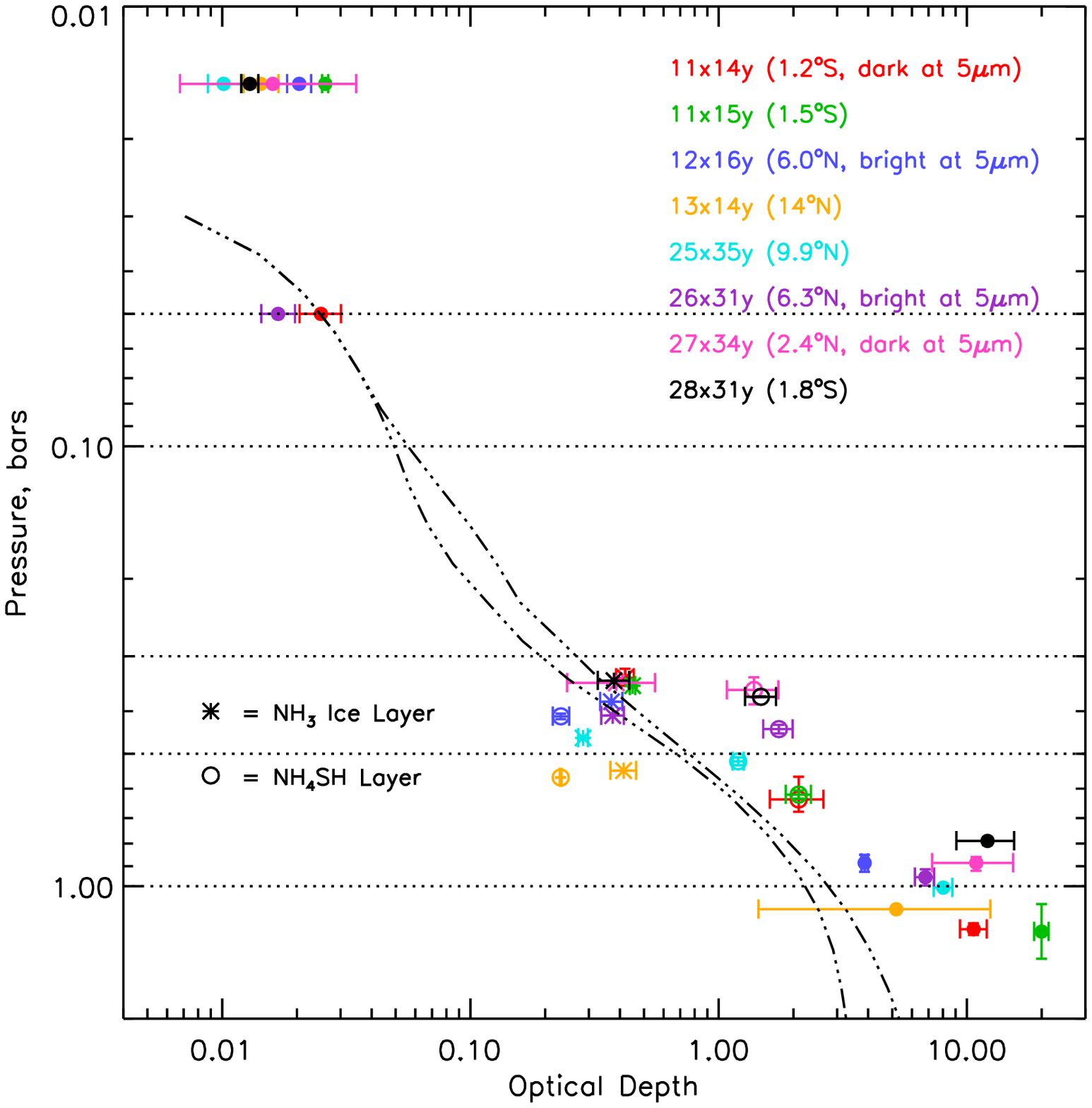}
\caption{Pressure vs. optical depth parameters for 4-layer
 multiple-scattering models that provide best fits to VIMS spectra, as
 given in Table\ \ref{Tbl:msfits}. 
The pair of triple-dot-dash curves are cumulative optical
 depth profiles derived from belt and zone opacity profiles of
 \cite{Irwin2001BZ}
}
\label{Fig:mspars}
\end{figure}

The vertical structure of the two-absorber models is visualized in
Fig.\ \ref{Fig:mspars} by plotting layer optical depth as a function
of pressure. Comparing this to Fig.\ \ref{Fig:rlfitpars}, we see that
the derived pressures are similar, with layer 3 of the reflecting-layer
model being split into two layers of different composition in the
current model.  For comparison we also show two independent results,
converted to cumulative vertical optical depth. Because most of our
layers have much higher optical depths than overlying layers, our
cumulative optical depths would be stair-steps that would be just
slightly to the right of our plotted layer values.   
The pair of triple-dot-dash curves are from belt and zone profiles of
\cite{Irwin2001BZ}.
These are well below most of our values, typically by factors of 2 or
more. Both our results and those of \cite{Irwin2001BZ} probably
contain far too much aerosol optical depth in the stratosphere because
both NIMS and VIMS observations near 2.3 \mum seem excessively high
compared to measurements of \cite{Banfield1998NIR} and to NICMOS
observations (Fig.\ \ref{Fig:f237mcomp}). The high stratospheric
opacity from our VIMS analysis is a result of the VIMS artifact
previously discussed in Sect.\ \ref{Sec:artifact}, which is probably
caused by light scattered off the VIMS grating. At least part of the
high stratospheric opacity obtained by \cite{Irwin2001BZ} is due to
their assumption of a Henyey-Greenstein phase function with an
asymmetry parameter of 0.5, which provides far less backscatter than
the sub-micron particles which are the likely source of stratospheric
scattering.  To make up for this lack of backscatter, the inferred
optical depth for their model had to increase to match I/F
measurements.
A more detailed comparison with NIMS results can be found in Sec.\
\ref{Sec:nims}.

\begin{figure}[!htb]\centering
\includegraphics[width=3.5in]{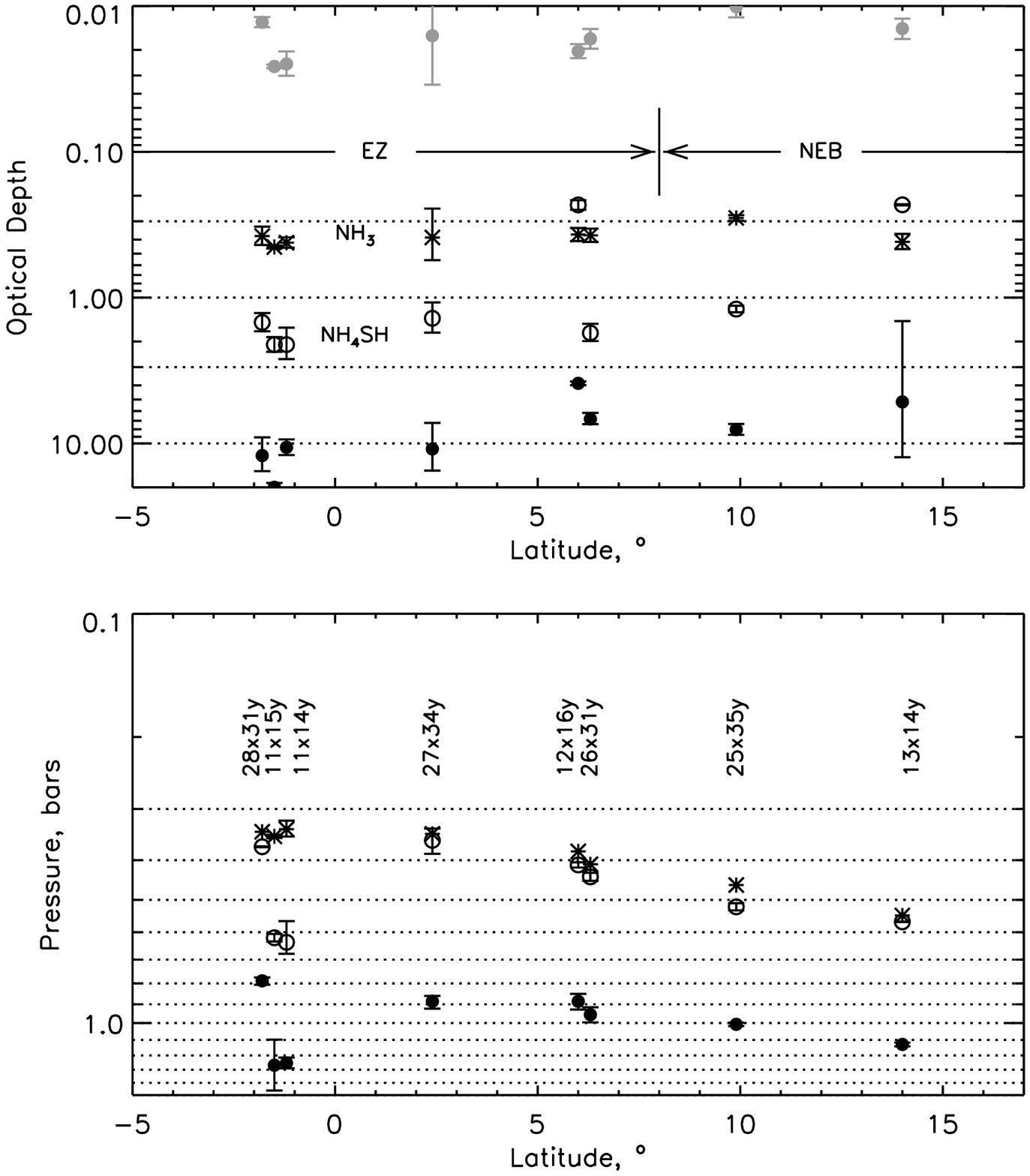}
\caption{Optical depth vs. latitude (top) and pressure vs latitude
(bottom) for the 4-layer model with \nht in layer 2 (plotted as
asterisks) and \nhfsh in layer 3 (plotted as open circles).  Layer 1
pressure are excluded from the bottom plot because they are not very
meaningful.}
\label{Fig:latvar}
\end{figure}

The variations of model parameters with respect to latitude are shown
in Fig.\ \ref{Fig:latvar}.  These results show some of the tendencies
observed in the reflecting layer parameters.  The cloud layers seem to
descend to higher pressures moving from zone to belt, but there is not
a clear tendency in optical depth.  The remarkable feature is the
close proximity of the layer of small \nht particles to the layer of
much larger \nhfsh particles, suggesting that both particle
populations might actually overlap in the same layer, perhaps with
different scale heights.  The descent to higher pressures in the NEB
also helps to hide the \nht particles from observations at thermal
wavelengths, helping to explain the latitudinal dependence of \nht
features detected by \cite{Wong2004}, as we discuss later in Sec.\
\ref{Sec:Wong}.

There are two counter examples to the general rule of near overlap of
\nht and \nhfsh layers. These are for two near equatorial spectra
(11x14y and 11x15y), for which these two layers are separated by
nearly 300 mb.  For these spectra the \nht layer stayed at low
pressures characteristic of the equatorial zone, but the \nhfsh layer
moved to higher pressures, as did the bottom cloud layer.  Optical
depths for these cases did not deviate much from those inferred for
the other spectra.  

\subsection{Sensitivity of spectra to model parameters.}

\begin{figure*}[!htb]\centering
\includegraphics[width=6in]{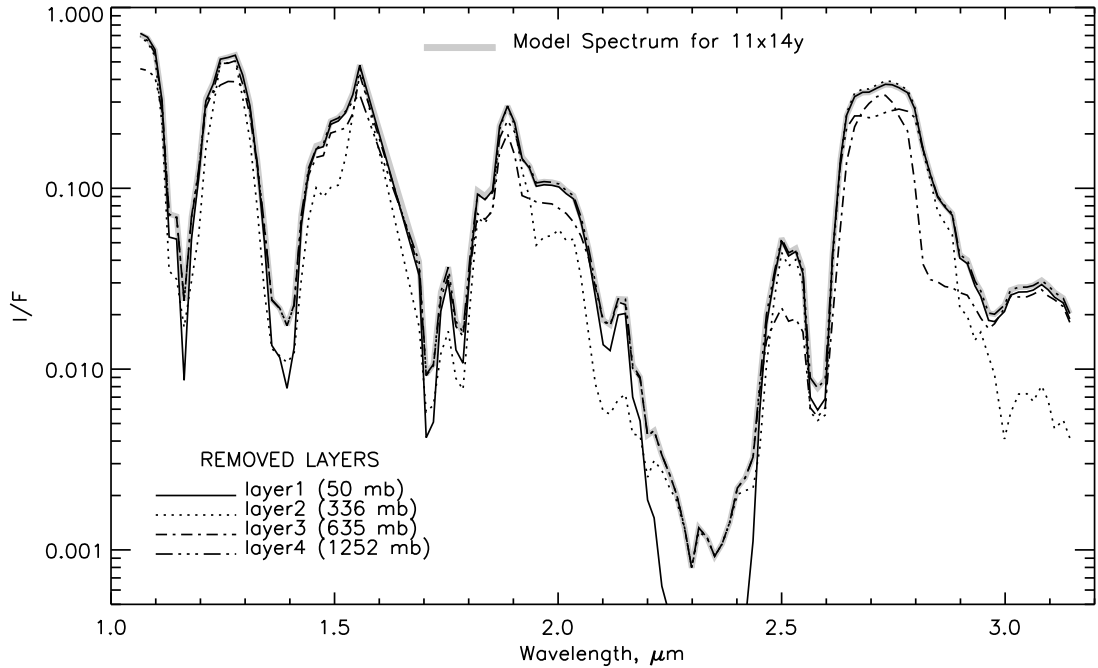}
\caption{Best-fit spectrum for location 11x14y (thick gray curve) compared to spectra
in which one layer is removed from the model, for each of the
layers. Layer 2 is the cloud of \nht-coated particles and layer 3 is the \nhfsh cloud. Note that
no values were computed between 1.58 \mum and 1.68 \mumx.}
\label{Fig:contributions}
\end{figure*}

The contributions of each model layer to the spectral features in the
11x14y spectrum are illustrated in Fig.\ \ref{Fig:contributions},
where the best-fit spectrum is compared to four spectra that each have
one layer removed. This shows that the \nhtx-coated particle layer
(336 mb) provides important contributions in several regions,
including near 1.5 \mum, near 2 \mumx, and from 2.95 \mum to 3.2
\mumx. The \nhfsh layer (635 mb) provides important contributions near
2 \mumx, 2.5 \mumx, and from 2.65 \mum to 2.95 \mumx, but has so much
absorption beyond 2.97 \mum that it contributes essentially nothing to
the I/F from 2.97 to 3.2 \mumx. Thus, these two different absorbers
are seen to make distinctly different contributions. The deepest cloud
layer (1252 mb) is seen to make significant contributions only to the
window regions.

\begin{figure*}[!htb]\centering
\includegraphics[width=6in]{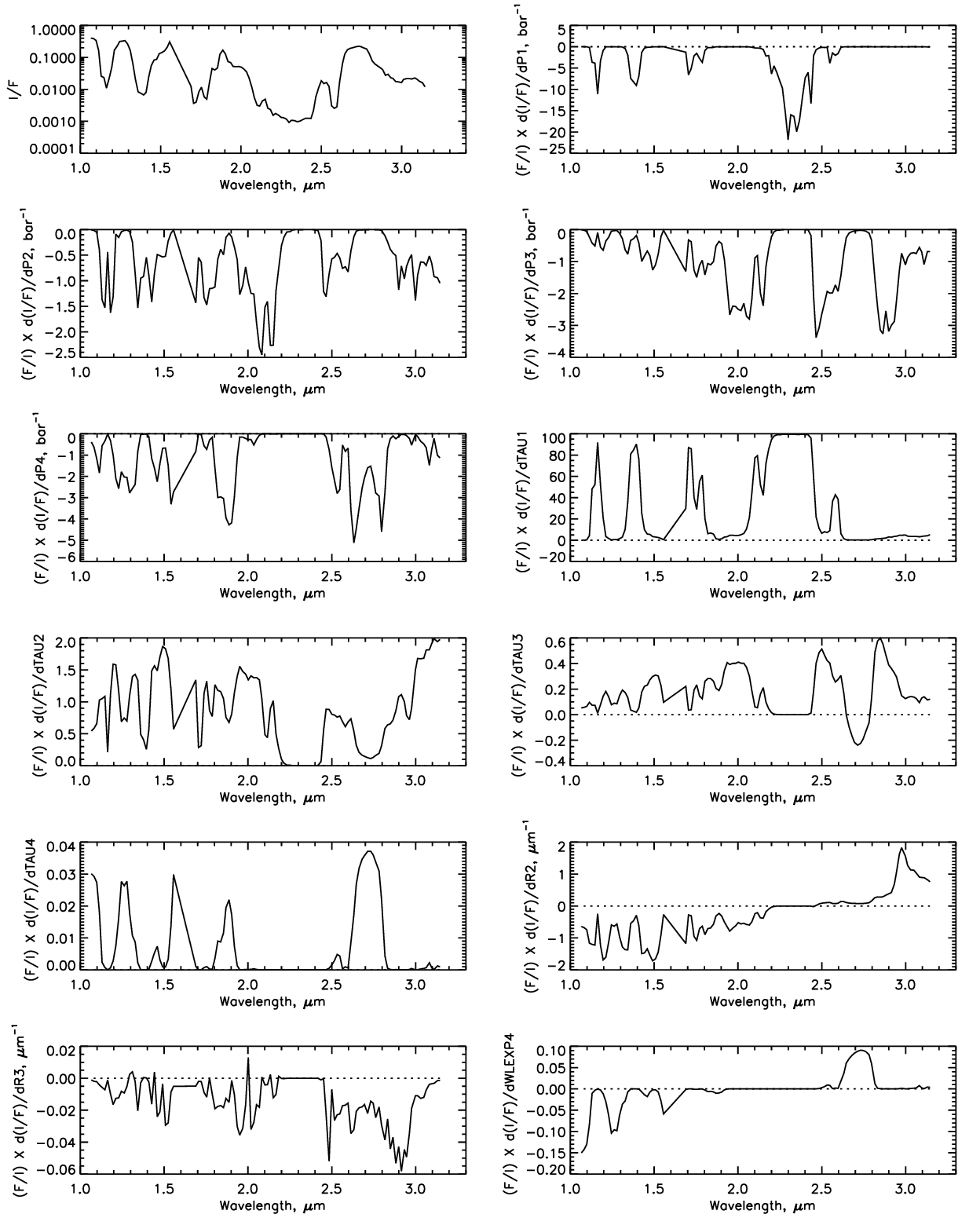}
\caption{Best-fit model spectrum for location 25x35y (upper left), and
fractional derivatives of that spectrum with respect to model
parameters for pressure (P1, P2, P3, P4), optical depth (TAU1, TAU2,
TAU3, TAU4), particle radius (R2, R3), and $\lambda$ exponent for
layer 4 (WLEXP4), in order from left to right and top to bottom. Here
model layer 2 is made of synthetic \nht-coated particles and layer 3 of \nhfshx.}
\label{Fig:derivatives}
\end{figure*}

The sensitivity of model spectra to specific model parameters can be
discerned from the fractional derivatives of a model spectrum with
respect to each of the model parameters.  This is shown in Fig.\
\ref{Fig:derivatives} for a model that is the best fit to the 25x35y
spectrum.  
%While the derivatives vary from model to model, 
This example illustrates the relative independence of the spectral
derivatives from each other, confirming the possibility of using
observations to independently constrain the parameters.
%providing a reasonable expectation that the parameters can be
%constrained by the observations.  
For example, although the upper layer of particles is only important
in the more opaque regions of the spectrum, the fractional derivatives
with respect to pressure (right, top row) and optical depth (right,
third row) are distinctly different.  A similar difference can be
found for other layers, and especially large differences exist between
different layers.  Note that particle size changes in layer 2 (the
\nhtx-coated layer) have the greatest effect at 2.97 \mum and from
there to 3.2 \mumx, while the size changes in layer 3 (the \nhfsh
layer) have the largest impact between 2.5 and 2.9 \mumx.  If we had
used as the example a fit at low phase angles, and included an
appropriately large number of quadrature points to capture the
backward phase function peak, then the large particle layer (layer 3)
would also have an important effect at short wavelengths.  But models
with such phase functions don't fit as well as models without strong
backward peaks.

In spite of the distinctive derivative spectra shown in Fig.\
\ref{Fig:derivatives}, there are many spectra where the best fit \nht
and \nhfsh layers are so close together that they become much more
highly correlated with respect to pressure and optical depth
variations. To fit these spectra and derive reasonable uncertainty
estimates, we revised the model to use a fixed pressure offset between
layers 2 and 3, so that the uncertainty for the layer 2 pressure then
becomes the same as that for the layer 3 pressure, except that the two
layers have correlated uncertainties. This tendency for both small
\nhtx-coated particles and much larger \nhfsh particles to reside at nearly
the same pressure level suggests that they might actually be members
of two different populations in the same layer.  It is also possible
that \nht forms as a coating on the larger \nhfsh particles, although
that would not work well for the entire \nht contribution, because the
small particle component is needed to raise the reflectivity at short
wavelengths, and the coated large particles don't seem capable of
producing the 2.97-\mum feature.

\subsection{Sensitivity of spectra to absorber gas profiles.}

\begin{figure*}[!htb]\centering
\includegraphics[width=6in]{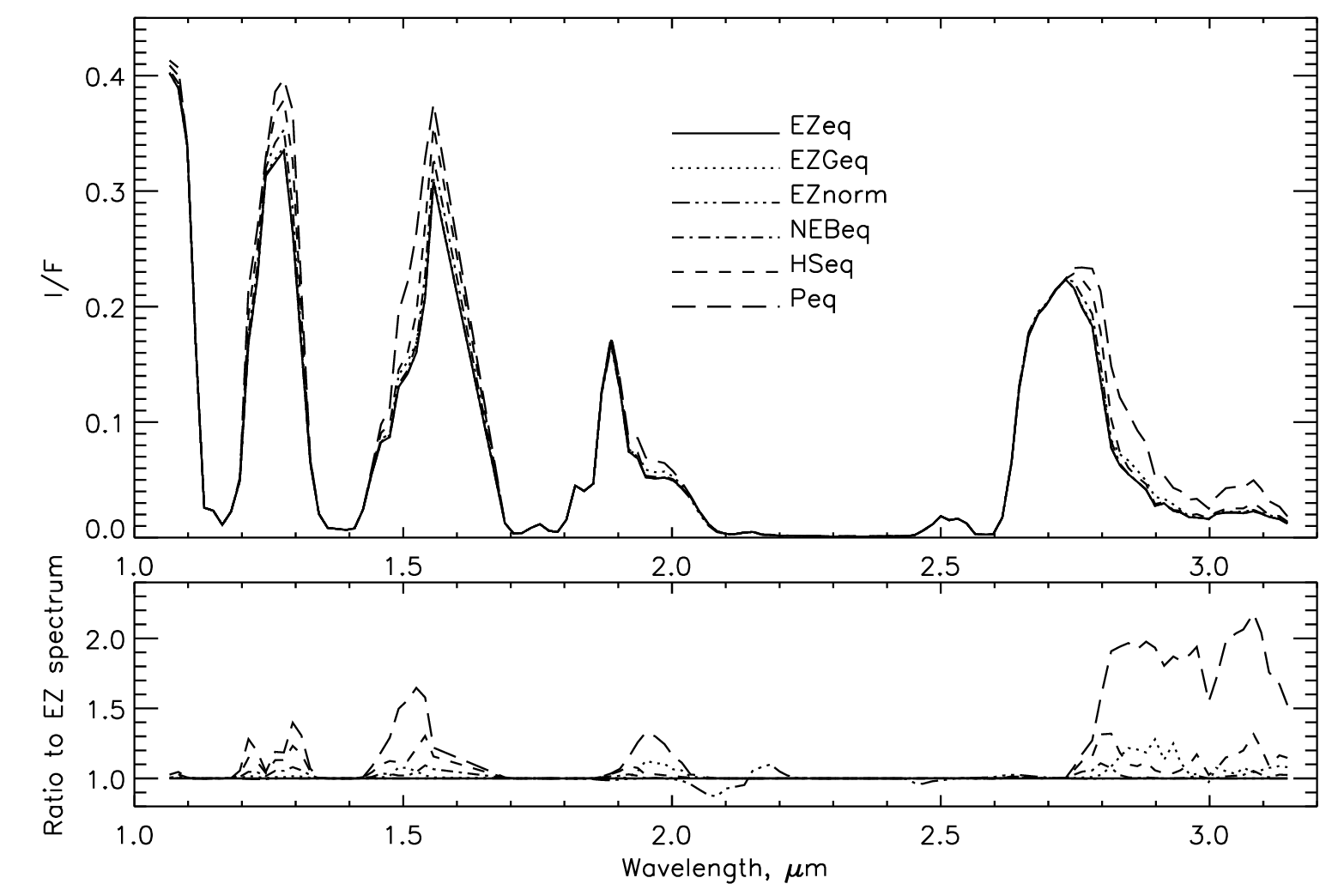}
\caption{Spectra (top) for different gas opacity profiles using the
same aerosol structure (the best fit structure for 25x35y and the EZ profile),
and ratio (bottom) of each spectrum to the EZ spectrum.  Note that
no values were computed between 1.58 \mum and 1.68 \mumx.}
\label{Fig:profilespecs}
\end{figure*}

The effects of different ammonia and para hydrogen profiles are
illustrated in Fig.\ \ref{Fig:profilespecs}.  The difference between
NEB and EZ profiles is relatively small, and our best fits were
obtained with the the EZ profile for all spectra for which we did fits
with both profiles, suggesting that the NEB profile we chose did not
contain quite enough \nht at high altitudes. The probe profile makes a
huge difference and is not a best fit with even the darkest spectrum
we observed (13x14y). While we cannot rule it out in a local region,
due to lack of spatial resolution, it is certainly not acceptable over
an extended region comparable to the area sampled by a VIMS pixel. The
difference between normal and equilibrium hydrogen profiles is also
relatively small and restricted to a region that does not provide a
significant constraint on the cloud models. Given that atmosphere
appears to be intermediate between these states at low latitudes
\citep{Conrath1984}, the actual difference is about half that between
the extreme cases shown in the figure.  Thus, we did not try to
account for potential variations in the ortho-para ratio.  Instead we
assumed equilibrium hydrogen in all the fits, confident that this
convenience would have no significant effect on our conclusions.

\subsection{Effects of calibration scale-factor errors}

We fit several spectra with the
nominal calibration and also with the VIMS I/F spectra multiplied by a
scaling factor to simulate the effect of a change in absolute calibration.
For a uniform fractional decrease in the measured I/F, we found that
derived pressures and optical depths also decreased, the former by
relatively little, just a fraction of the fractional calibration
change, and the latter by much more, changing by one (at low optical
depths) to several times (at higher optical depths) the fractional
change in calibration (large optical depths require a larger
fractional change to produce the same fractional change in
reflectivity).  This is not too surprising, since the overall
brightness (which is changed by the scaling) is a strong function of
the aerosol loading, while the shape of the spectrum (which is not
changed) is strongly controlled by the vertical distribution of the
aerosols. These results suggest that opacities we derive have
additional uncertainties contributed by calibration uncertainties of
about 5\% to 20\%, while the pressures have additional uncertainties
0-4\%.  We also found about a 10\% affect on the wavelength dependence
exponent for the bottom layer. The particle radius error contribution
was not significant for layer 2 but 5-10\% for layer 3.  All these
contributions should be combined with prior estimates using the
root-sum-square method.

\section{Discussion}

\subsection{Comparison with Galileo Probe results}

The Galileo Probe entered the atmosphere of Jupiter on 7 December 1995
at a latitude of 6.6\degx N, near the boundary between the Equatorial
Zone and NEB, and at the edge of a 5-\mum hot spot, which is a region
of unusually low aerosol opacity (in the deeper layers). The
Nephelometer \citep{Ragent1998} and the Net Flux Radiometer
\citep{Sro1998} both detected a cloud near the 1.2-bar level (base
near 1.35 bars) at the edge of a hot spot, with roughly unit opacity
at a wavelength of 5 \mumx. This pressure and optical depth are
roughly comparable to that of the bottom cloud derived from several of
the VIMS spectra, but there is not a good match to any obtained close
to the probe entry latitude.  This mismatch is not too surprising; given the large
size of the VIMS pixels and the small scale of a typical hot spot, a
close agreement cannot be expected.

An additional cloud layer was detected indirectly. A cloud above the
level at which probe measurements began (440 mb) was inferred from Net
Flux Radiometer observations of spin-induced modulations of the direct
solar beam signal at multiple wavelengths \citep{Sro1998}. The optical
depth was estimated to be $\sim$1.5-2 at a wavelength of 0.5 \mumx,
and the particle size between 0.5 and 0.75 \mumx. This upper cloud
seems reasonably consistent with our VIMS results for nearby latitudes
(12x16y and 26x31y). As clouds close to this level are found
throughout the equatorial zone, it is plausible that these might also
be present at the edge or even over a hot spot. Although we find a
combination of small and large particles, their spectral effects on
the solar beam intensity in the NFR passbands might be similar to
those of a single population of particles of intermediate size. Our
lowest opacity model (for the 12x16y spectrum) has two contributions
that would come from above the probe deployment level.  At 2.0 \mum
the \nhfsh layer has a small optical depth of 0.23, which would not
increase much at 0.5 \mum because of the large particle size in this
layer.  The layer of \nhtx-coated particles, with an optical depth of
0.37 at 2 \mumx, would provide most of the opacity at 0.5 \mumx, which
could plausibly 4-5 times greater than its 2-\mum value.  This would
yield a total 0.5-\mum opacity comparable to the probe estimate, which
is for a layer that does not vary much in optical depth for any of our
spectral models, making it more likely that the probe comparison is a
meaningful one.

Our VIMS model results also agree with a lack of cloud particles in
the 700 mb region, and we do find lower cloud layers close to the
probe cloud density maximum near 1.2 bars, although the optical depth
we find there is generally much higher than obtained by the probe,
which was in an unusually transparent region of the atmosphere.  We
don't agree with the probe-derived ammonia profile however; when we
try to fit the least cloudy region (13x14y) with that ammonia profile,
the best fit $\chi^2$ value is nearly double what we obtained using
either equatorial zone or NEB ammonia profiles. The part of the
profile that presents problems is the low level of \nht in the 400 mb
to 1.5 bar pressure range, which is the part derived from NFR net flux
measurements.

\subsection{Comparison with Galileo NIMS results}\label{Sec:nims}

\begin{figure*}[!htb]\centering
\includegraphics[width=6in]{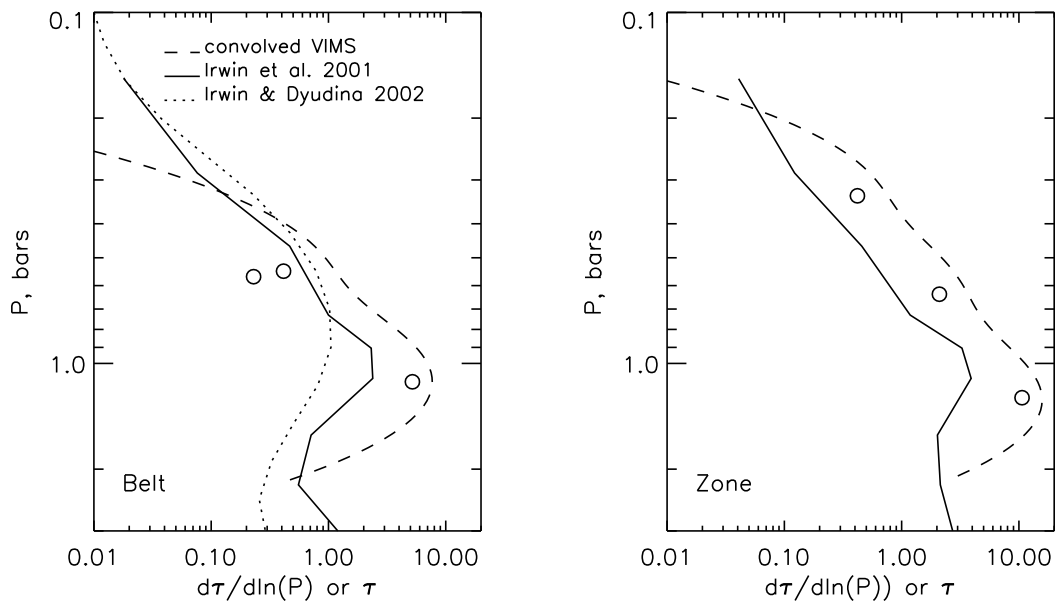}
\caption{Extinction optical depth per unit per unit $\ln{P}$ interval
 for a belt sample (left) and for a zone sample (right) derived by
 \cite{Irwin2001BZ}(solid lines). The over-plotted circles are optical depths from
 VIMS fits for the 13x14y spectrum (left) and the 11x14y spectrum (right),
 which represent belt and zone cloud structures respectively. 
The dashed lines are the d$\tau$/d$\ln{P}$ values derived from the
smoothing of VIMS scaled layer optical depths, computed as described
in the text. The dotted curves (left panel) is the opacity profile
from \cite{Irwin2002PCA} scaled to a wavelength of 2 \mumx, as
described in the text.}
\label{Fig:nims}
\end{figure*}

In Fig.\ \ref{Fig:nims} we compare our opacity results at a reference
wavelength of 2 \mum to those of \cite{Irwin2001BZ} and
\cite{Irwin2002PCA}. These comparisons are made in terms of the
extinction optical depth per unit $\ln{P}$ interval. This is roughly
equivalent to optical depth per unit altitude.  We estimated that
function for our results by convolving our opacities at discrete
layers with a Gaussian smoothing function of the form
$\exp{(-x^2/w^2)}$, where the width $w$ was set to roughly simulate
the vertical correlation used in the NIMS retrievals. We didn't scale
the \cite{Irwin2001BZ} results because they assumed
wavelength-independent parameters over the entire 1-2.5 \mum interval.
We did scale the \cite{Irwin2002PCA} results from their 1-\mum
reference wavelength to our 2-\mum reference using their derived
particle radius profile and calculations of the wavelength dependence
of Mie scattering efficiency (using their refractive index of 1.4+0i).
We find a rough agreement on the fractional increase of opacity with
depth and agreement on the pressure level of the local peak in
opacity, but don't agree well on the absolute values of opacity.  Our
results contain several times as much optical depth as the obtained by
\cite{Irwin2001BZ}, and disagree even more with the profile of
\cite{Irwin2002PCA}, which has nearly an order of magnitude less
opacity when scaled to a wavelength of 2 \mumx. At that wavelength it
also disagrees with \cite{Irwin2001BZ}.

A factor that might introduce disagreements with the NIMS analysis of
\cite{Irwin2001BZ} and \cite{Irwin2002PCA} is the difference in
treatments of \nht absorption.  \cite{Sro2010iso} show that
%In Fig.\ \ref{Fig:absorbers}, for example, we see that 
prior ammonia absorption models provided too
little absorption at 1.59, 1.9, and 2.3 \mumx.  As can be seen from
Fig.\ \ref{Fig:profilespecs}, variations in the amount of ammonia
absorption have a significant impact on the I/F observed in the deeply
penetrating windows at 1.28 and 1.59 \mumx, which provide important
constraints on the vertical location of aerosols.  Likewise,
variations in \nht absorption models might affect the pressure derived
from fitting the spectra.

Another difference that might affect the comparisons is that our fits
result in different particle sizes, with the \nhfsh layer having
much larger particles than were derived by \cite{Irwin2002PCA}. 
Our \nht and \nhfsh layers are made with refractive indexes of specific materials,
while \cite{Irwin2002PCA} used conservative particles.  
have a strong influence on scaling results to other wavelengths.

\subsection{Comparison with Galileo SSI results}

Our results are also in partial agreement with results based on
Galileo Solid State Imager (SSI) observations.
% \citep{Banfield1998, Simon-Miller2001}.
\cite{Banfield1998}, who analyzed SSI data in three channels at 756
nm, 727 nm, and 889 nm, conclude that below an ubiquitous upper
tropospheric haze, which exhibits little small scale variation, there
is a highly variable cloud component (0 to 20 in optical depth),
usually at p = 750$\pm$200 mb, less than a scale height in vertical
extent, and the principal cause of features seen at red and longer
wavelengths. At first, this appears to be in serious conflict with our
VIMS results, but it is only the lower end of the pressure range that
is most in conflict, as a comparison of cumulative opacities will soon
demonstrate.  The subsequent analysis by \cite{Simon-Miller2001},
reached similar conclusions, though somewhat less discrepant with
near-IR results, specifically that a white cloud exists at pressures
of 880-970 mb in belts and 620-720 mb in zones. These models use a
relatively simple cloud structure with a small number of parameters,
befitting the small number of spectral constraints provided by the
observations.  The structure consists of a stratospheric haze
extending to a pressure that marks the top of the main cloud layer
with a bottom pressure at which (generally) is located a physically
thin sheet cloud. The main cloud is assumed to have the same scale
height as the gas, leaving two pressures and three optical depths to
be constrained by the models.

\begin{figure*}[!htb]\centering
\includegraphics[width=6in]{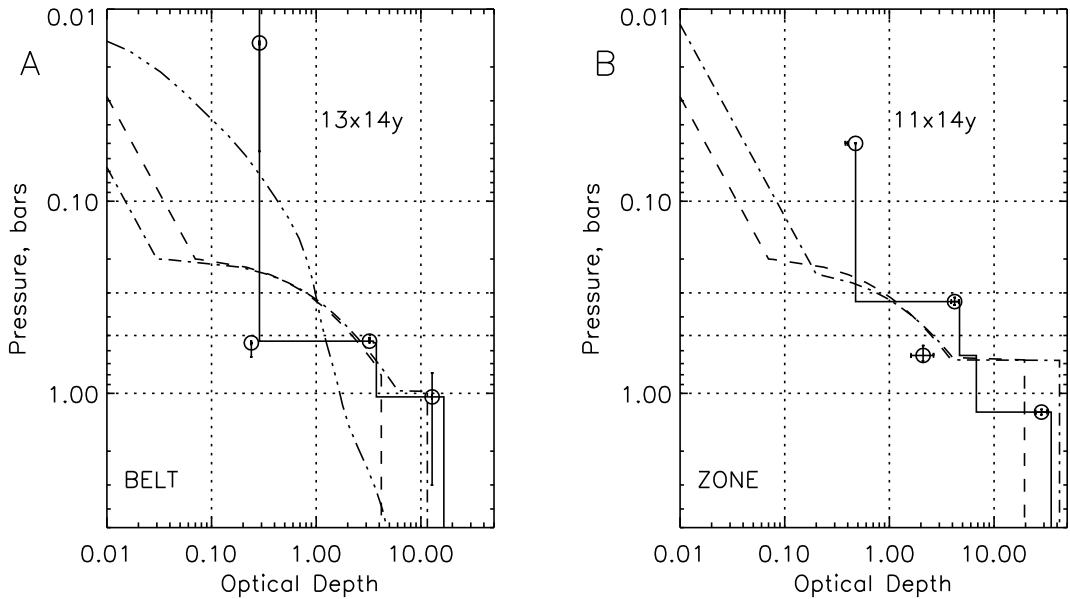}
\caption{Cumulative optical depth for a belt sample (A) and for a zone
sample (B) derived from Galileo SSI observations by
\cite{Banfield1998} (dashed curves) and by
\cite{Simon-Miller2001}(dot-dash curves). The over-plotted circles are
optical depths from VIMS fits for the 13x14y spectrum (A) and the
11x14y spectrum (B), which represent belt and zone cloud structures
respectively.  VIMS layer optical depths (circles) and cumulative
optical depths (solid lines) have been scaled to a wavelength of 0.75
\mum as described in the text. The triple-dot-dash curve is the
cumulative opacity profile from \cite{Irwin2002PCA} scaled to a
wavelength of 0.75 \mumx, using their particle radius profile.}
\label{Fig:Simon}
\end{figure*}

In Fig.\ \ref{Fig:Simon} we provide comparisons for a belt (panel A)
and a zone (panel B) between our VIMS cumulative optical depths and
those of \cite{Banfield1998NIR} and \cite{Simon-Miller2001}. Here our
VIMS-derived opacities are scaled to their wavelength of 0.75 \mumx.
We scaled up the stratospheric haze opacity by a factor of 19
(assuming a $\lambda^{-3}$ dependence) and the layer near 300 mb by a
factor of ten, which is appropriate for our composite particle
with a radius of 0.3 \mumx.
The bottom layer was scaled up by a factor of 2.67, to account for the
approximate $\lambda^{-1}$ dependence we inferred for that layer. The
main \nhfsh 500-600 mb layer, was not scaled because its opacity is
relatively flat for the best-fit particle sizes, although its
reflectivity can increase significantly at shorter wavelengths due to
its high real index and reduced absorption.  At SSI wavelengths this
layer appears to make a relatively small contribution compared to the
layers surrounding it.

Turning to the SSI-VIMS comparison itself, it is not surprising that
our stratospheric haze opacity is much too high because of the
excessive I/F values in the VIMS spectra near 2.3 \mum.  Otherwise,
there is gross agreement on the opacity levels deeper in the
atmosphere.  For the belt comparison (panel A) the cumulative opacity
maximum is in close agreement with the \cite{Simon-Miller2001} result,
while for the zone comparison (panel B) the best agreement is with the
\cite{Banfield1998} result, although there is a significant
disagreement in the vertical location of the high-opacity bottom
cloud. Our VIMS analysis puts it close to 1.1 bars, while the SSI
results place it closer to 700 mb.  On the other hand, we do find a
significant cloud layer close to 700 mb in this region, and perhaps
with a different scaling from 2 \mum to 0.75 \mum the disagreement
might be substantially reduced, suggesting that the particle size is
smaller or the inferred wavelength dependence is greater than what we
obtain from our best fit over the 1-3.2 \mum region.  Another point
worth remembering is that we are not comparing a large statistical
sample, and some of the differences observed might be attributable to
spatial and temporal variations.  In the belt region, for example we
obtain somewhat better agreement with the 25x35y spectrum. Clearly, a
larger sampling of variations would allow for a more meaningful
comparison.

Features of the SSI analysis that might contribute the disagreement in
bottom cloud locations include changes made in the methane absorption
coefficients of \cite{Kark1998Icar}, failing to include hydrogen
collision-induced opacity from the third overtone band, which provides
significant opacity in the 756-nm window, and uncertainties in the
actual transmissions of the Galileo SSI filters. According to
\cite{Simon-Miller2001}, the 756-nm absorption coefficient was
increased by a factor of 10 and the 727-nm coefficient from 0.17 to
0.22.  The latest constraints on these coefficients, given by
\cite{Kark2009IcarusSTIS}, have resulted in only a factor of two
increase in the 756-nm coefficient and no change in the 727-nm
coefficient. The two Galileo SSI analyses thus place more absorption at
higher altitudes than current best estimates would imply, and thus the
pressures of the deeper cloud layers might be underestimated
somewhat. Particle size differences are also a potential source of
disagreements on opacity.  \cite{Banfield1998} use a fixed particle
size of 0.2 \mumx, while \cite{Simon-Miller2001} made particle size a
fitted parameter, but didn't report specific values because they were
poorly constrained.

\subsection{Comparison with \cite{Matcheva2005} CIRS results}

The \cite{Matcheva2005} analysis of 2000 Cassini CIRS observations of
Jupiter is unusual in relying on emission rather than reflected
sunlight and in not requiring the use of band models, but instead
relying on line-by-line calculations. This work also provides an important
constraint on cloud composition because they are sensitive primarily
to absorption, which varies greatly from one material to another
(Fig.\ \ref{Fig:indexplot}).  \cite{Matcheva2005} chose a narrow
spectral window centered at 7.18 \mumx, which is free of NH$_3$
absorption and thus independent of the NH$_3$ mixing ratio profile.
Their analysis ignores scattering, based on the assumption that all
particle radii are less than 1 $\mu$m, although this is contradicted
by other modeling efforts, e.g. this work as well as that of \cite{Brooke1998}
and \cite{Wong2004}, which generally find a significant component of
large particles ($\sim$10 \mum in radius) within the region relevant
to the Matcheva et al. analysis.  At latitudes with relatively low
cloud opacity, such as the NEB, they find a cloud absorption peak at
1100$\pm$50 mb, and a cumulative optical depth near 1 down to 1400
mb. At latitudes with thick overcast such as the EZ, they find peak
absorptions at p$\geq$ 900 mb and optical depths up to $\tau \ge 4$.
Their technique is limited to total optical depth (cloud plus gas) of
$\tau <$ 4. In the EZ, this limits retrievals to p $<$ 980 mb, and in
the NEB to p $<$ 1.5 bars.

\begin{figure*}[!htb]\centering
\includegraphics[width=6in]{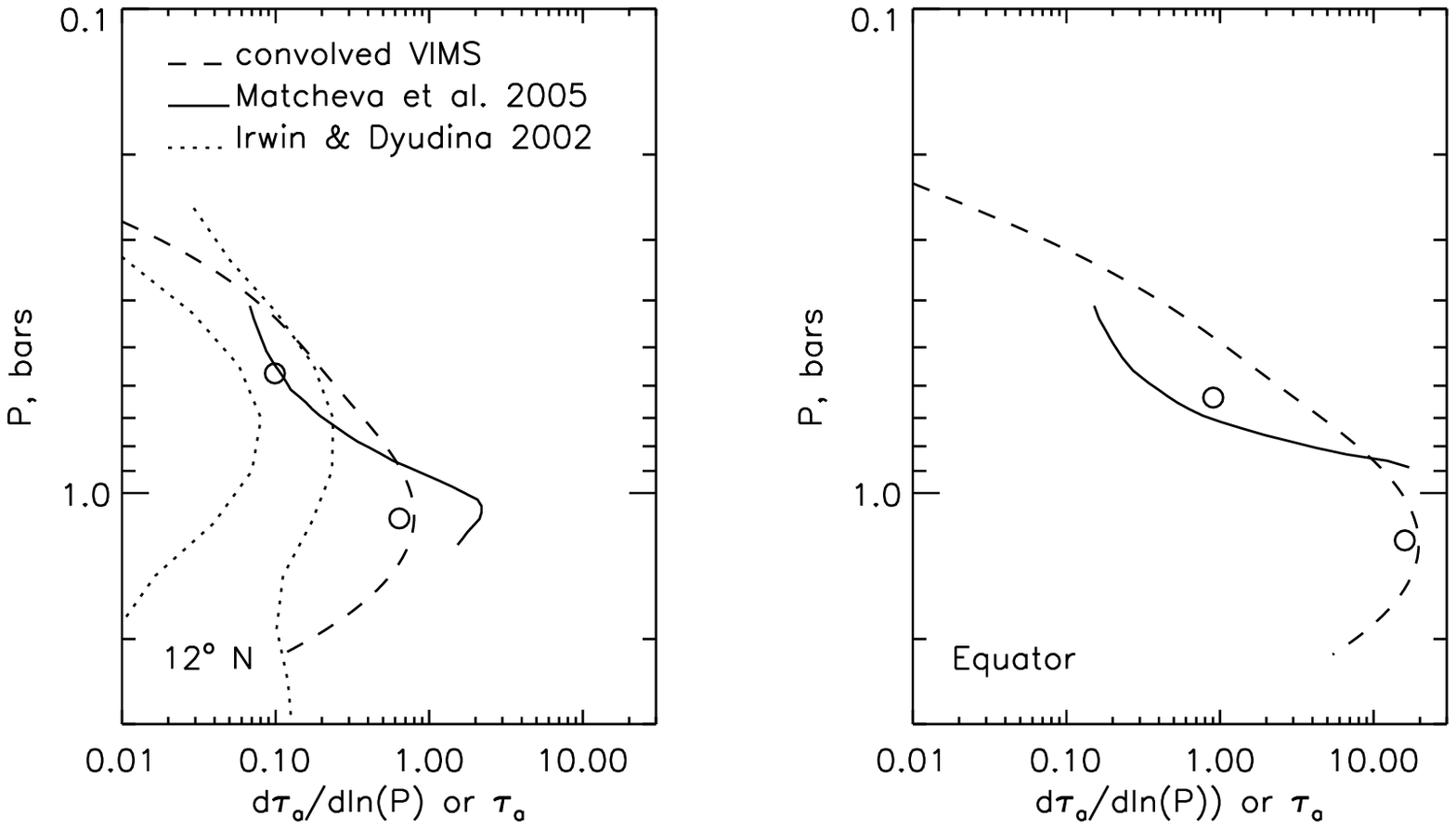}
\caption{Cloud absorption optical depth per unit $\ln{P}$ interval for
12\deg N (left) and the equator (right) derived by \cite{Matcheva2005}
from CIRS thermal emission observations at 7.18 \mumx. Above and below
the plotted pressures, the retrieval is dominated by guess profiles
and thus not shown.  The over-plotted circles are optical depths from
VIMS fits for the 13x14y spectrum (left) and the 11x14y spectrum
(right), which represent belt and zone cloud structures respectively.
The scaling of VIMS results to 7.18 \mum is described
in the text.
The dashed lines are the VIMS results convolved with a
smoothing function that crudely simulates the vertical resolution of the
Matcheval et al. retrieval.  The $d\tau_a/d\ln{P}$ values for
\cite{Irwin2002PCA} (dotted lines at left) are scaled according to particle
size using imaginary indexes of \nht ($N_i$=0.02) and \nhfsh ($N_i$=0.2)
to simulate \nht and \nhfsh absorption respectively.}
\label{Fig:Matcheva}
\end{figure*}

In Fig.\ \ref{Fig:Matcheva} we compare our absorption optical depth
profiles inferred from the 13x14y and 11x14y VIMS spectra with the
\cite{Matcheva2005} absorption optical depth profiles for a region
near 12\deg N (left panel) and the equator (right panel),
respectively.  The comparison is made in terms of the parameter
$d\tau_a/d\ln P$, which can be thought of as a crude measure of
optical depth per unit height.  Our model opacities from VIMS are
scaled to a wavelength of 7.18 \mum to make a more appropriate
comparison to the Matcheva et al. results.  The VIMS stratospheric
opacities, scaled using a $\lambda^{-3}$ dependence, are too small to
be relevant at 7.18 \mumx. For the 300 mb layer we scaled from 2-\mum
values to 7.18-\mum values using the factor of 57 reduction that
applies for our 0.3-\mum composite particle, and a further reduction
to account for the absorption optical depth being 80\% of the
extinction value. These scalings make this layer ignorable as
well. The optical depth of the main 3-\mum absorbing layer has a
relatively flat wavelength dependence and was used without adjustment,
except that we accounted for absorption optical depth being only 43\%
of the total. At 7.18 \mum the particles in this layer have a single
scattering albedo of $\sim$0.57 if they are pure \nhfsh compared to
$\sim$0.7-0.85 if they are pure \nhtx.
% see page 21 Jup Log L for relevant plots.

The appropriate scaling for the VIMS bottom cloud is unclear.  The
wavelength exponents for the bottom layer in the reflecting layer fits
(Table\ \ref{Tbl:rlfits}) are generally small and slightly negative
for most models. However, the exponents obtained from the multiple
scattering fits (Table\ \ref{Tbl:msfits}) vary widely, being mostly
negative for the low phase angle fits and mostly positive for the
medium phase angle fits.  We decided to use $n$=-1 for the low opacity
example and $n$=0 for the high-opacity equatorial example, mainly
because they provided a crude agreement with the Matcheva et
al. results at higher pressures. For this bottom cloud we also assumed
the same absorption to extinction optical depth ratio as for \nhfshx.
After scaling our opacities to the appropriate wavelength, we
distributed them vertically using the same Gaussian function described
for the NIMS comparison, this time adjusting the vertical width to
simulate the vertical resolution of the Matcheva et
al. retrieval. This was done to facilitate comparison and not because we
believe that the opacity is so distributed.

At 12\deg N, the VIMS and CIRS opacity profiles are in crude agreement
on vertical gradient, total absorption, pressure location of the
absorption peak (near 1.1 bars), but disagree on the magnitude of the
peak. This is probably about as good an agreement as might be expected, given the
very different models for vertical distribution.  For the 12\deg N
case, we also included a scaled version of the \cite{Irwin2002PCA}
profile, but added an imaginary index of 0.02 to simulate \nht
absorption and 0.2 to simulate \nhfsh absorption.  Here we see that
the NIMS result does not scale well to the 7-\mum region, no matter
which absorber we assume. The vertical gradient, peak location, and
magnitude are in strong disagreement with the other results.

The equatorial comparison in Fig.\ \ref{Fig:Matcheva} tells a somewhat
different story. Our results provide a weaker vertical opacity (after
smoothing), while the CIRS analysis yields a very strong gradient.  It
is worth noting that the layers in disagreement are those in which we
find significant scattering contributions, suggesting that the
Matcheva et al. analysis might need to be revised to take account of
scattering effects. The comparison might also be affected by local
variations, and might be better if we had a larger statistical sample.
It may also be a problem for one or the other analyses to obtain
accurate results for regions with very high cloud opacity, since
sensitivity beyond the first few optical depths is very diminished for
both VIMS and CIRS analyses.  Perhaps the greatest uncertainty in this
comparison is associated with the uncertainty in how to scale the
bottom cloud results to the same wavelength.  Currently we can only
say that the two results might be consistent.

\subsection{Cloud composition}

Models that fit the VIMS spectra have some ammonia ice in the 330-550
mb region, probably in the form of ammonia combined with conservative
material (near 3 \mum) of relatively high real refractive index.  \nht
could be present either as a coating or as the core material, in
either case representing less than half the total particle
volume. These particles have sub-micron particle radii, probably near
0.3 \mumx. The non-\nht component might be sedimented hydrocarbons
generated by photolysis of stratospheric methane, although it does not
appear likely that hydrazine is a major component.  The main layer
responsible for the 3-\mum absorption feature is generally in the
500-600 mb region, and the best-fit composition of this layer is
\nhfshx, although this layer could also contain a contribution from
\nht ice, either in the form of a separate particle population or
perhaps as a coating. In fact, the close proximity of the \nht and
\nhfsh layers in many cases, suggest that they might be co-mingled.

It is possible within the constraints of the spectra that low-opacity
regions have 3-\mum absorption provided entirely by \nht and that
\nhfsh is only needed in the higher-opacity regions. But it seems more
sensible that the deeper clouds are all composed of similar materials,
and that \nht contributions appear in a different layer at generally
lower pressures, although the fact that the two layers are virtually
on top of each other in the low opacity regions raises questions about
that possibility as well. Perhaps the entire particle distribution is
coated with \nhtx, but that only the small-particle component actually
allows the \nht absorption features to become visible.  Perhaps the
top layer of clouds is the small-particle component that displays
these features, while the bulk of the cloud is larger particles that
don't show the features.

We tried a number of models for the bottom cloud.  The one that seemed
to work best is one in which the phase function is wavelength
independent and matches that of \cite{Sro2002jup}, while the optical
depth varies as $\lambda^n$, with $n$ ranging from -1.3 to
+1.6. This resulted in pressures from 800 mb to 1.3 bars and optical
depths from 5 to 20 (Table \ref{Tbl:msfits}). This cloud might be
composed of \nhfshx, but using that material in spherical particles
degraded the fit quality.  If we used a large number of quadrature
points to capture the wavelength dependence of the phase function with
higher accuracy, we found the details of that variation did not match
the observations.  On the other hand, if we used a small number of
quadrature points to suppress the backscatter peak and its variations,
this also made the lower cloud insufficiently bright, resulting in
poor fits to the window regions.  It appears that if \nhfsh is the
main component of this cloud, it must have a lower imaginary index in
the near-IR window regions, or have increased backscatter with a wavelength
dependence different from that of spherical particles.  There may also
be accompanying variations in deeper regions of the atmosphere that
are not constrained by the observations we used.

\subsection{Dynamical considerations}

At first glance, a scenario in which \nhfsh provides the main
component of cloud particles in vicinity of 500 mb seems implausible.
Such a scenario is certainly not consistent with a uniform upwelling
in zones and descending air in belts, as commonly suggested in early
studies of Jovian dynamics (e.g. \cite{Hess1969};
\cite{Stone1976}). However, there is an observational conflict with
that simple belt-zone dynamical model, namely the Galileo discovery
that most of the lightning occurs within belts, where air was thought
to be descending, and is associated with rapidly growing cloud
clusters reaching pressures of a few hundred millibars \citep{Ingersoll2000}.
These events would certainly dredge up condensible material from
deeper levels, including \nhtx, H$_2$S, water, and likely particles of
\nhfshx.  If these events provided a major source of condensibles in
the upper troposphere, then it is less difficult to understand how
\nhfsh particles might contribute a major source of cloud material
near 500 mb.  The potentially important role of these thunderstorms in
controlling the composition of the upper troposphere is also
highlighted by \cite{Showman2005}, who provide a dynamical explanation
for another non-intuitive observation, namely the global depletion of
\nht vapor between 500 mb and 5 bars in both belts and zones.  Their
explanation is based on the idea ``(1) that the majority of air that
ascends across the 5 bar interface resides in isolated thunderstorms,
and (2) that 50\% of the ammonia within these storms is lost through
the 4-6 bar level either by direct ammonia rainout through the base of
the storms or downward transport of ammonia vapor in convective
downdrafts moistened by evaporation of rainfall.'' They speculate that
``air undergoing large scale ascent above the 0.5-bar cloud tops in
zones is probably supplied horizontally from the belts (where
thunderstorms predominantly occur) rather than from below.'' The
belt-to-zone mass flux could be provided by eddies
\citep{Ingersoll2000}. It also seems plausible that during the strong
convective events, composite particles could form, including
ammonia-coated \nhfshx.  When distributed horizontally, such particles
would be mixed with less ammonia-rich air, promoting evaporation of
the coating and leaving behind the \nhfsh core, which might also
contain some component of H$_2$O, although the VIMS spectra would not
be consistent with a large fraction of water ice in such composite
particles.  A quantitative microphysical investigation of this
scenario is needed to provide a better understanding of its
plausibility.

\subsection{Relation to ammonia ice detected at thermal wavelengths}\label{Sec:Wong}

Detections of \nht spectral features near 10 $\mu$m were first
obtained by \cite{Wong2004}. They used the brightness temperature
difference between 1040 \icm and 1060 \icm to measure the strength of
the ammonia ice signature.  This parameter peaked at the equator and
at 23\deg N, and was much more widely distributed than the very rare
discrete ammonia ice clouds identified by \cite{Baines2002Icar}, but was
clearly not as omnipresent as the 3-\mum absorption feature.  They
modeled their observations with 4:1 prolate spheroids of a volume
equivalent radius of 0.79 \mumx, a cloud base of 790 mb, and extending
up to 100 mb with a 1:1 particle to gas scale height, accompanied by a
much more compact (1/8 the gas scale height) gray absorber, which they
modeled as 10-\mum \nht ice spheres.  For average spectra from two
regions (22-25\deg N, covering 140-240\deg W, and 14-17\deg N,
covering 10-70\deg W) they obtained for the small particle component
optical depths of 0.75 and an upper limit of 0.2 respectively.  The
larger optical depth is likely roughly comparable to optical depths in
the equatorial region where they also found a peak in the ice
signature.

Although the vertical variation in opacity of the \cite{Wong2004}
cloud models is not a good match to ours, the latitudinal variation in
ammonia ice they infer is at least qualitatively consistent with our
results in Fig.\ \ref{Fig:latvar}. In our models the \nhtx-containing
layer descends from 400 mb to 600 mb from the equator to 14\degx N.
This change is especially relevant because of the \cite{Wong2004}
conclusion that ammonia ice present deeper than 500 mb would not be
detectable at 10 \mum because it would be too far below the peak of
the contribution function of the observations. This implies that
their observations would detect much less \nht in the NEB. Their broad
vertical distribution of the small particle component for the
cloudiest region, which should be similar to the equatorial region,
would place about half its optical depth at pressures less than 450
mb, which would be about 0.38 extinction optical depths, comparable to
what we would estimate for the layer of small ammonia-coated
particles.  Their more compact cloud could easily be composed of
\nhfshx.

\subsection{Relation to SIACs}

\cite{Baines2002Icar} identified anomalous discrete cloud features in
the wake of the GRS, which they termed spectrally identifiable ammonia
cloud (SIACs) based on low reflectivity at 2.7 \mum relative to high
reflectivity at 1.6 \mumx, and a local dip in reflectivity at 2 \mum
relative to 1.94 \mumx, both depressions characteristic of ammonia ice
absorption (the longer wavelength absorption is only apparent in
optically thick clouds, and not noticeable in Fig.\
\ref{Fig:reflections}). These features were observed in Galileo NIMS
spectra, taken at very high spatial resolution, and appear to occupy
less than 1\% of Jupiter's cloud-top surface.  The VIMS spectra we
analyzed (or any other VIMS spectra) have far too low a spatial
resolution to detect SIACs. Unlike SIACs, the 3-\mum absorption
feature we analyzed appears to be very widely distributed on Jupiter.
Although we focused on low-latitude spectra for detailed analysis in
this paper, we did look at VIMS spectra over a wide range of latitudes
and longitudes, and found obvious strong absorption at 3 \mum
everywhere we looked.  The SIACs appear to be the result of unusually
strong localized vertical transport that causes significant
condensation of ammonia vapor, such that the ammonia spectral features
become significantly enhanced.  It appears from our results, that most
locations on Jupiter don't have very much condensed \nhtx, so that the
the spectral character of SIACs is rarely observed.  The alternative
possibility that \nht is in fact the major cloud component, but has
its spectral signature significantly altered by some photochemical
``tanning'' process remains to be quantitatively investigated, and
whether it could satisfy known spectral and photochemical constraints
seems doubtful.

\subsection{Comparison with Saturn}

\begin{figure}[!htb]\centering
\includegraphics[width=3.5in]{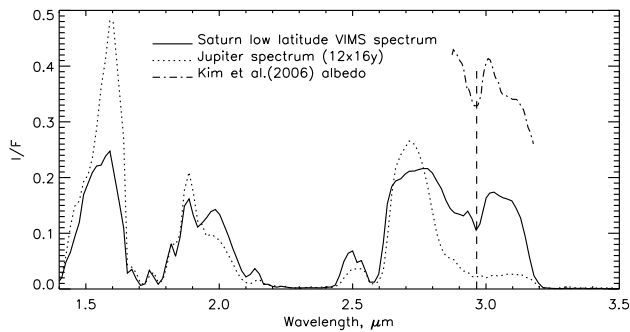}
\caption{A typical low-latitude VIMS spectrum of Saturn (from x=21,
y=2 in data cube CM\_1587635989) reveals absorption by PH$_3$ between
2.8 \mum and 3.0 \mumx, but does not display the broad absorption from
3.0 \mum to 3.2 \mum that is characteristic of the typical Jovian
spectrum, such as the plotted 12x16y VIMS spectrum (dashed).  The
cloud albedo (dot-dash) inferred by \cite{Kim2006} from analysis of a
groundbased high-resolution spectrum contains a local absorption
qualitatively consistent with a layer of small particles containing a
minor fraction of ammonia ice, either as a core or coating. The
vertical dotted line is at 2.965 \mumx.}
\label{Fig:saturn}
\end{figure}

Saturn's near-IR spectrum does not contain the broad and strong
3-\mum absorption feature so apparent in Jupiter's spectrum \citep{Encrenaz1999, Baines2005}. 
The stark difference between spectra of these two planets is
illustrated in Fig.\ \ref{Fig:saturn}, where we compare  a typical low-latitude Saturn
spectrum from VIMS data cube CM\_1587635989
%and CM\_1587636728 
with a VIMS Jupiter spectrum (12x16y).  The high reflectance of Saturn's
clouds at 3.0 \mum ($\sim$0.2), just where Jupiter's I/F is quite low ($\sim$0.02),
implies that the main cloud layer contributing to Saturn's reflection
spectrum in the near IR is not composed of medium sized pure ammonia
ice particles (nor \nhfsh particles).  
This is likely the layer of upper troposheric haze that \cite{Perez-Hoyoz2005}
found to extend from a base at 300 mb - 500 mb to a top near 40 mb - 80 mb,
varying with latitude and season, but generally of substantial optical depth
($\approx$ 6-20 at $\sim$1 \mumx), with a weak dependence on
wavelength.  This layer is probably thick enough at 3 \mum to obscure
the much deeper ammonia cloud thought to reside between 1.4 and 1.8
bars.

However, there is some evidence for at least a component of ammonia
ice in Saturn's tropospheric haze.  From analysis of a high-resolution
groundbased spectrum of Saturn, \cite{Kim2006} developed evidence for
a weak cloud absorption feature at 2.96 \mumx, where their inferred
model cloud layer at 460 mb had its albedo drop by $\sim$20\%, which
they suggested might be due to ammonia ice.  A small absorption
feature at the same wavelength is also seen in many spatially resolved
VIMS spectra of Saturn (also evident in Fig.\ \ref{Fig:saturn}), and
is not a characteristic of the phosphine spectrum that provides the
main absorption in this region.  This very well may be due to a
population of ammonia-coated particles within the main tropospheric
haze layer of Saturn.  A small composite particle of this type with
$\sim$30\% ammonia content, would lack the broad absorption feature of
larger pure ammonia ice, and have a relatively narrow, but subdued,
peak absorption displaced towards the ammonia imaginary index peak at
2.965 \mumx, as is shown in Fig.\ \ref{Fig:coatex}.  This shifted and
attenuated feature could also be produced by particles with a small
core of ammonia ice coated by a thick layer of hydrocarbons, as
demonstrated by the benzene-coated example of \cite{Kalogerakis2008}.
It seems likely, but remains to be demonstrated, that a moderately
thick tropospheric haze containing a component of such particles could
provide accurate fits to Saturn's near-IR spectra.

\section{Summary and conclusions}

From an analysis of VIMS low and medium phase angle spatially resolved
spectra obtained during the Cassini 2000 flyby of Jupiter, we obtained
the following main conclusions regarding the VIMS
instrument performance, the source of Jupiter's 3-\mum absorption, and
Jupiter's low-latitude cloud structure.

\begin{enumerate}

\item By comparing VIMS spectra with groundbased spectra, selected
NICMOS observations, and an ISO spectrum, we were able to confirm that
the VIMS absolute calibration is relatively consistent with other
observations, but that VIMS measurements have two significant
artifacts, one in the 1.60-1.68 \mum region where a huge responsivity
correction turns out to be insufficiently accurate, and the other is
an excessively high I/F measurements at low signal levels, which
caused model fits to yield stratospheric haze opacity $\sim$10 times
that derived from groundbased spectra .

\item We found that VIMS spectra contain evidence for a prominent and
widespread 3-\mum absorber in Jupiter's clouds and a much weaker
absorber active near 2 \mumx. This was made obvious by reflecting
layer fits constrained outside the absorbing region, which produce
more than the observed reflectivity in the absorbing regions, by
$\sim$20\% at 2.04 \mum and by 300-500\% at 3-3.1 \mum.  

\item We were able to fit a low-opacity VIMS
spectrum quite well using \nht ice as the only 3-\mum absorber, 
but were not able to fit very well the spectra from much cloudier
regions, the main defect being an
absorption feature near 2 \mum in the model spectra that was much
larger than observed.  Models with \nhfsh as the only absorber in a
similar layer structure were more successful in fitting high
opacity spectra and almost as good fitting low opacity spectra, the
main defect being lack of a 2.97-\mum spectral feature which appears
to require a contribution from \nhtx.

\item The best fits to VIMS spectra were obtained with both \nht and
\nhfsh absorbers present in Jupiter's clouds, but the \nht
contribution appears to help most in the form of a
coating or as a core material comprising less than half the
total particle volume.  We roughly simulated \nhtx-coated particles using a
synthetic refractive index with absorption similar to that of \nht but
significantly weaker.  A layer of these small (r$\sim$0.3 \mumx)
particles above a layer of much larger (r$\sim$10 \mumx) \nhfsh
particles provided the best fits to all the spectra.  The \nhtx-coated
cloud layer (modeled as physically thin but ranging from 330 to 550
mb) was found to have the least variability in optical depth among the
eight low latitude spectra, with an average of 0.38 and a standard
deviation of only 0.05.  The main 3-\mum absorbing layer varies in
optical depth from 0.23 to 2.1, a factor of nine, with pressures
ranging from 358 to 635 mb.  In most cases the \nhtx-containing layer
was found in such close proximity to the \nhfsh layer that they could be
intermingled. A caveat in these results is that the simulation of the
\nhtx-coated particle does not closely follow the size and wavelength
dependence of a true \nhtx-coated particle, and thus some other
structure may be required to provide physically consistent fits.

\item Comparison with Galileo Probe results provided
crude agreement on the vertical location of clouds, the upper level
cloud opacity, the pressure of the lower cloud, and the relative 
lack of cloud particles near 700 mb, but did not agree
with the NFR-derived vertical profile of \nhtx.

\item Comparison with the analysis of Galileo NIMS observations by
\cite{Irwin2001BZ} shows a rough agreement with the vertical trend of
opacity and the location of the peaks in opacity, but we find greater
absolute opacity values.  We cannot confirm (or rule out) the large
belt-zone differences they found in the 1.5-2 bar region. When the
results of \cite{Irwin2002PCA}, which apply to belt regions, are
scaled to 2 \mumx, there is an even larger discrepancy than with
\cite{Irwin2001BZ}. 

\item Comparison of our results scaled to a wavelength of 0.75 \mum
with Galileo SSI results by \cite{Banfield1998} and
\cite{Simon-Miller2001} finds reasonable agreement with cumulative
opacity profiles, with the exception of the location of the lower
high-opacity cloud in zone models, which the SSI analysis places near
700 mb and our analysis places near 1100 mb.  

\item Comparison of our absorption profiles with the analysis of
Cassini CIRS observations by \cite{Matcheva2005}, with plausible
scalings with wavelength, reveals a rough agreement, but with
CIRS opacities less than ours by a factor of two or more. 
However, the validity of the comparison is uncertain because of the
large uncertainty in appropriate scaling factors to use for the lower
cloud.

\item The presence of a layer of sub-micron \nht ice coated particles
at pressures of 300-500 mb does not appear to conflict with
thermal spectra in the 9-10 \mum region, where only a weak ammonia
signature is observed. This is possible because the coated
particle can produce the reflective contribution needed to match the
near-IR spectra, but ten times less extinction optical depth at 9.4
\mum than would be obtained from pure \nht particles of the same
size. Our estimated extinction optical depth for this layer is crudely
compatible with the optical depth estimate of \cite{Wong2004} in the
latitude band we analyzed. The increased pressure we find for the \nht
layer in the NEB is qualitatively consistent with the low signature
found by Wong et al. in a similar region.

\item Saturn's spectra do not exhibit the broad 3-\mum absorption feature
characteristic of Jovian spectra, perhaps because the cloud layer containing
that absorber is obscured by the moderately thick tropospheric haze that overlies it.
However, Saturn spectra do exhibit a small absorption feature near 2.965 \mumx,
which is qualitatively consistent with \nht absorption when diluted by combining
\nht with a substantial conservative core or shell.

\end{enumerate}

\noindent

\section*{Acknowledgments.} \addcontentsline{toc}{section}{Acknowledgments}

Support for this work was
provided by NASA through its Planetary Atmospheres Program,
under grant NAG5-13297, the Outer Planet Data Analysis Program under grant
NNG05GG93G, the Cassini Data Analysis Program under grant NNX07AJ82G, and 
the Jupiter Data Analysis Program under grant NNX09AE07G. We thank two anonymous
reviewers for prompt, detailed, and constructive comments.

%\newpage

%\bibliographystyle{/home/home2/sro/uranus/paper3/elsart-harv-nomonth}
%\bibliography{/home/home2/sro/uranus/paper3/outerplanets}

\begin{thebibliography}{69}
\expandafter\ifx\csname natexlab\endcsname\relax\def\natexlab#1{#1}\fi
\expandafter\ifx\csname url\endcsname\relax
  \def\url#1{\texttt{#1}}\fi
\expandafter\ifx\csname urlprefix\endcsname\relax\def\urlprefix{URL }\fi

\bibitem[{{Achterberg} et~al.(2006){Achterberg}, {Conrath}, and
  {Gierasch}}]{Achterberg2006}
{Achterberg}, R.~K., {Conrath}, B.~J., {Gierasch}, P.~J., 2006. {Cassini CIRS
  retrievals of ammonia in Jupiter's upper troposphere}. Icarus 182, 169--180.

\bibitem[{{Baines} et~al.(2002){Baines}, {Carlson}, and
  {Kamp}}]{Baines2002Icar}
{Baines}, K.~H., {Carlson}, R.~W., {Kamp}, L.~W., 2002. {Fresh Ammonia Ice
  Clouds in Jupiter I. Spectroscopic Identification, Spatial Distribution, and
  Dynamical Implications}. Icarus 159, 74--94.

\bibitem[{{Baines} et~al.(2005){Baines}, {Drossart}, {Momary}, {Formisano},
  {Griffith}, {Bellucci}, {Bibring}, {Brown}, {Buratti}, {Capaccioni},
  {Cerroni}, {Clark}, {Coradini}, {Combes}, {Cruikshank}, {Jaumann},
  {Langevin}, {Matson}, {McCord}, {Mennella}, {Nelson}, {Nicholson}, {Sicardy},
  and {Sotin}}]{Baines2005}
{Baines}, K.~H., {Drossart}, P., {Momary}, T.~W., {Formisano}, V., {Griffith},
  C., {Bellucci}, G., {Bibring}, J.~P., {Brown}, R.~H., {Buratti}, B.~J.,
  {Capaccioni}, F., {Cerroni}, P., {Clark}, R.~N., {Coradini}, A., {Combes},
  M., {Cruikshank}, D.~P., {Jaumann}, R., {Langevin}, Y., {Matson}, D.~L.,
  {McCord}, T.~B., {Mennella}, V., {Nelson}, R.~M., {Nicholson}, P.~D.,
  {Sicardy}, B., {Sotin}, C., 2005. {The Atmospheres of Saturn and Titan in the
  Near-Infrared First Results of Cassini/vims}. Earth Moon and Planets 96,
  119--147.

\bibitem[{{Banfield} et~al.(1998{\natexlab{a}}){Banfield}, {Conrath},
  {Gierasch}, {Nicholson}, and {Matthews}}]{Banfield1998NIR}
{Banfield}, D., {Conrath}, B.~J., {Gierasch}, P.~J., {Nicholson}, P.~D.,
  {Matthews}, K., 1998{\natexlab{a}}. {Near-IR Spectrophotometry of Jovian
  Aerosols-Meridional and Vertical Distributions}. Icarus 134, 11--23.

\bibitem[{{Banfield} et~al.(1998{\natexlab{b}}){Banfield}, {Gierasch}, {Bell},
  {Ustinov}, {Ingersoll}, {Vasavada}, {West}, and {Belton}}]{Banfield1998}
{Banfield}, D., {Gierasch}, P.~J., {Bell}, M., {Ustinov}, E., {Ingersoll},
  A.~P., {Vasavada}, A.~R., {West}, R.~A., {Belton}, M.~J.~S.,
  1998{\natexlab{b}}. {Jupiter's Cloud Structure from Galileo Imaging Data}.
  Icarus 135, 230--250.

\bibitem[{{Birnbaum} et~al.(1996){Birnbaum}, {Borysow}, and
  {Orton}}]{Birnbaum1996}
{Birnbaum}, G., {Borysow}, A., {Orton}, G.~S., 1996. {Collision-Induced
  Absorption of H\_2-H\_2 and H\_2-He in the Rotational and Fundamental Bands
  for Planetary Applications}. Icarus 123, 4--22.

\bibitem[{{Borysow}(1991)}]{Borysow1991h2h2f}
{Borysow}, A., 1991. {Modeling of collision-induced infrared absorption spectra
  of H2-H2 pairs in the fundamental band at temperatures from 20 to 300 K}.
  Icarus 92, 273--279.

\bibitem[{{Borysow}(1992)}]{Borysow1992h2he}
{Borysow}, A., 1992. {New model of collision-induced infrared absorption
  spectra of H2-He pairs in the 2-2.5 micron range at temperatures from 20 to
  300 K - an update}. Icarus 96, 169--175.

\bibitem[{{Borysow}(1993)}]{Borysow1993errat}
{Borysow}, A., 1993. {Erratum}. Icarus 106, 614.

\bibitem[{{Bowles} et~al.(2008){Bowles}, {Calcutt}, {Irwin}, and
  {Temple}}]{Bowles2008}
{Bowles}, N., {Calcutt}, S., {Irwin}, P., {Temple}, J., 2008. {Band parameters
  for self-broadened ammonia gas in the range 0.74 to 5.24 {$\mu$}m to support
  measurements of the atmosphere of the planet Jupiter}. Icarus 196, 612--624.

\bibitem[{{Brooke} et~al.(1998){Brooke}, {Knacke}, {Encrenaz}, {Drossart},
  {Crisp}, and {Feuchtgruber}}]{Brooke1998}
{Brooke}, T.~Y., {Knacke}, R.~F., {Encrenaz}, T., {Drossart}, P., {Crisp}, D.,
  {Feuchtgruber}, H., 1998. {Models of the ISO 3-{$\mu$}m Reflection Spectrum
  of Jupiter}. Icarus 136, 1--13.

\bibitem[{{Brown} et~al.(2004){Brown}, {Baines}, {Bellucci}, {Bibring},
  {Buratti}, {Capaccioni}, {Cerroni}, {Clark}, {Coradini}, {Cruikshank},
  {Drossart}, {Formisano}, {Jaumann}, {Langevin}, {Matson}, {McCord},
  {Mennella}, {Miller}, {Nelson}, {Nicholson}, {Sicardy}, and
  {Sotin}}]{Brown2004SSR}
{Brown}, R.~H., {Baines}, K.~H., {Bellucci}, G., {Bibring}, J.-P., {Buratti},
  B.~J., {Capaccioni}, F., {Cerroni}, P., {Clark}, R.~N., {Coradini}, A.,
  {Cruikshank}, D.~P., {Drossart}, P., {Formisano}, V., {Jaumann}, R.,
  {Langevin}, Y., {Matson}, D.~L., {McCord}, T.~B., {Mennella}, V., {Miller},
  E., {Nelson}, R.~M., {Nicholson}, P.~D., {Sicardy}, B., {Sotin}, C., 2004.
  {The Cassini Visual And Infrared Mapping Spectrometer (Vims) Investigation}.
  Space Science Reviews 115, 111--168.

\bibitem[{{Carlson} et~al.(1993){Carlson}, {Lacis}, and
  {Rossow}}]{Carlson1993neb}
{Carlson}, B.~E., {Lacis}, A.~A., {Rossow}, W.~B., 1993. {Tropospheric gas
  composition and cloud structure of the Jovian North Equatorial Belt}. \jgr
  98, 5251--5290.

\bibitem[{{Clapp} and {Miller}(1996)}]{Clapp1996}
{Clapp}, M.~L., {Miller}, R.~E., 1996. {Complex Refractive Indices of
  Crystalline Hydrazine from Aerosol Extinction Spectra}. Icarus 123, 396--403.

\bibitem[{{Clark} and {McCord}(1979)}]{Clark1979jup}
{Clark}, R.~N., {McCord}, T.~B., 1979. {Jupiter and Saturn - Near-infrared
  spectral albedos}. Icarus 40, 180--188.

\bibitem[{{Colina} et~al.(1996){Colina}, {Bohlin}, and {Castelli}}]{Colina1996}
{Colina}, L., {Bohlin}, R.~C., {Castelli}, F., 1996. {The 0.12-2.5 micron
  Absolute Flux Distribution of the Sun for Comparison With Solar Analog
  Stars}. \aj 112, 307--315.

\bibitem[{{Conrath} and {Gierasch}(1984)}]{Conrath1984}
{Conrath}, B.~J., {Gierasch}, P.~J., 1984. {Global variation of the para
  hydrogen fraction in Jupiter's atmosphere and implications for dynamics on
  the outer planets}. Icarus 57, 184--204.

\bibitem[{{de Pater}(1986)}]{DePater1986Icar}
{de Pater}, I., 1986. {Jupiter's zone-belt structure at radio wavelengths. II -
  Comparison of observations with model atmosphere calculations}. Icarus 68,
  344--365.

\bibitem[{{de Pater} et~al.(2001){de Pater}, {Dunn}, {Romani}, and
  {Zahnle}}]{DePater2001Icar}
{de Pater}, I., {Dunn}, D., {Romani}, P., {Zahnle}, K., 2001. {Reconciling
  Galileo Probe Data and Ground-Based Radio Observations of Ammonia on
  Jupiter}. Icarus 149, 66--78.

\bibitem[{{Edgington} et~al.(1999){Edgington}, {Atreya}, {Trafton}, {Caldwell},
  {Beebe}, {Simon}, and {West}}]{Edgington1999}
{Edgington}, S.~G., {Atreya}, S.~K., {Trafton}, L.~M., {Caldwell}, J.~J.,
  {Beebe}, R.~F., {Simon}, A.~A., {West}, R.~A., 1999. {Ammonia and Eddy Mixing
  Variations in the Upper Troposphere of Jupiter from HST Faint Object
  Spectrograph Observations}. Icarus 142, 342--356.

\bibitem[{{Encrenaz} et~al.(1999){Encrenaz}, {Drossart}, {Feuchtgruber},
  {Lellouch}, {B{\'e}zard}, {Fouchet}, and {Atreya}}]{Encrenaz1999}
{Encrenaz}, T., {Drossart}, P., {Feuchtgruber}, H., {Lellouch}, E.,
  {B{\'e}zard}, B., {Fouchet}, T., {Atreya}, S.~K., 1999. {The atmospheric
  composition and structure of Jupiter and Saturn from ISO observations: a
  preliminary review}. Plan. \& Sp. Sci. 47, 1225--1242.

\bibitem[{{Fanale} et~al.(1977){Fanale}, {Johnson}, and {Matson}}]{Fanale1977}
{Fanale}, F.~P., {Johnson}, T.~V., {Matson}, D.~L., 1977. {Io's surface and the
  histories of the Galilean Satellites}. In: Burns, J.~A. (Ed.), {Planetary
  Satelliites}. {University of Arizona, Tucson}, pp. 379--405.

\bibitem[{{Ferraro} et~al.(1980){Ferraro}, {Sill}, and {Fink}}]{Ferraro1980}
{Ferraro}, J.~R., {Sill}, G., {Fink}, U., 1980. {Infrared Intensity
  Measurements of Cryodeposited Thin Films of NH$_{3}$, NH$_{4}$HS, H$_{2}$S,
  and Assignments of Absorption Bands}. Appl. Spectr. 34, 525--533.

\bibitem[{{Folkner} et~al.(1998){Folkner}, {Woo}, and {Nandi}}]{Folkner1998}
{Folkner}, W.~M., {Woo}, R., {Nandi}, S., 1998. {Ammonia abundance in Jupiter's
  atmosphere derived from the attenuation of the Galileo probe's radio signal}.
  \jgr 103, 22847--22856.

\bibitem[{{Fouchet} et~al.(2000){Fouchet}, {Lellouch}, {B{\'e}zard},
  {Encrenaz}, {Drossart}, {Feuchtgruber}, and {de Graauw}}]{Fouchet2000}
{Fouchet}, T., {Lellouch}, E., {B{\'e}zard}, B., {Encrenaz}, T., {Drossart},
  P., {Feuchtgruber}, H., {de Graauw}, T., 2000. {ISO-SWS Observations of
  Jupiter: Measurement of the Ammonia Tropospheric Profile and of the
  $^{15}$N/$^{14}$N Isotopic Ratio}. Icarus 143, 223--243.

\bibitem[{{Gibson} et~al.(2005){Gibson}, {Welch}, and {de
  Pater}}]{Gibson2005Icar}
{Gibson}, J., {Welch}, W.~J., {de Pater}, I., 2005. {Accurate jovian radio flux
  density measurements show ammonia to be subsaturated in the upper
  troposphere}. Icarus 173, 439--446.

\bibitem[{{Hess} and {Panofsky}(1969)}]{Hess1969}
{Hess}, S.~L., {Panofsky}, H.~A., 1969. {The atmospheres of the other planets}.
  In: {T.~F.~Malone} (Ed.), Compendium of Meteorology. {Boston: American
  Meteorological Society}, pp. 391--400.

\bibitem[{{Howett} et~al.(2007){Howett}, {Carlson}, {Irwin}, and
  {Calcutt}}]{Howett2007}
{Howett}, C.~J.~A., {Carlson}, R.~W., {Irwin}, P.~G.~J., {Calcutt}, S.~B.,
  2007. {Optical constants of ammonium hydrosulfide ice and ammonia ice}.
  Journal of the Optical Society of America B Optical Physics 24, 126--136.

\bibitem[{{Hu} et~al.(2000){Hu}, {Wielicki}, {Lin}, {Gibson}, {Tsay},
  {Stamnes}, and {Wong}}]{Hu2000}
{Hu}, Y.~X., {Wielicki}, B., {Lin}, B., {Gibson}, G., {Tsay}, S.~C., {Stamnes},
  K., {Wong}, T., 2000. {delta-fit: a fast and accurate treatment of particle
  scattering phase functions with weighted singular-value decomposition least
  squares fitting}. J. Quant. Spectr. and Rad. Trans. 65, 681--690.

\bibitem[{{Ingersoll} et~al.(2000){Ingersoll}, {Gierasch}, {Banfield},
  {Vasavada}, and {Galileo Imaging Team}}]{Ingersoll2000}
{Ingersoll}, A.~P., {Gierasch}, P.~J., {Banfield}, D., {Vasavada}, A.~R.,
  {Galileo Imaging Team}, 2000. {Moist convection as an energy source for the
  large-scale motions in Jupiter's atmosphere}. Nature 403, 630--632.

\bibitem[{{Irwin} and {Dyudina}(2002)}]{Irwin2002PCA}
{Irwin}, P.~G.~J., {Dyudina}, U., 2002. {The Retrieval of Cloud Structure Maps
  in the Equatorial Region of Jupiter Using a Principal Component Analysis of
  Galileo/NIMS Data}. Icarus 156, 52--63.

\bibitem[{{Irwin} et~al.(2006){Irwin}, {Sromovsky}, {Strong}, {Sihra},
  {Bowles}, {Calcutt}, and {Remedios}}]{Irwin2006ch42e}
{Irwin}, P.~G.~J., {Sromovsky}, L.~A., {Strong}, E.~K., {Sihra}, K., {Bowles},
  N., {Calcutt}, S.~B., {Remedios}, J.~J., 2006. {Improved near-infrared
  methane band models and k-distribution parameters from 2000 to 9500 cm-1 and
  implications for interpretation of outer planet spectra}. Icarus 181,
  309--319.

\bibitem[{{Irwin} et~al.(2001){Irwin}, {Weir}, {Taylor}, {Calcutt}, and
  {Carlson}}]{Irwin2001BZ}
{Irwin}, P.~G.~J., {Weir}, A.~L., {Taylor}, F.~W., {Calcutt}, S.~B., {Carlson},
  R.~W., 2001. {The Origin of Belt/Zone Contrasts in the Atmosphere of Jupiter
  and Their Correlation with 5-{$\mu$}m Opacity}. Icarus 149, 397--415.

\bibitem[{{Kalogerakis} et~al.(2008){Kalogerakis}, {Marschall}, {Oza}, {Engel},
  {Meharchand}, and {Wong}}]{Kalogerakis2008}
{Kalogerakis}, K.~S., {Marschall}, J., {Oza}, A.~U., {Engel}, P.~A.,
  {Meharchand}, R.~T., {Wong}, M.~H., 2008. {The coating hypothesis for ammonia
  ice particles in Jupiter: Laboratory experiments and optical modeling}.
  Icarus 196, 202--215.

\bibitem[{{Karkoschka}(1998)}]{Kark1998Icar}
{Karkoschka}, E., 1998. {Methane, Ammonia, and Temperature Measurements of the
  Jovian Planets and Titan from CCD-Spectrophotometry}. Icarus 133, 134--146.

\bibitem[{{Karkoschka} and {Tomasko}(2009)}]{Kark2009IcarusSTIS}
{Karkoschka}, E., {Tomasko}, M., 2009. {The haze and methane distributions on
  Uranus from HST-STIS spectroscopy}. Icarus 202, 287--309.

\bibitem[{{Kim} et~al.(2006){Kim}, {Kim}, {Geballe}, {Kim}, and
  {Brown}}]{Kim2006}
{Kim}, J.~H., {Kim}, S.~J., {Geballe}, T.~R., {Kim}, S.~S., {Brown}, L.~R.,
  2006. {High-resolution spectroscopy of Saturn at 3 microns: CH$_{4}$,
  CH$_{3}$D, C$_{2}$H$_{2}$, C$_{2}$H$_{6}$, PH$_{3}$, clouds, and haze}.
  Icarus 185, 476--486.

\bibitem[{{Lara} et~al.(1998){Lara}, {Bezard}, {Griffith}, {Lacy}, and
  {Owen}}]{Lara1998}
{Lara}, L.-M., {Bezard}, B., {Griffith}, C.~A., {Lacy}, J.~H., {Owen}, T.,
  1998. {High-Resolution 10-micronmeter Spectroscopy of Ammonia and Phosphine
  Lines on Jupiter}. Icarus 131, 317--333.

\bibitem[{{Martonchik} et~al.(1984){Martonchik}, {Orton}, and
  {Appleby}}]{Martonchik1984}
{Martonchik}, J.~V., {Orton}, G.~S., {Appleby}, J.~F., 1984. {Optical
  properties of NH3 ice from the far infrared to the near ultraviolet}. Appl.
  Optics 23, 541--547.

\bibitem[{{Matcheva} et~al.(2005){Matcheva}, {Conrath}, {Gierasch}, and
  {Flasar}}]{Matcheva2005}
{Matcheva}, K.~I., {Conrath}, B.~J., {Gierasch}, P.~J., {Flasar}, F.~M., 2005.
  {The cloud structure of the jovian atmosphere as seen by the Cassini/CIRS
  experiment}. Icarus 179, 432--448.

\bibitem[{{McCord} et~al.(2004){McCord}, {Coradini}, {Hibbitts}, {Capaccioni},
  {Hansen}, {Filacchione}, {Clark}, {Cerroni}, {Brown}, {Baines}, {Bellucci},
  {Bibring}, {Buratti}, {Bussoletti}, {Combes}, {Cruikshank}, {Drossart},
  {Formisano}, {Jaumann}, {Langevin}, {Matson}, {Nelson}, {Nicholson},
  {Sicardy}, and {Sotin}}]{McCord2004}
{McCord}, T.~B., {Coradini}, A., {Hibbitts}, C.~A., {Capaccioni}, F., {Hansen},
  G.~B., {Filacchione}, G., {Clark}, R.~N., {Cerroni}, P., {Brown}, R.~H.,
  {Baines}, K.~H., {Bellucci}, G., {Bibring}, J.-P., {Buratti}, B.~J.,
  {Bussoletti}, E., {Combes}, M., {Cruikshank}, D.~P., {Drossart}, P.,
  {Formisano}, V., {Jaumann}, R., {Langevin}, Y., {Matson}, D.~L., {Nelson},
  R.~M., {Nicholson}, P.~D., {Sicardy}, B., {Sotin}, C., 2004. {Cassini VIMS
  observations of the Galilean satellites including the VIMS calibration
  procedure}. Icarus 172, 104--126.

\bibitem[{{Miller} et~al.(1996){Miller}, {Klein}, {Juergens}, {Mehaffey},
  {Oseas}, {Garcia}, {Giandomenico}, {Irigoyen}, {Hickok}, {Rosing}, {Sobel},
  {Bruce}, {Flamini}, {Devidi}, {Reininger}, {Dami}, {Soufflot}, {Langevin},
  and {Huntzinger}}]{Miller1996SPIE}
{Miller}, E.~A., {Klein}, G., {Juergens}, D.~W., {Mehaffey}, K., {Oseas},
  J.~M., {Garcia}, R.~A., {Giandomenico}, A., {Irigoyen}, R.~E., {Hickok}, R.,
  {Rosing}, D., {Sobel}, H.~R., {Bruce}, C.~F., {Flamini}, E., {Devidi}, R.,
  {Reininger}, F.~M., {Dami}, M., {Soufflot}, A., {Langevin}, Y., {Huntzinger},
  G., 1996. {The Visual and Infrared Mapping Spectrometer for Cassini}. In:
  {Horn}, L. (Ed.), Society of Photo-Optical Instrumentation Engineers (SPIE)
  Conference Series. Vol. 2803 of Society of Photo-Optical Instrumentation
  Engineers (SPIE) Conference Series. pp. 206--220.

\bibitem[{{Niemann} et~al.(1998){Niemann}, {Atreya}, {Carignan}, {Donahue},
  {Haberman}, {Harpold}, {Hartle}, {Hunten}, {Kasprzak}, {Mahaffy}, {Owen}, and
  {Way}}]{Niemann1998}
{Niemann}, H.~B., {Atreya}, S.~K., {Carignan}, G.~R., {Donahue}, T.~M.,
  {Haberman}, J.~A., {Harpold}, D.~N., {Hartle}, R.~E., {Hunten}, D.~M.,
  {Kasprzak}, W.~T., {Mahaffy}, P.~R., {Owen}, T.~C., {Way}, S.~H., 1998. {The
  composition of the Jovian atmosphere as determined by the Galileo probe mass
  spectrometer}. \jgr 103, 22831--22846.

\bibitem[{{Orton} et~al.(1982){Orton}, {Appleby}, and {Martonchik}}]{Orton1982}
{Orton}, G.~S., {Appleby}, J.~F., {Martonchik}, J.~V., 1982. {The effect of
  ammonia ice on the outgoing thermal radiance from the atmosphere of Jupiter}.
  Icarus 52, 94--116.

\bibitem[{{P{\'e}rez-Hoyos} et~al.(2005){P{\'e}rez-Hoyos},
  {S{\'a}nchez-Lavega}, {French}, and {Rojas}}]{Perez-Hoyoz2005}
{P{\'e}rez-Hoyos}, S., {S{\'a}nchez-Lavega}, A., {French}, R.~G., {Rojas},
  J.~F., 2005. {Saturn's cloud structure and temporal evolution from ten years
  of Hubble Space Telescope images (1994 2003)}. Icarus 176, 155--174.

\bibitem[{{Press} et~al.(1992){Press}, {Teukolsky}, {Vetterling}, and
  {Flannery}}]{Press1992}
{Press}, W.~H., {Teukolsky}, S.~A., {Vetterling}, W.~T., {Flannery}, B.~P.,
  1992. {Numerical recipes in FORTRAN. The art of scientific computing, 2nd
  ed.} Cambridge: University Press.

\bibitem[{{Ragent} et~al.(1998){Ragent}, {Colburn}, {Rages}, {Knight}, {Avrin},
  {Orton}, {Yanamandra-Fisher}, and {Grams}}]{Ragent1998}
{Ragent}, B., {Colburn}, D.~S., {Rages}, K.~A., {Knight}, T.~C.~D., {Avrin},
  P., {Orton}, G.~S., {Yanamandra-Fisher}, P.~A., {Grams}, G.~W., 1998. {The
  clouds of Jupiter: Results of the Galileo Jupiter mission probe nephelometer
  experiment}. \jgr 103, 22891--22910.

\bibitem[{{Sault} et~al.(2004){Sault}, {Engel}, and {de Pater}}]{Sault2004}
{Sault}, R.~J., {Engel}, C., {de Pater}, I., 2004. {Longitude-resolved imaging
  of Jupiter at {$\lambda$}=2 cm}. Icarus 168, 336--343.

\bibitem[{{Schaeidt} et~al.(1996){Schaeidt}, {Morris}, {Salama},
  {Vandenbussche}, {Beintema}, {Boxhoorn}, {Feuchtgruber}, {Heras}, {Lahuis},
  {Leech}, {Roelfsema}, {Valentijn}, {Bauer}, {van der Bliek}, {Cohen}, {de
  Graauw}, {Haser}, {van der Hucht}, {Huygen}, {Katterloher}, {Kessler},
  {Koornneef}, {Luinge}, {Lutz}, {Planck}, {Spoon}, {Waelkens}, {Waters},
  {Wieprecht}, {Wildeman}, {Young}, and {Zaal}}]{Schaeidt1996}
{Schaeidt}, S.~G., {Morris}, P.~W., {Salama}, A., {Vandenbussche}, B.,
  {Beintema}, D.~A., {Boxhoorn}, D.~R., {Feuchtgruber}, H., {Heras}, A.~M.,
  {Lahuis}, F., {Leech}, K., {Roelfsema}, P.~R., {Valentijn}, E.~A., {Bauer},
  O.~H., {van der Bliek}, N.~S., {Cohen}, M., {de Graauw}, T., {Haser}, L.~N.,
  {van der Hucht}, K.~A., {Huygen}, E., {Katterloher}, R.~O., {Kessler}, M.~F.,
  {Koornneef}, J., {Luinge}, W., {Lutz}, D., {Planck}, M., {Spoon}, H.,
  {Waelkens}, C., {Waters}, L.~B.~F.~M., {Wieprecht}, E., {Wildeman}, K.~J.,
  {Young}, E., {Zaal}, P., 1996. {The photometric calibration of the ISO Short
  Wavelength Spectrometer.} Astron. \& Astrophys. 315, L55--L59.

\bibitem[{{Seiff} et~al.(1998){Seiff}, {Kirk}, {Knight}, {Young}, {Mihalov},
  {Young}, {Milos}, {Schubert}, {Blanchard}, and {Atkinson}}]{Seiff1998}
{Seiff}, A., {Kirk}, D.~B., {Knight}, T.~C.~D., {Young}, R.~E., {Mihalov},
  J.~D., {Young}, L.~A., {Milos}, F.~S., {Schubert}, G., {Blanchard}, R.~C.,
  {Atkinson}, D., 1998. {Thermal structure of Jupiter's atmosphere near the
  edge of a 5-{$\mu$}m hot spot in the north equatorial belt}. \jgr 103,
  22857--22890.

\bibitem[{{Showman} and {de Pater}(2005)}]{Showman2005}
{Showman}, A.~P., {de Pater}, I., 2005. {Dynamical implications of Jupiter's
  tropospheric ammonia abundance}. Icarus 174, 192--204.

\bibitem[{{Simon-Miller} et~al.(2001){Simon-Miller}, {Banfield}, and
  {Gierasch}}]{Simon-Miller2001}
{Simon-Miller}, A.~A., {Banfield}, D., {Gierasch}, P.~J., 2001. {Color and the
  Vertical Structure in Jupiter's Belts, Zones, and Weather Systems}. Icarus
  154, 459--474.

\bibitem[{{Sromovsky}(2005{\natexlab{a}})}]{Sro2005raman}
{Sromovsky}, L.~A., 2005{\natexlab{a}}. {Accurate and approximate calculations
  of Raman scattering in the atmosphere of Neptune}. Icarus 173, 254--283.

\bibitem[{{Sromovsky}(2005{\natexlab{b}})}]{Sro2005pol}
{Sromovsky}, L.~A., 2005{\natexlab{b}}. {Effects of Rayleigh-scattering
  polarization on reflected intensity: a fast and accurate approximation method
  for atmospheres with aerosols}. Icarus 173, 284--294.

\bibitem[{{Sromovsky} et~al.(1998){Sromovsky}, {Collard}, {Fry}, {Orton},
  {Lemmon}, {Tomasko}, and {Freedman}}]{Sro1998}
{Sromovsky}, L.~A., {Collard}, A.~D., {Fry}, P.~M., {Orton}, G.~S., {Lemmon},
  M.~T., {Tomasko}, M.~G., {Freedman}, R.~S., 1998. {Galileo probe measurements
  of thermal and solar radiation fluxes in the Jovian atmosphere}. \jgr 103,
  22929--22978.

\bibitem[{{Sromovsky} and {Fry}(2002)}]{Sro2002jup}
{Sromovsky}, L.~A., {Fry}, P.~M., 2002. {Jupiter's Cloud Structure as
  Constrained by Galileo Probe and HST Observations}. Icarus 157, 373--400.

\bibitem[{{Sromovsky} and {Fry}(2007)}]{Sro2007struc}
{Sromovsky}, L.~A., {Fry}, P.~M., 2007. {Spatially resolved cloud structure on
  Uranus: Implications of near-IR adaptive optics imaging}. Icarus 192,
  527--557.

\bibitem[{{Sromovsky} and {Fry}(2010)}]{Sro2010iso}
{Sromovsky}, L.~A., {Fry}, P.~M., 2010. {The source of 3-\mum absorption in
  Jupiter's clouds: Reanalysis of ISO observations with new NH$_3$ absorption
  models}. Icarus 210, 211-229.

\bibitem[{{Sromovsky} et~al.(2006){Sromovsky}, {Irwin}, and {Fry}}]{Sro2006ch4}
{Sromovsky}, L.~A., {Irwin}, P.~G.~J., {Fry}, P.~M., 2006. {Near-IR methane
  absorption in outer planet atmospheres: Improved models of temperature
  dependence and implications for Uranus cloud structure}. Icarus 182,
  577--593.

\bibitem[{{Stone}(1976)}]{Stone1976}
{Stone}, P.~H., 1976. {The meteorology of the Jovian atmosphere}. In:
  {T.~Gehrels} (Ed.), IAU Colloq. 30: Jupiter: Studies of the Interior, Atmosp
  here, Magnetosphere and Satellites. pp. 586--618.

\bibitem[{{Toon} and {Ackerman}(1981)}]{Toon1981}
{Toon}, O.~B., {Ackerman}, T.~P., 1981. {Algorithms for the calculation of
  scattering by stratified spheres}. \ao 20, 3657--3660.

\bibitem[{{von Zahn} et~al.(1998){von Zahn}, {Hunten}, and
  {Lehmacher}}]{VonZahn1998}
{von Zahn}, U., {Hunten}, D.~M., {Lehmacher}, G., 1998. {Helium in Jupiter's
  atmosphere: Results from the Galileo probe helium interferometer experiment}.
  J. Geophys. Res. 103, 22815--22830.

\bibitem[{{Warren}(1984)}]{Warren1984}
{Warren}, S.~G., 1984. {Optical constants of ice from the ultraviolet to the
  microwave}. Appl. Optics 23, 1206--1225.

\bibitem[{{West}(1991)}]{West1991}
{West}, R.~A., 1991. {Optical properties of aggregate particles whose outer
  diameter is comparable to the wavelength}. Appl. Opt. 30, 5316--5324.

\bibitem[{{West} et~al.(1989){West}, {Orton}, {Draine}, and
  {Hubbell}}]{West1989}
{West}, R.~A., {Orton}, G.~S., {Draine}, B.~T., {Hubbell}, E.~A., 1989.
  {Infrared absorption features for tetrahedral ammonia ice crystals}. Icarus
  80, 220--223.

\bibitem[{{West} et~al.(1986){West}, {Strobel}, and {Tomasko}}]{West1986}
{West}, R.~A., {Strobel}, D.~F., {Tomasko}, M.~G., 1986. {Clouds, aerosols, and
  photochemistry in the Jovian atmosphere}. Icarus 65, 161--217.

\bibitem[{{Wong} et~al.(2004){Wong}, {Bjoraker}, {Smith}, {Flasar}, and
  {Nixon}}]{Wong2004}
{Wong}, M.~H., {Bjoraker}, G.~L., {Smith}, M.~D., {Flasar}, F.~M., {Nixon},
  C.~A., 2004. {Identification of the 10-{$\mu$}m ammonia ice feature on
  Jupiter}. Plan. \& Space Sci. 52, 385--395.

\bibitem[{{Yang} et~al.(2000){Yang}, {Liou}, {Mishchenko}, and
  {Gao}}]{Yang2000}
{Yang}, P., {Liou}, K.~N., {Mishchenko}, M.~I., {Gao}, B., 2000. {Efficient
  Finite-Difference Time-Domain Scheme for Light Scattering by Dielectric
  Particles: Application to Aerosols}. Appl. Opt. 39, 3727--3737.

\bibitem[{{Zheng} and {Borysow}(1995)}]{Zheng1995h2h2o1}
{Zheng}, C., {Borysow}, A., 1995. {Modeling of collision-induced infrared
  absorption spectra of H2 pairs in the first overtone band at temperatures
  from 20 to 500 K}. Icarus 113, 84--90.

\end{thebibliography}

\end{document}